\documentclass[epsf]{article}
\usepackage[twocolumn]{emulateapj}
\slugcomment{To appear in the Astrophysical Journal
2002, 565; Astro-ph/0103241}

\newcommand\etal{{ et al. }}
\def\lsim{\mathrel{\rlap{\lower 4pt \hbox{\hskip 1pt $\sim$}}\raise 1pt \hbox
        {$<$}}}
\def\gsim{\mathrel{\rlap{\lower 4pt \hbox{\hskip 1pt $\sim$}}\raise 1pt \hbox
        {$>$}}}
 
\lefthead{Umeda \& Nomoto 2001}
\righthead{Nucleosynthesis of Zinc and Iron-Peak Elements 
in Pop III Type II Supernovae}

\begin{document}
\title{Nucleosynthesis of Zinc and Iron-Peak Elements 
in Pop III Type II Supernovae: Comparison with 
Abundances of Very Metal-Poor Halo Stars} 

\author{Hideyuki Umeda and Ken'ichi Nomoto }

\affil{
Research Center for the Early Universe and Department of Astronomy,  
School of Science, \\University of Tokyo, Bunkyo-ku, Tokyo,
113-0033, Japan\\ umeda@astron.s.u-tokyo.ac.jp, nomoto@astron.s.u-tokyo.ac.jp}

\begin{abstract}
     We calculate nucleosynthesis in core-collapse explosions of
massive Pop III stars, and compare the results with abundances of
metal-poor halo stars to constrain the parameters of Pop III
supernovae.  We focus on iron-peak elements and, in particular, we
try to reproduce the large [Zn/Fe] observed in
extremely metal-poor stars.  The interesting trends of the observed
ratios [Zn, Co, Mn, Cr, V/Fe] can be related to the variation of the
relative mass of the complete and incomplete Si-burning regions in
supernova ejecta.  We find that [Zn/Fe] is larger for deeper
mass-cuts, smaller neutron excess, and larger explosion energies.  The
large [Zn/Fe] and [O/Fe] observed in the very metal-poor halo
stars suggest deep mixing of complete Si-burning material and a
significant amount of fall-back in Type II supernovae.  Furthermore,
large explosion energies
($E_{51} \gsim 2$ for $M \sim 13 M_\odot$ and $E_{51} \gsim 20$ for $M
\gsim 20 M_\odot$) are required to reproduce [Zn/Fe] $\sim 0.5$.
The observed trends of the
abundance ratios among the iron-peak elements
are better explained with
this high energy (``Hypernova'') models rather than the simple ``deep'' 
mass-cut effect, because 
the overabundance of Ni can be avoided in the hypernova models.
We also present the yields of pair-instability supernova
explosions of $M \simeq 130 - 300 M_\odot$ stars, and
discuss that the abundance features of very metal-poor stars
cannot be explained by pair-instability supernovae.
\end{abstract}


\section{Introduction}

     The abundance pattern of metal-poor stars with [Fe/H] $< -2$
([A/B]$\equiv$ log$_{10}$(A/B)$-$log$_{10}$(A/B)$_\odot$) 
provides us with very important information on the formation,
evolution, and explosions of massive stars in the early evolution of
the galaxy (e.g., Wheeler, Sneden \& Truran 1989; Matteucci 2001).  
Those metal-poor stars may have been formed just a few
generations after the first generation Population (Pop) III stars or they may
even represent the second generation
(see, e.g., Weiss, Abel \& Hill 2000 for recent reviews).
Their abundance patterns may be the result of nucleosynthesis in even 
one single Type II supernova (SN II) (Audouze \& Silk 1995; Ryan, Norris,
\& Beers 1996; Shigeyama \& Tsujimoto 1998; Nakamura et al. 1999).
Therefore comparisons with nucleosynthesis patterns in 
massive metal-poor stars may help
constrain the explosion mechanism of SNe II, which is still quite
uncertain, the initial mass function (IMF) of Pop III stars, and the
mixing of ejected material in the interstellar medium.

     With the use of high resolution spectroscopic devices attached to
large telescopes, abundance measurements of extremely metal-poor stars
have become possible (e.g., McWilliam et al. 1995; Ryan et al. 1996).
The number and quality of the data is expected to increase with new
large telescopes such as SUBARU and VLT.  The observed abundances of
metal-poor halo stars show quite interesting patterns.  There are
significant differences between the abundance patterns in the iron-peak
elements below and above [Fe/H]$ \sim -2.5$.  
For [Fe/H]$\lsim -2.5$, the mean values of [Cr/Fe] and [Mn/Fe]
decrease toward smaller metallicity, while [Co/Fe] increases.

     For Zn, early observations have shown that [Zn/Fe]$ \sim 0$ for
[Fe/H] $\simeq -3$ to $0$ (Sneden, Gratton, \& Crocker 1991).
Recently 
Primas et al. (2000) have suggested that [Zn/Fe] increases toward
smaller metallicity as seen in Figure 1, and Blake et al. (2001) has 
one with [Zn/Fe] $\simeq 0.6$ at [Fe/H] = $-3.3$ (see Ryan 2001).

\begin{figure}
\hskip 3cm
\epsfxsize=8cm
\epsfbox{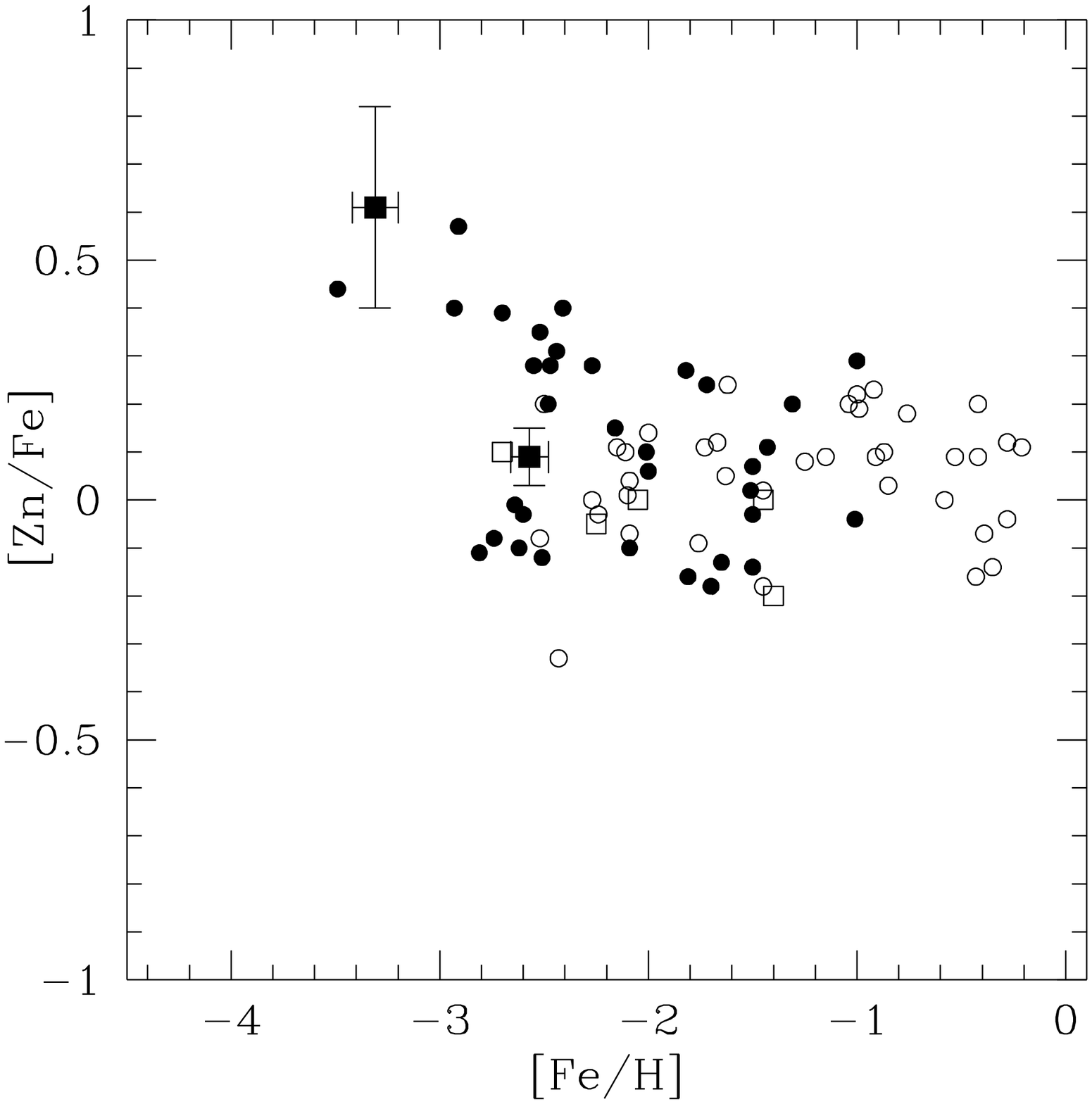}
\caption{Observed abundance ratios of [Zn/Fe].
These data are taken from Primas et al. (2000) (filled
circles), Blake et al. (2001) (filled square) and from Sneden et al. 
(1991) (others).
\label{FIG1}}
\end{figure}

\placefigure{FIG1}

     These trends could be explained with SNe II nucleosynthesis, but
progenitors and supernova explosion models are significantly
constrained.  In SNe II, stellar material undergoes shock heating and the
subsequent explosive nucleosynthesis. Iron-peak elements including Cr,
Mn, Co, and Zn are produced in two distinct regions, which are
characterized by the peak temperature, $T_{\rm peak}$, of the shocked
material. For $T_{\rm peak} > 5\times 10^9$K, material undergoes
complete Si burning whose products include Co, Zn, V, and some Cr
after radioactive decays.  For $4\times 10^9$K $<T_{\rm peak} <
5\times 10^9$K, incomplete Si burning takes place and its after decay
products include Cr and Mn (e.g., Hashimoto, Nomoto, Shigeyama 1989; 
Woosley \& Weaver 1995, WW95 hereafter; Arnett 1996;
Thielemann, Nomoto \& Hashimoto 1996).

     We have discussed, using the progenitor models for solar
metallicity (Nomoto \& Hashimoto 1988), that the decreasing trend of Mn,
Cr and the increasing trend of Co toward the lower metallicity can be
explained simultaneously if the mass-cut that divides the ejecta and
the compact remnant tends to be deeper for more massive core-collapse
SNe (Nakamura et al. 1999). This is because Mn and Cr are produced
mainly in the incomplete explosive Si-burning region, while
Co is produced in the deeper complete explosive Si-burning region. The
mass-cut is typically located somewhere close to the border of
complete and incomplete Si-burning regions. Therefore, the deeper mass-cut
leads to larger Co/Mn.

     As for Zn, its main production site has not been clearly
identified.  If it is mainly produced by s-processes, the abundance ratio
[Zn/Fe] should decrease with [Fe/H].  This is not consistent with the
observations of [Zn/Fe]$ \sim 0$ for [Fe/H] $\simeq -2.5$ to $0$ and the
increase in [Zn/Fe] toward lower metallicity for [Fe/H]$\lsim -2.5$.
 Another possible site of Zn production is explosive burning in 
SNe II.  However, previous nucleosynthesis calculations in 
SNe II appears to predict too small Zn/Fe ratio 
(WW95; Thielemann et al. 1996). 

In the application to the Galactic
chemical evolution model, it has been know that the Fe yield of the
WW95 model was too large, and thus only the Fe yield (not Zn, etc)
has been reduced by a factor of three to better fit the observations 
(Timmes, Woosley \& Weaver 1995;
Goswami \& Prantzos 2000). This makes the [Zn/Fe] ratio larger, but
still  [Zn/Fe]$\lsim 0$. Also, this procedure is not justified if
Zn and Fe are produced in the same site. Hoffman et al. (1996)
proposed another site for Zn, i.e., the neutrino-driven neutron
star wind
following the delayed explosion of a core-collapse SNe. Such a site
have been considered for a r-process site
(e.g., Woosley et al. 1994), and the Zn synthesis was found
to be very sensitive to the wind condition such as neutron excess.

     Understanding the origin of the variation in [Zn/Fe] is very important
especially for studying the abundance of Damped Ly-$\alpha$ systems (DLAs),
because [Zn/Fe] $=0$ is usually assumed, after the
work by Sneden et al. (1991), to determine their abundance pattern.
In DLAs super-solar [Zn/Fe] ratios have often been observed, but they have
been explained by assuming dust depletion is 
larger for Fe than for Zn 
(e.g., Lu et al 1996; Pettini et al. 1999; Prochaska \& 
Wolfe 1999; Molaro et al. 2000; Hou, Boissier, \& Prantzos 2001)
However, recent observations (Primas et al. 2000; Blake et al. 2001) 
suggest that the
assumption [Zn/Fe] $=0$ may not always be correct.

     In this paper, using recently calculated presupernova models
(\S2; Umeda \& Nomoto 2002; Umeda et al. 2000) we study the nucleosynthesis
pattern of iron-peak elements, focusing on Zn, in massive Pop III 
stars. We show that depending on stellar masses, mass-cuts, $Y_e$, 
\footnote{Here $Y_e \equiv \Sigma Z_j X_j/A_j $ is the 
electron mole number;
$Z_j, X_j$, and $A_j$ are respectively the atomic number, mass fraction and
mass number for each species. With $Y_e$, neutron excess is defined as
$\eta\equiv 1-2Y_e$.} 
and explosion energies, large Zn/Fe can be achieved in 
the explosive nucleosynthesis yields 
of our SNe II models of Pop III stars (\S3, 4).
Since Zn and Co are mainly produced in the same
region, their enhancement, the reduction of Mn and Cr,
and the near constancy of Ni can be
understood simultaneously with the same mechanism (\S4, 5).  
We also discuss the
reasons why previous models (Hashimoto et al. 1989; 
WW95; Thielemann et al. 1996; Nomoto et al. 1997) underproduce
Zn (\S4). Finally, we present the yields of the pair-instability
supernova explosions of $M \simeq 130 - 300 M_\odot$ stars, and
discuss that the abundance features of very metal-poor stars
cannot be explained by pair-instability supernovae (\S5, Appendix).

\section{Evolution of Pop III Massive Stars}

     The elemental abundances in metal-poor halo stars may preserve
nucleosynthesis patterns of the SN explosions of Pop III stars because
of the following reasons.  First, the IMF of very metal-poor stars ($Z
\lsim 10^{-2}-10^{-3} Z_\odot$) can be similar to Pop III stars, since
the dominant cooling mechanism for interstellar matter is almost the
same (B\"ohringer \& Hensler 1989).  Second, the nucleosynthesis
pattern in the explosions of very metal-poor Pop II stars ($Z < 10^{-2}
Z_\odot $) are similar to those of Pop III supernovae (Umeda et
al. 2000).  Also a single SN event is likely to induce star formation
for metallicity [Fe/H] $\sim -4$ to $-2$, so that the observed
metal-poor stars can be the second or very early generation (Ryan et
al. 1996; Shigeyama \& Tsujimoto 1998; Nakamura et al. 1999).

     For these reasons, we use our Pop III (metal-free) models to
compare their explosive nucleosynthesis with the observed abundances
of very low-metal stars.  We calculate the evolution of massive Pop
III stars for a mass range of $M = 13-30M_\odot$ from
pre-main sequence to Type II SN explosions.  Stellar
evolution is calculated with a Henyey-type stellar evolution code
(Umeda et al. 2000), which runs a large nuclear reaction network with
240 isotopes to calculate detailed nucleosynthesis and nuclear energy
generation. For the Pop III models we assumed no mass-loss.
SN explosions are simulated with a piecewise parabolic method code.  
The detailed nucleosynthesis during the explosion is calculated by
post-processing as in Nakamura et al. (2001), using a code of
Hix \& Thielemann (1996).

     In the metal-free star evolution, CNO elements are absent during
the early stage of hydrogen burning. Therefore, the CNO cycle does not
operate initially and the star contracts until the central temperature
rises sufficiently high for the 3$\alpha$ reaction to produce $^{12}$C
with mass fraction $\sim 10^{-10}$. Then Pop III stars undergo the CNO
cycle at a much higher central temperature ($T_{\rm c} \sim 1.5\times
10^8$ K) than metal-poor Pop II stars
(e.g., Ezer, \& Cameron 1971; Castellani, Chieffi, \& Tornamb\'e 1983).
On the other hand, the late
core evolution and the resultant Fe core masses of Pop III stars are
not significantly different from Pop II stars (e.g., WW95; Limongi,
Chieffi, \& Straniero 1998; Umeda et al. 2000).
In Table 1, we show
the ``Fe''-core masses of our model defined as a region with
the electron mole number
$Y_e \leq 0.49$ for various Z and initial masses.  
Figure 2 shows the distribution of
neutron excess, $\eta \equiv 1-2 Y_e$,
in the inner core at the beginning of collapse ($\rho_{\rm c}
\sim 3\times 10^{10}$ g cm$^{-3}$).

\begin{figure}
\plottwo{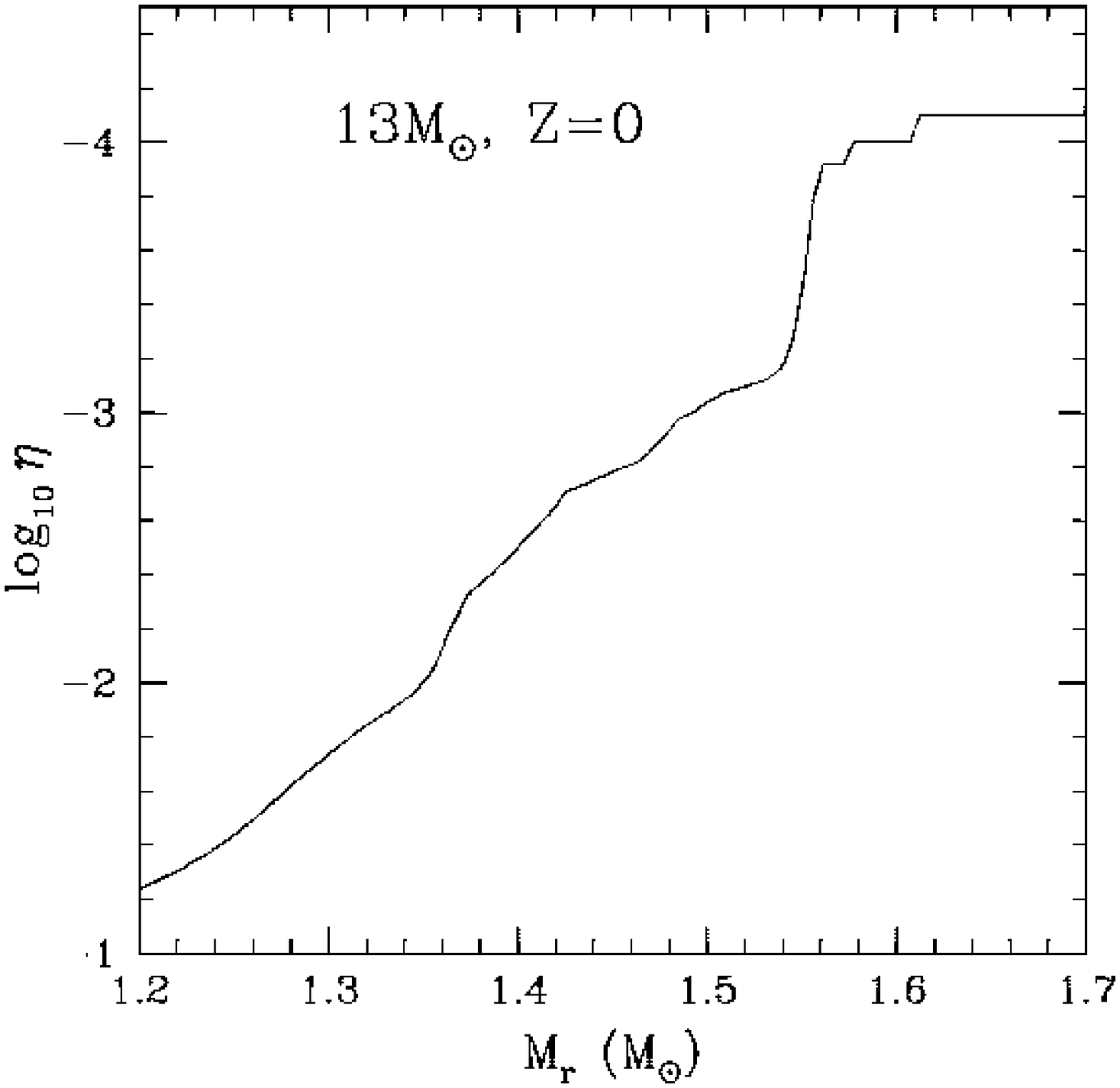}{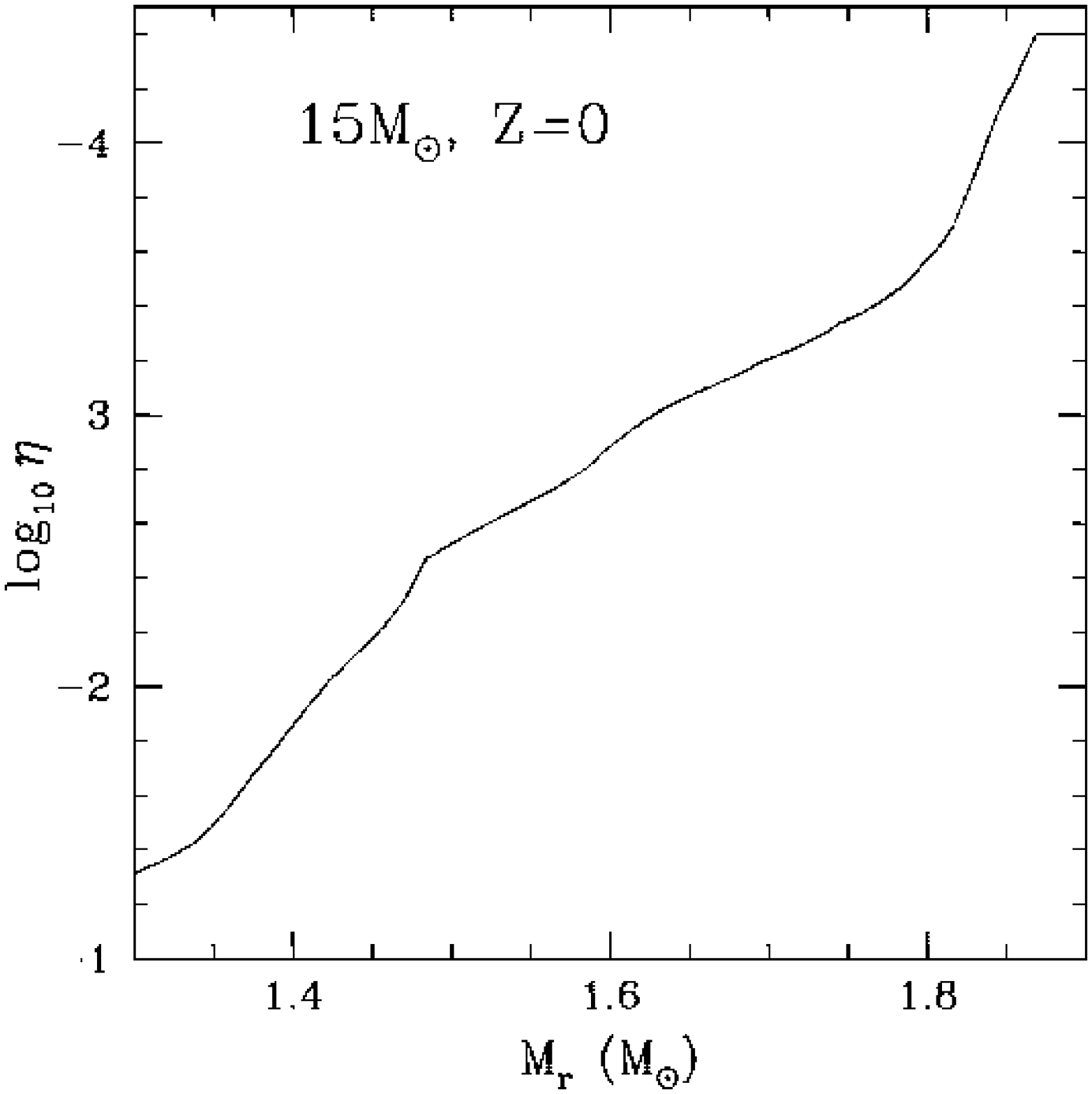}

\plottwo{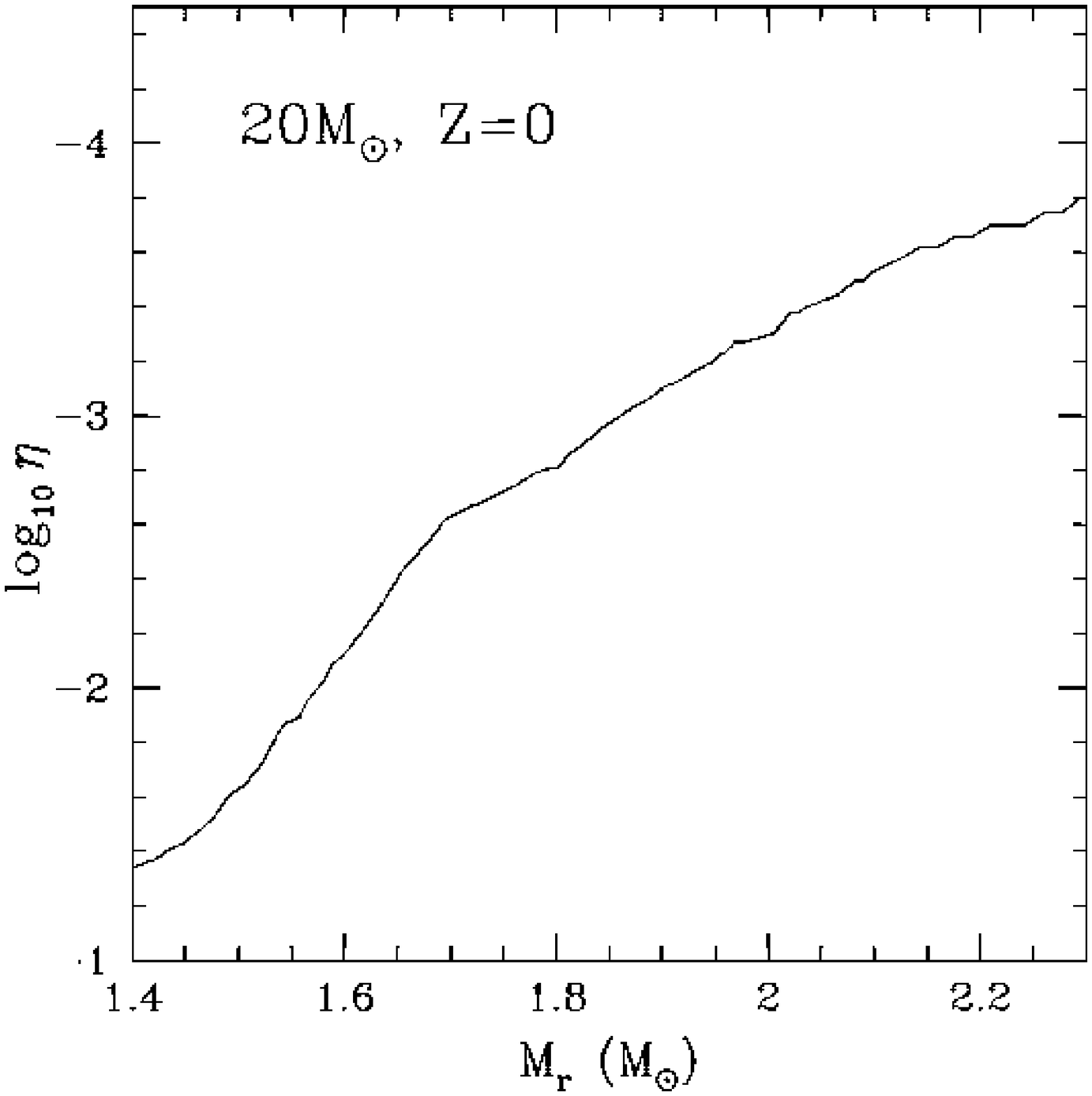}{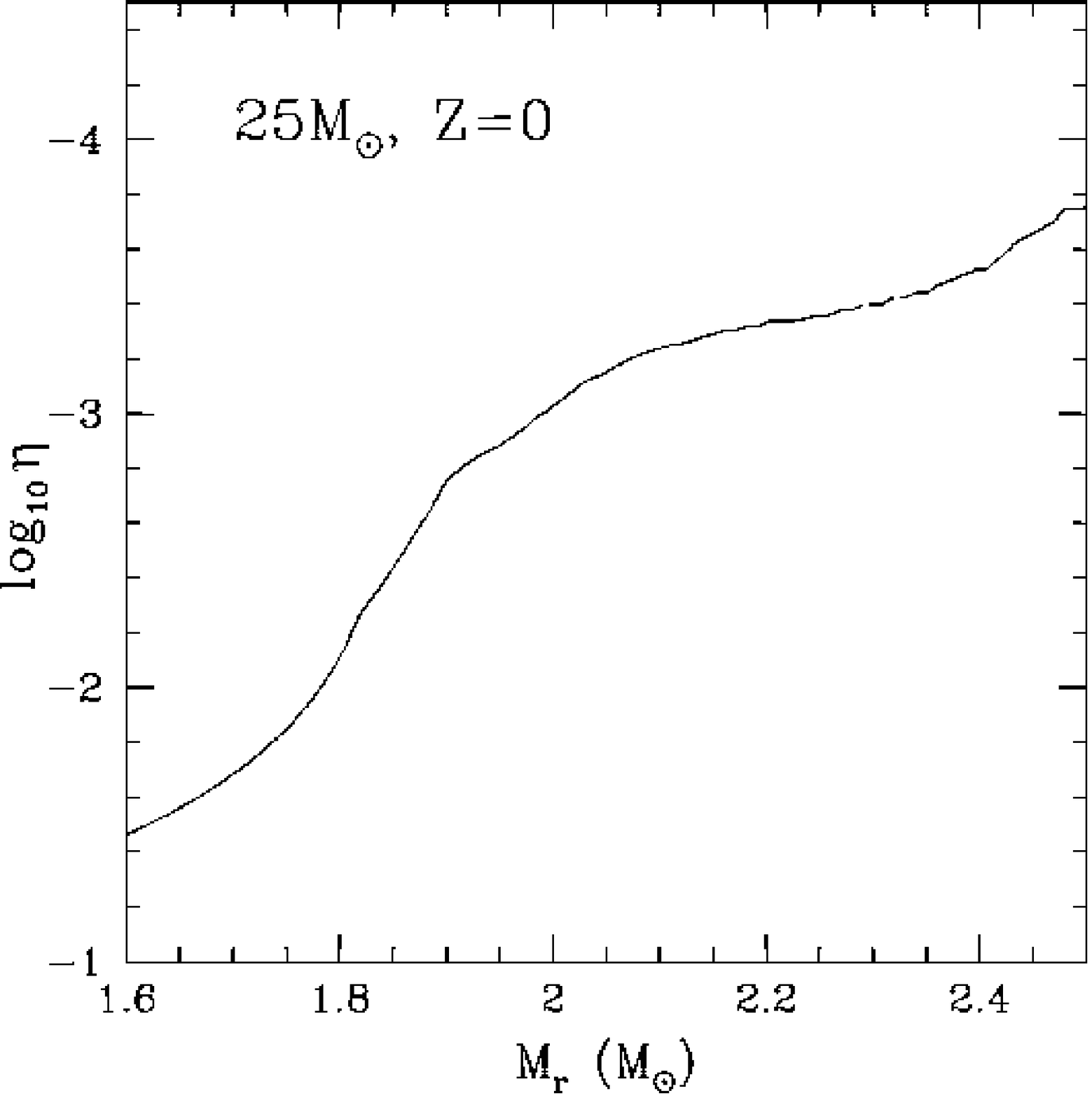}

\caption{ Distribution of 
neutron excess $\eta \equiv 1-2 Y_e$ 
in the inner core at the beginning of collapse ($\rho_{\rm c}
\sim 3\times 10^{10}$ g cm$^{-1}$)
of the Pop III pre-supernova progenitor
models with $13-25M_\odot$.
\label{FIG2}}
\end{figure}

\placefigure{FIG2}

     One of the major
uncertainties in the calculations of stellar evolution
is the treatment of convection.  
Here we use the models calculated with
relatively slow convective mixing ($f_k=0.05$ in the parameter
described in Umeda et al. 2000).  Larger $f_k$ leads to stronger
mixing of nuclear fuel, thus resulting in stronger convective
shell-burning, which leads to smaller mass core and smaller
mass compact remnants.  

\section{Synthesis of Zinc in Pop III Supernovae}

     We briefly summarize general nucleosynthesis of Pop III supernovae
obtained in Umeda et al. (2000).  The mass fraction ratio of odd- and even-Z
elements (e.g., Al/Mg), and the inverse ratio of $\alpha$-elements and
their isotopes (e.g. $^{13}$C/$^{12}$C) decrease for lower
metallicity.  However, the former ratios almost saturate for low
metallicity ($Z  \lsim 10^{-3}$) and the latter ones are difficult to
observe.  Therefore, the differences in the nucleosynthesis pattern
between the very metal-poor Pop II and Pop III stars are difficult to
observe.  The abundance ratios among the even-Z elements are almost
independent of the metallicity.  
    Dependencies on the mass and explosion energy are more important.
These results suggest that in discussing the abundance pattern of very
metal-poor stars, effects other than the metallicity are likely to
be more important.  

In the following subsections,
we discuss the dependence of the production of Zn and other iron-peak
elements on the progenitor mass, mass-cut, and explosion energy
separately. In order to clarify the site of Zn synthesis in our
models, we first show in Figure 3, the ratio of the integrated pre-supernova 
yield divided by the post-explosion yield in our 20$M_\odot$ Pop III
model ($E_{51}=E_{\rm exp}/(10^{51}$ erg) =1).  
This indicates how much of each element is made by 
s-processes before explosion or explosive Si burning.
This figure shows that elements heavier than Si are mostly
synthesized in the explosion. In particular, Zn is 
almost entirely produced by explosive Si burning.

\begin{figure}
\plotone{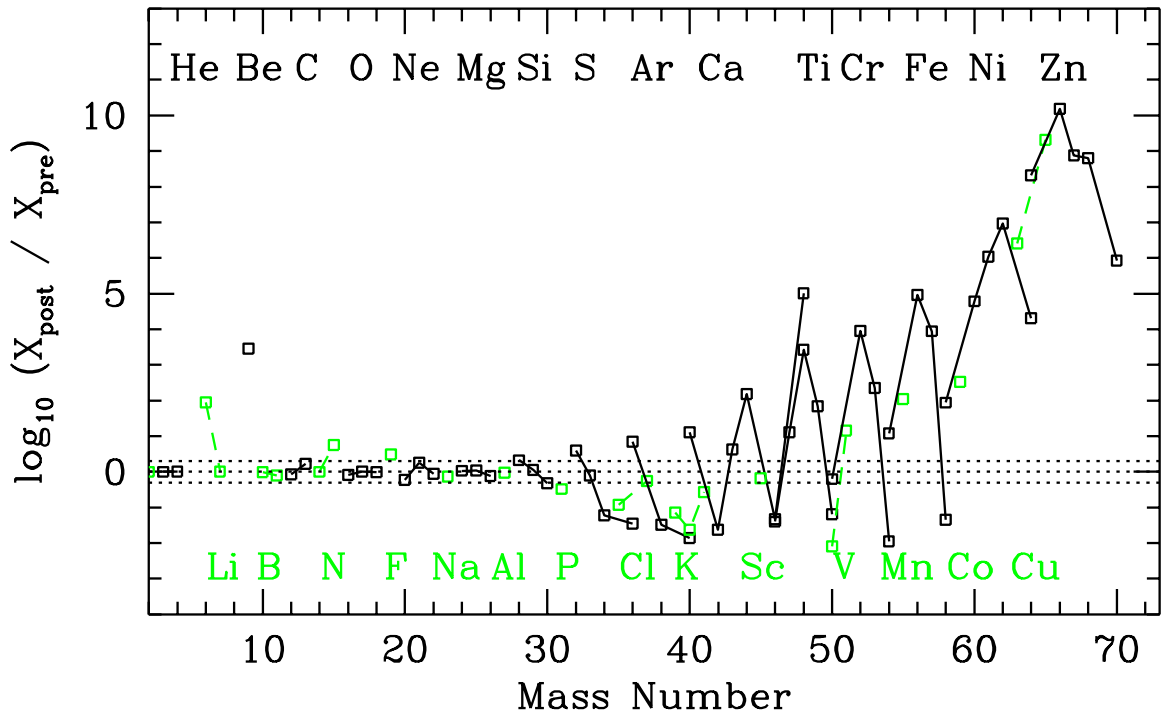}

\caption{ The ratio of the integrated pre-supernova 
yield divided by the post-explosion yield in our 20$M_\odot$ PopIII
model ($E_{51}=1$).  
\label{FIG3}}
\end{figure}

\placefigure{FIG3}

\subsection{Dependence on Mass-Cut ($M_{\rm cut}$)}

     Here we discuss the dependence of yields on the mass-cut,
$M_r=M_{\rm cut}$.  The explosion energy is assumed to be $E_{51}=
1$.  The abundance distribution 
for select species after explosion
for 13 and 15 $M_\odot$ models is shown in Figure 4.  
Also, in this figure, the regions for the complete and
incomplete Si-burnings are indicated.
The upper bounds of the complete Si-burning region and the incomplete
Si-burning region are defined by $X(^{56}$Ni) = $10^{-3}$ and
$X(^{28}$Si) = $10^{-4}$, respectively.  
The labeled elements V, Cr, Mn,
Co and Zn are the decay products of
unstable $^{51}$Mn,
$^{52}$Fe, $^{55}$Co, $^{59}$Cu, and $^{64}$Ge, respectively.

\begin{figure}
\plottwo{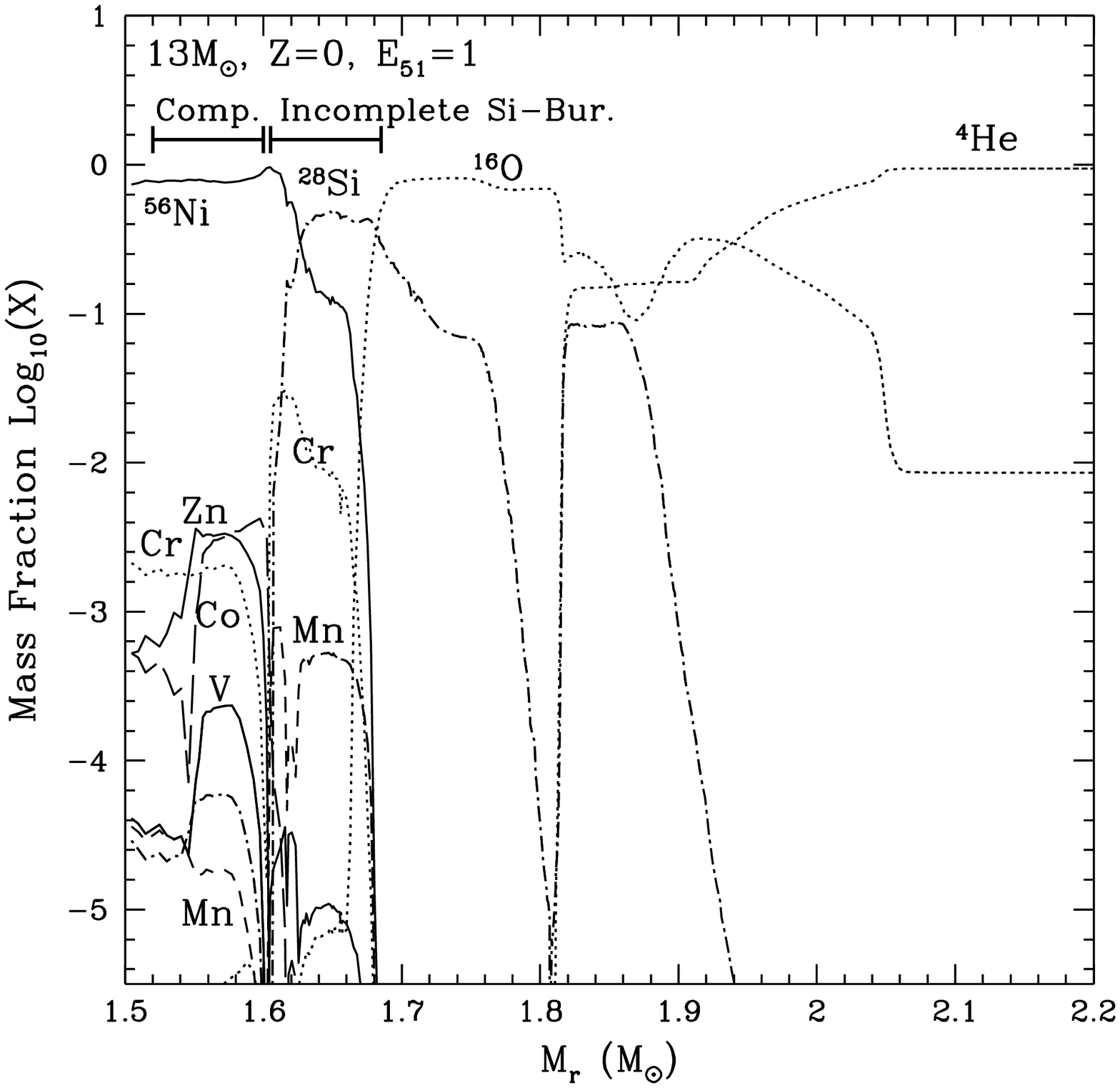}{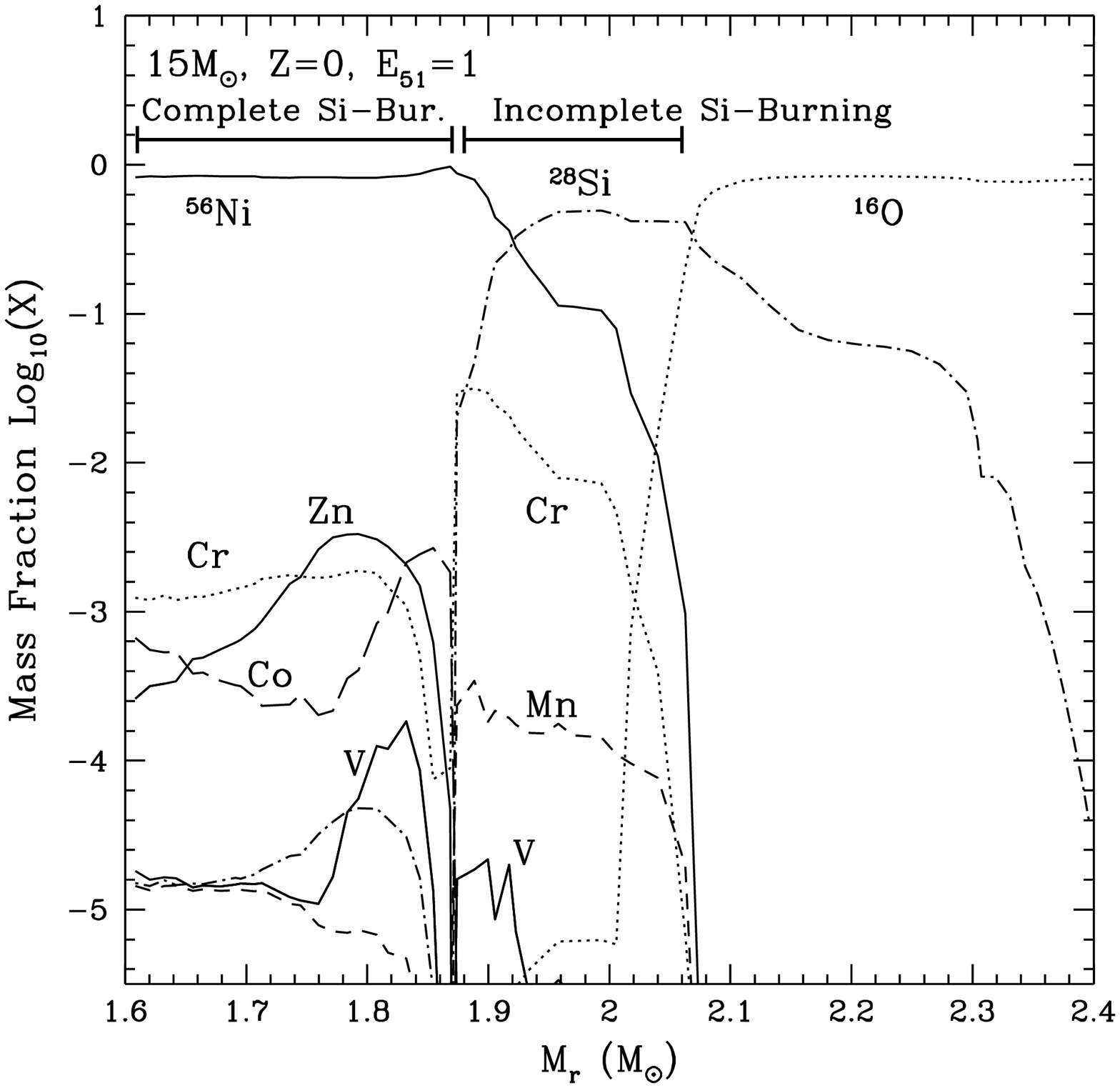}
\caption{Abundance distribution just after supernova explosion.  The
left and right panels are Pop III models with masses 13$M_\odot$ and
15$M_\odot$, respectively. The labeled elements V, Cr, Mn,
Co and Zn are the decay products of
unstable $^{51}$Mn,
$^{52}$Fe, $^{55}$Co, $^{59}$Cu, and $^{64}$Ge, respectively.
\label{FIG4}}
\end{figure}

\placefigure{FIG4}

     In the ejecta, the mass fraction of the complete Si burning
products is larger if the mass-cut is deeper (i.e., $M_{\rm cut}$ is
smaller).  Mn and Cr are produced mainly in the incomplete explosive
Si-burning region, while Co and Zn are mainly produced in the deeper
complete explosive Si-burning region.  Therefore, if the mass-cut is
deeper, the abundance ratios Co/Fe and Zn/Fe increase, while the
ratios Mn/Fe and Cr/Fe decrease as seen in Figure 5.

\begin{figure}
\plottwo{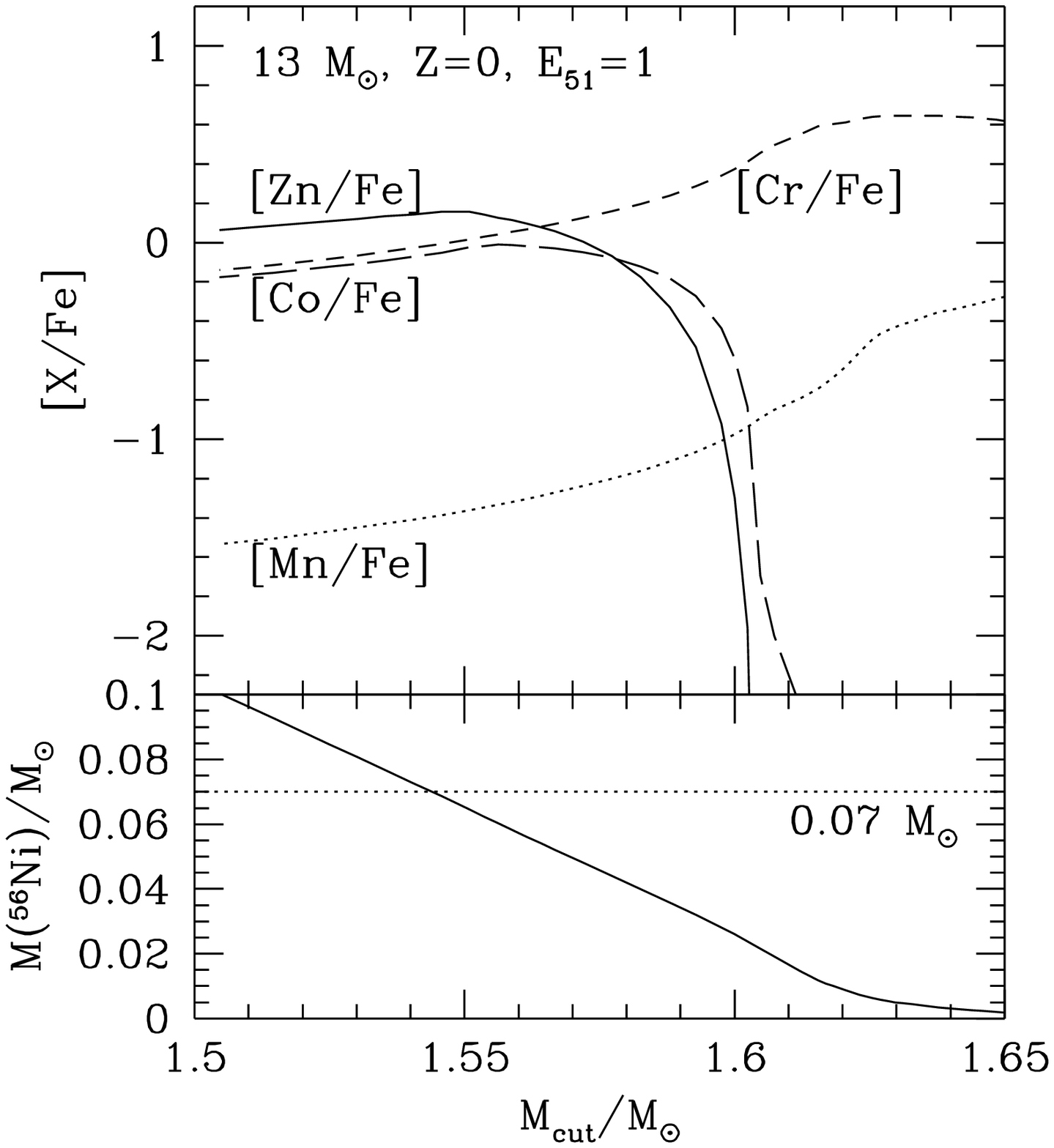}{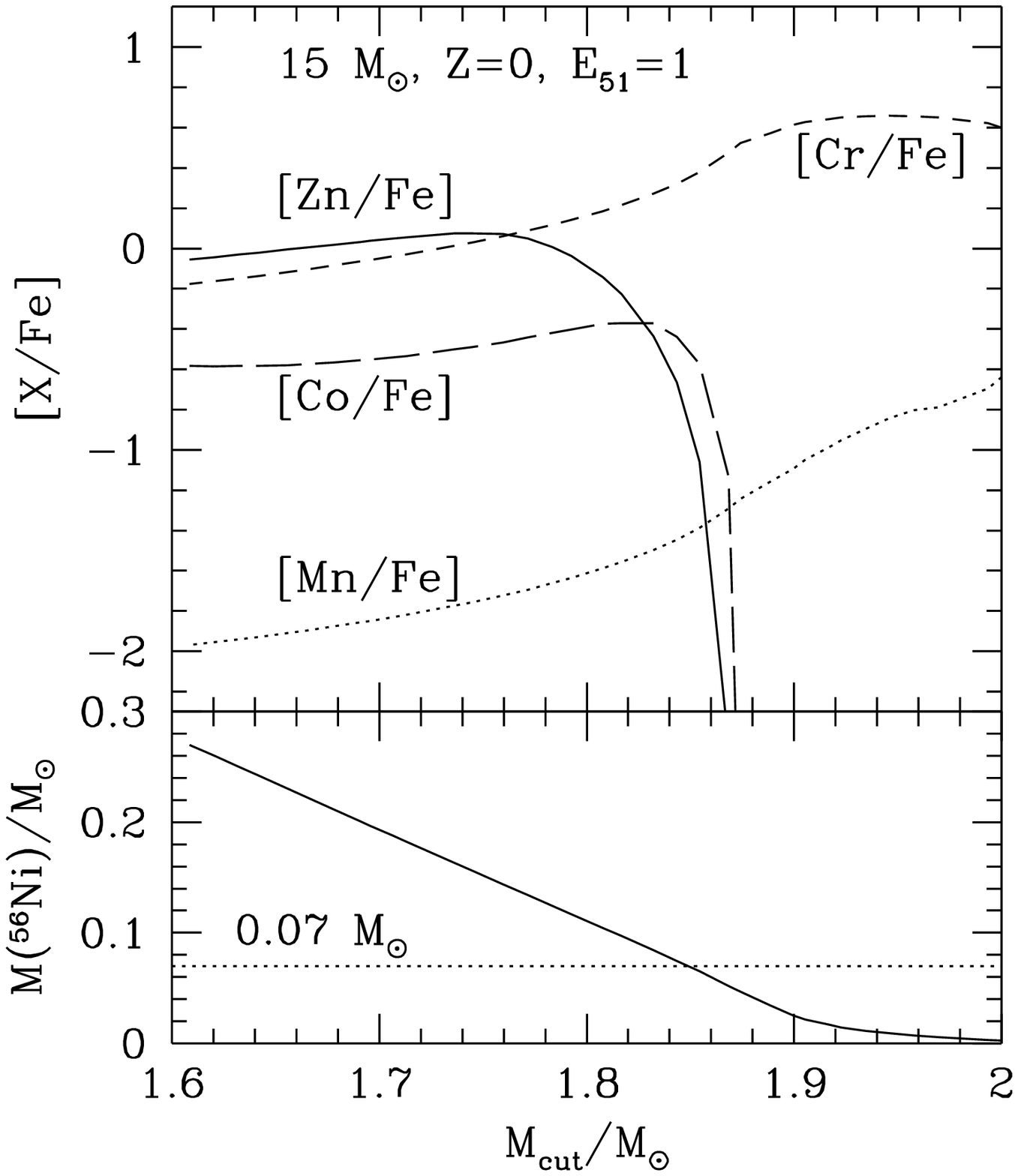}

\plottwo{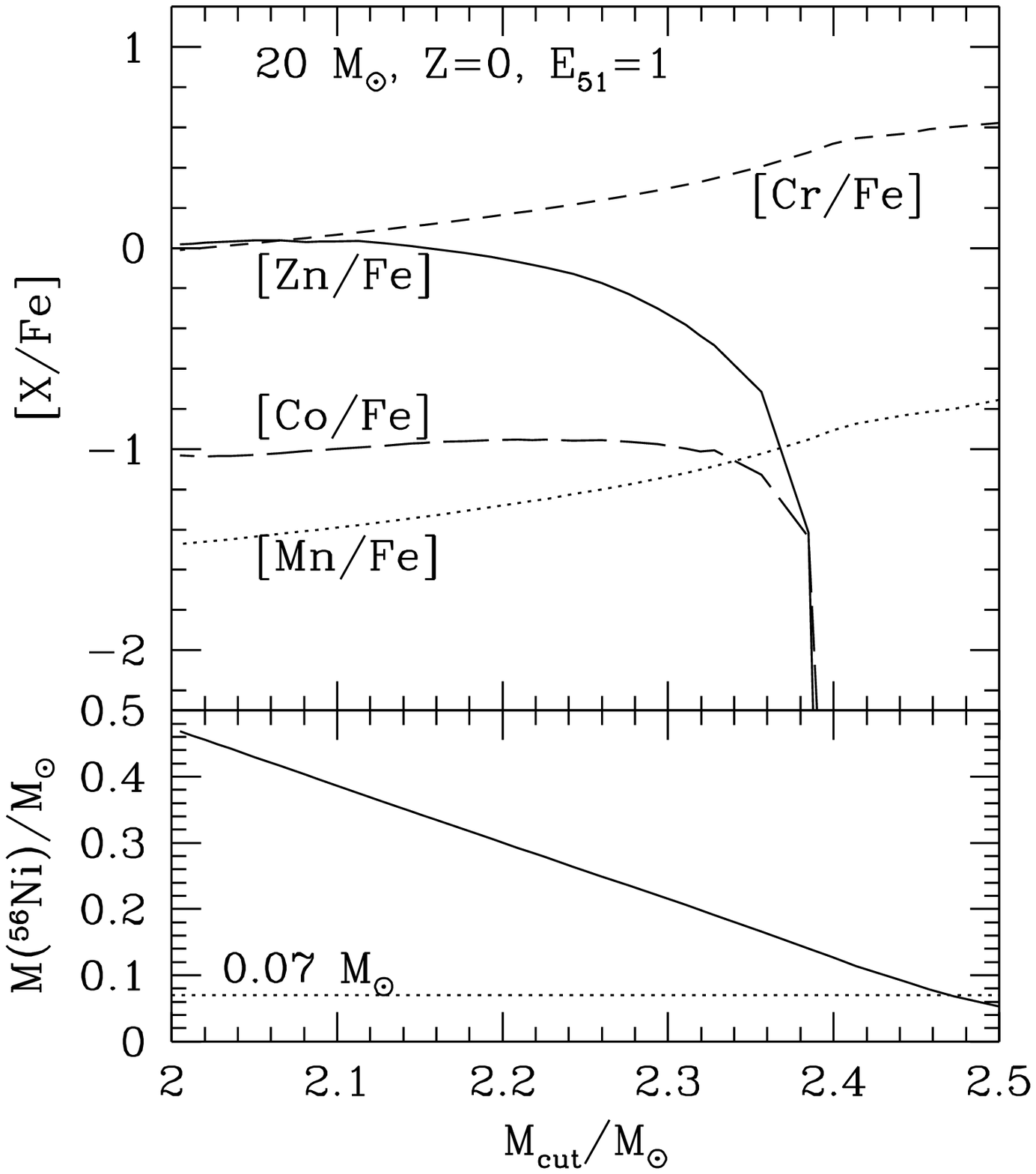}{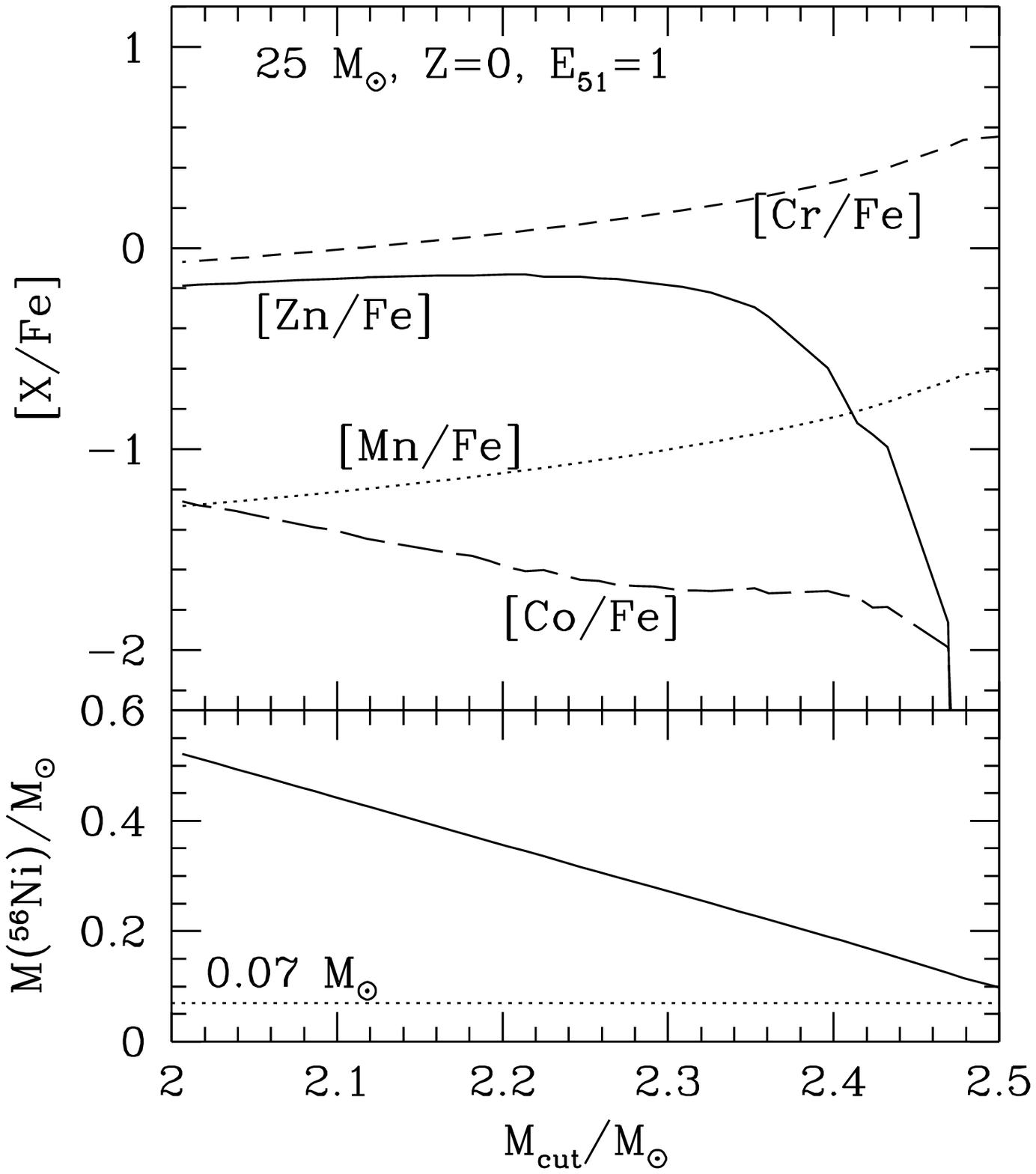}
\caption{
Dependence of abundance ratios on the mass coordinate at the mass-cut,
$M_{\rm cut}$, for Pop III SNe II with $E_{51}=1$.
\label{FIG5}}
\end{figure}

\placefigure{FIG5}

     Figure 5 shows that larger [Zn, Co/Fe] and smaller [Cr, Mn/Fe] can be
achieved simultaneously for a smaller $M_{\rm cut}$.  In the 13 $M_\odot$
model for $M_{\rm cut} \sim 1.55M_\odot$, for example, [Zn/Fe] is
large enough to be consistent with the observed ratio 
[Zn/Fe]$\sim 0.1$ for stars with [Fe/H]$\gsim -2$.
More massive stars can also yield [Zn/Fe]$\gsim 0$ if the mass-cut is deep
enough. The observed large ratio [Zn/Fe]$\sim 0.5$ in very
metal-poor stars can only be achieved with the combination of deep
mass-cut and large explosion energy (\S 3.4). 

     Another effect of the mass cut on the yield is the $^{56}$Ni mass
in the ejecta $M(^{56}$Ni), which is larger for smaller $M_{\rm cut}$
as shown in Figure 5.  Ejection of the large amount of radioactive
$^{56}$Ni has actually been seen in such bright supernovae as SNe
1997ef and 1998bw (e.g., Nomoto et al. 2000).  We note that for
$M\gsim 15M_\odot$, the
$M(^{56}$Ni) required to get [Zn/Fe]$\sim 0.5$ appears to be too large
to be compatible with observations of [O/Fe]$ \sim 0 - 0.5$ in
metal-poor stars.  However, if fall-back 
of a large enough amount of iron-peak
elements occurs after mixing, $M(^{56}$Ni) can be smaller without
changing the [Zn/Fe] ratio, as will be discussed in \S 3.5.

\bigskip
\bigskip
\bigskip

\subsection{Dependence on Stellar Mass}

     Figure 5 shows that the relation between [X/Fe] and $M(^{56}$Ni)
is sensitive to the progenitor mass.  To understand this behavior, we
summarize in Table 2 the location of the incomplete Si-burning region
in $M_r$ and $M(^{56}$Ni) contained in this region for several models.
The thickness in mass of the incomplete Si-burning region
(and thus $M(^{56}$Ni) there) is larger for larger progenitor masses.
Suppose that all SNe II eject the same amount of $^{56}$Ni. Then the
fraction of complete Si burning products is larger for less massive
stars because the incomplete Si-burning region is thinner.  This is why
the 13$M_\odot$ model yields larger [Zn, Co/Fe] than more massive
models if for example $M(^{56}$Ni) = $0.07M_\odot$ (Fig. 5).  
Models heavier than 15$M_\odot$ are consistent with very metal-poor
star data if $M_{\rm cut}$ is small (i.e., $M(^{56}$Ni) is
large).

      The maximum values of [X/Fe] as a function of 
$M_{\rm cut}$ also depend on
the stellar mass because of the different density - temperature
histories and $Y_e$ distribution of the progenitors.  There is a
tendency that the maximum [Zn, Co/Fe] decreases with increasing stellar
mass.  One of the reasons for this trend is that more massive stars
have thicker incomplete Si-burning region, and thus deeper material
must be ejected to make [Zn, Co/Fe] large.  As discussed in the next
subsection, in the deeper regions $Y_e$ is typically smaller,
which leads also to smaller Zn and Co mass fractions.
On the other hand, there is no clear
mass dependence for [Mn, Cr/Fe].

\subsection{Dependence on $Y_e$}

     In order to see the dependence on $Y_e$, we compare
the post-explosive abundance distribution 
of the 25$M_\odot$ ($E_{51}$=1) models in Figure 6.
The left panel shows the original model, whose $Y_e$ distribution is
enlarged in Figure 7.  In the right panel $Y_e$ is modified to be 0.5
at $M_r \leq 2.5M_\odot$.  As shown in these figures, Zn and Co
abundances are very sensitive to $Y_e$.  If $Y_e$ decreases 
below $Y_e\simeq 0.4998$, the Zn
abundance becomes significantly smaller, because
Zn is the decay product of the symmetric species $^{64}$Ge.  The
dependence of the Co abundance on $Y_e$ is not monotonic but rather
complicated, though the Co abundance is larger in the $Y_e=0.5$ model
of this example. [X(Co) is relatively large for $Y_e=0.5$.  
With lowering $Y_e$, it
decreases once but increases again toward $Y_e \simeq
0.49$.]  Previous progenitor models in Nomoto \& Hashimoto (1988) have
much lower $Y_e$ than our current models in the complete Si-burning
region because of the different initial metallicity and
treatment of convection, and this is
the main reason why those models significantly underproduce Zn even
for the 13$M_\odot$ model. 

\begin{figure}
\plottwo{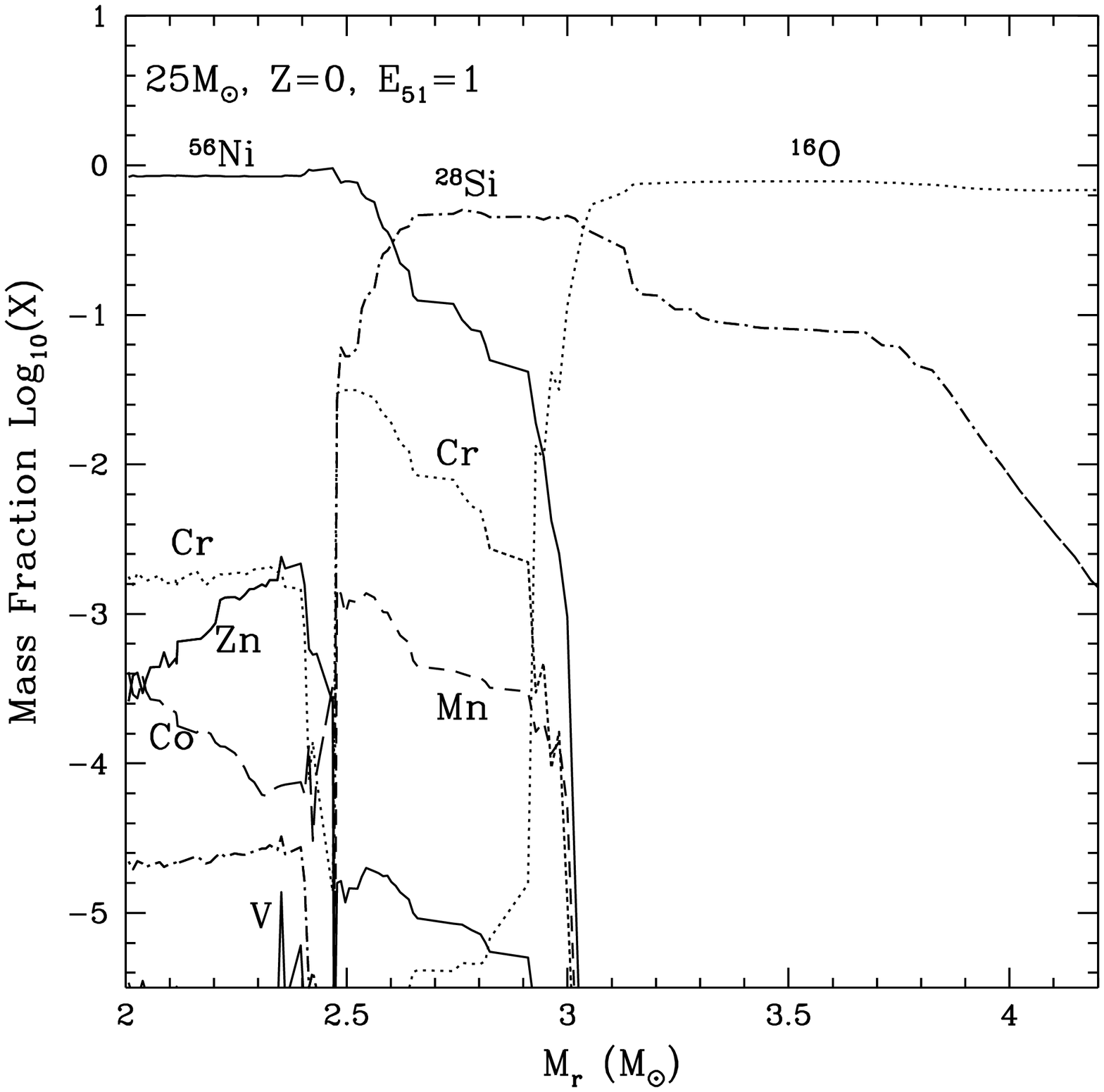}{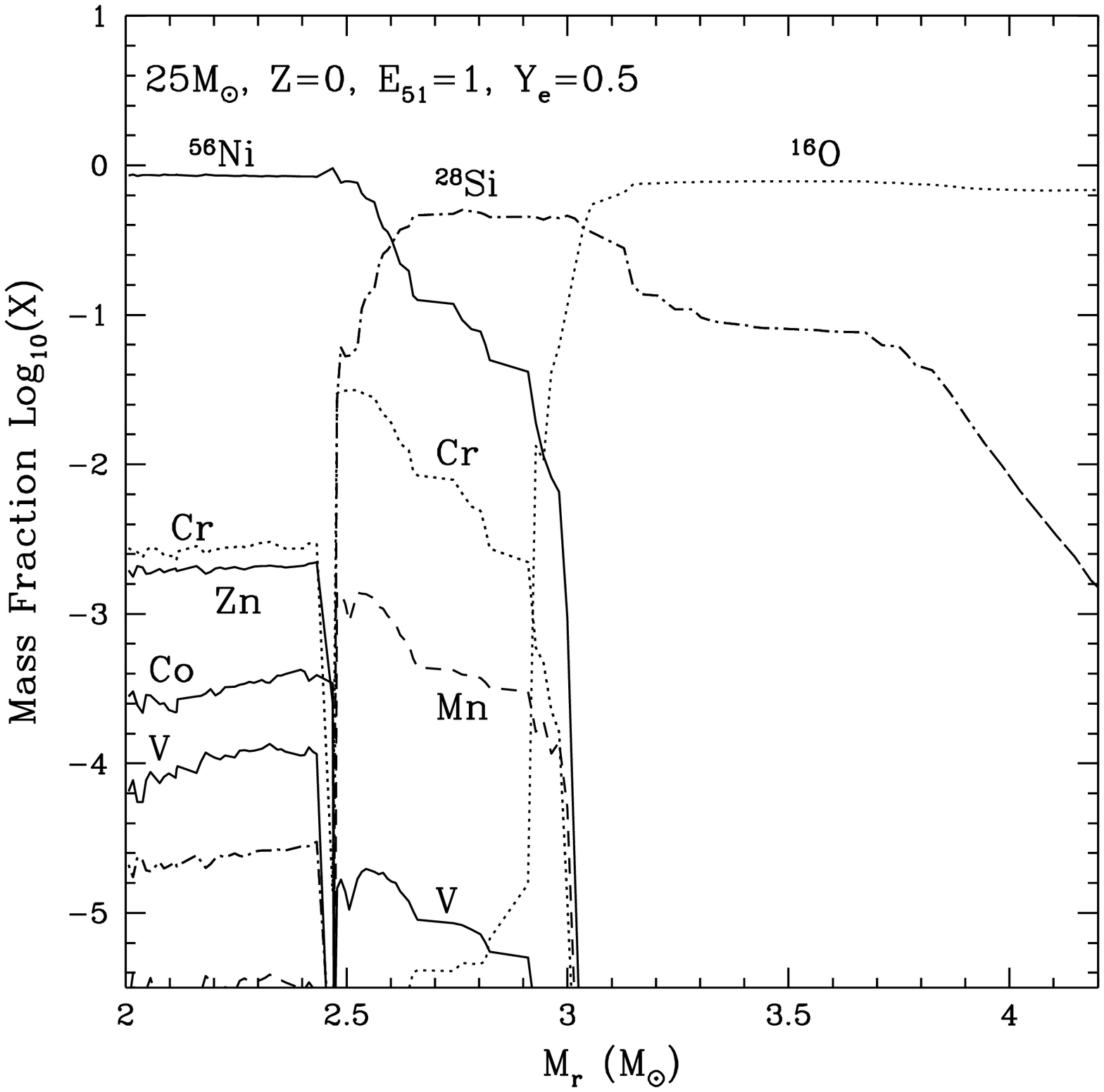}
\caption{Abundance distribution after supernova
explosions of Pop III 25$M_\odot$ stars with $E_{51}=1$. 
The left panel is the original model.  In the
right panel $Y_e$ is modified to be 0.5 below $M_r=2.5M_\odot$.
\label{FIG6}}
\end{figure}

\placefigure{FIG6}

\begin{figure}
\plottwo{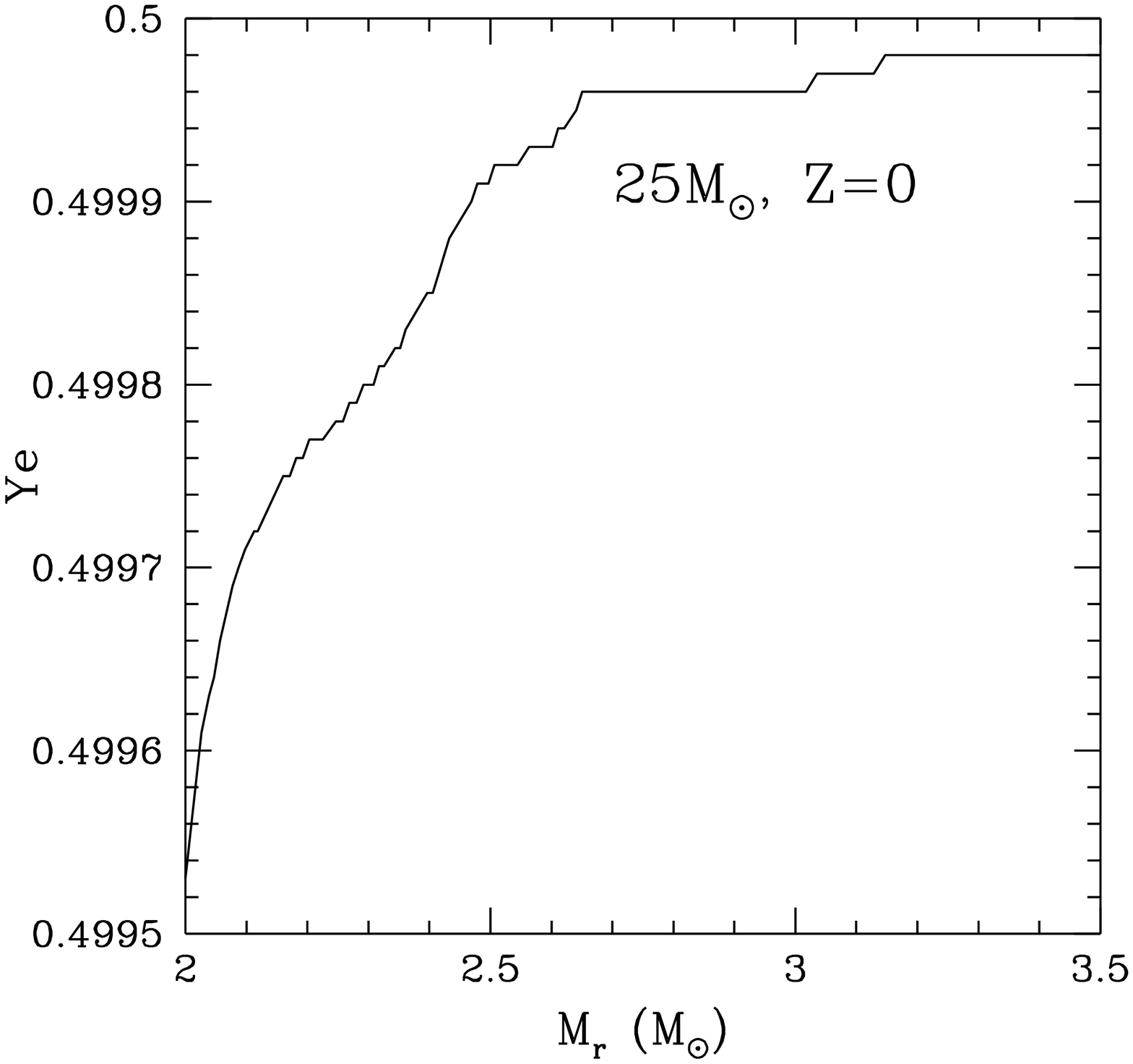}{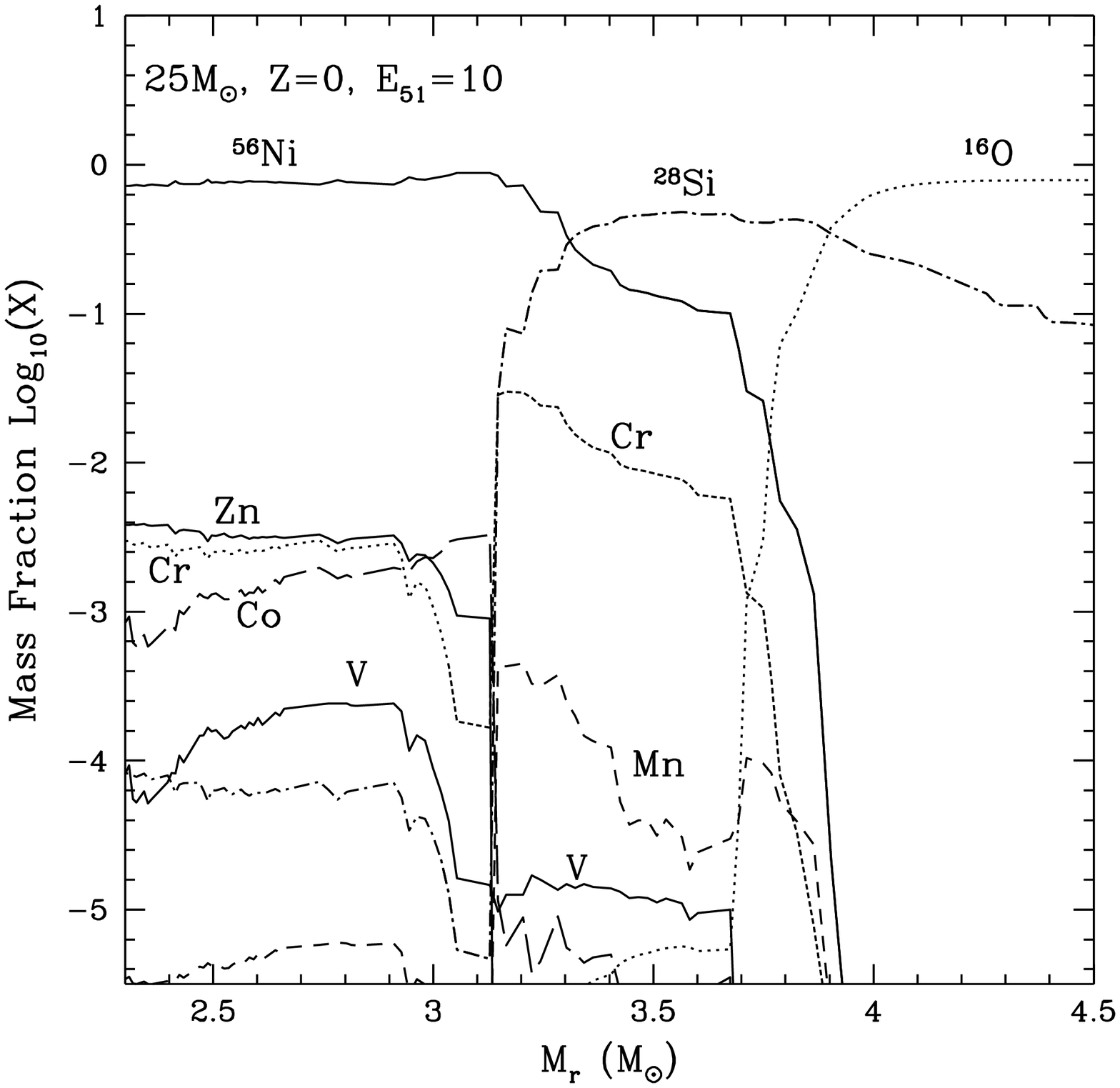}
\caption{(Left) The distribution of $Y_e$ in the Pop III 25$M_\odot$
model shown in Figure 6. 
\label{FIG7}}

\caption{(Right) Abundance distributions after the supernova explosion of a
Pop III 25$M_\odot$ star with $E_{\rm exp}=10^{52}$ erg. 
\label{FIG8}}

\end{figure}

\placefigure{FIG7}

     In our present progenitor models (as well as in Nomoto \& Hashimoto
1988), we have applied electron-capture rates by Fuller, Fowler, \&
Newman (1980, 1982).  The use of the recent rates by Langanke \&
Mart\'inez-Pinedo (2000), which are lower than those of Fuller et al. (1980,
1982) would lead to larger $Y_e$ (Heger et al. 2000). 
Mezzacappa et al. (2000) and Rampp \& Janka (2000) have recently 
performed simulations of SNe II with full Boltzmann neutrino
transports, and they have 
shown that $Y_e$ in the deep layers may be enhanced up to $\sim$ 0.5 by
neutrino processes.  
In this case, Zn production may be significantly enhanced over
what has been calculated in previous models.
We should note that $X$(Zn) can be large even if
$Y_e\lsim 0.4995$ for energetic explosion as will be discussed in the
next subsection.

\subsection{Dependence on Explosion Energy}

     Recent observations suggest that at least some core collapse
SNe explode with large explosion energies, which may be called
``Hypernovae'' (e.g., Galama et al. 1998;
Iwamoto et al. 1998, 2000; Nomoto et al. 2000).  
These SNe likely originate
from relatively massive SNe ($M \gsim 25M_\odot$).

     In Figure 8 we show nucleosynthesis in the 25$M_\odot$ star with
the explosion energy of $10^{52}$ erg. By comparing with Figure
6, we find that for larger explosion energies, the boundaries of both the
complete and incomplete Si burning regions move outward in mass
coordinates.  We also find that for a larger explosion energy the
incomplete Si burning region is thicker in mass, thus containing a
larger amount of $^{56}$Ni (Table 2).  

\placefigure{FIG8}

   So, we may expect that hypernova explosions, which eject a large amount of
complete Si burning products, also produce a larger amount of
$^{56}$Ni than ordinary SNe II, unless significant fall-back takes
place after mixing (see \S 3.5).  This is seen in Figure 9, which shows
that [Zn/Fe] $\gsim 0.3$ only for $M(^{56}$Ni) $\gsim 0.7 M_\odot$ (i.e.,
very bright supernovae).

\begin{figure}

\plottwo{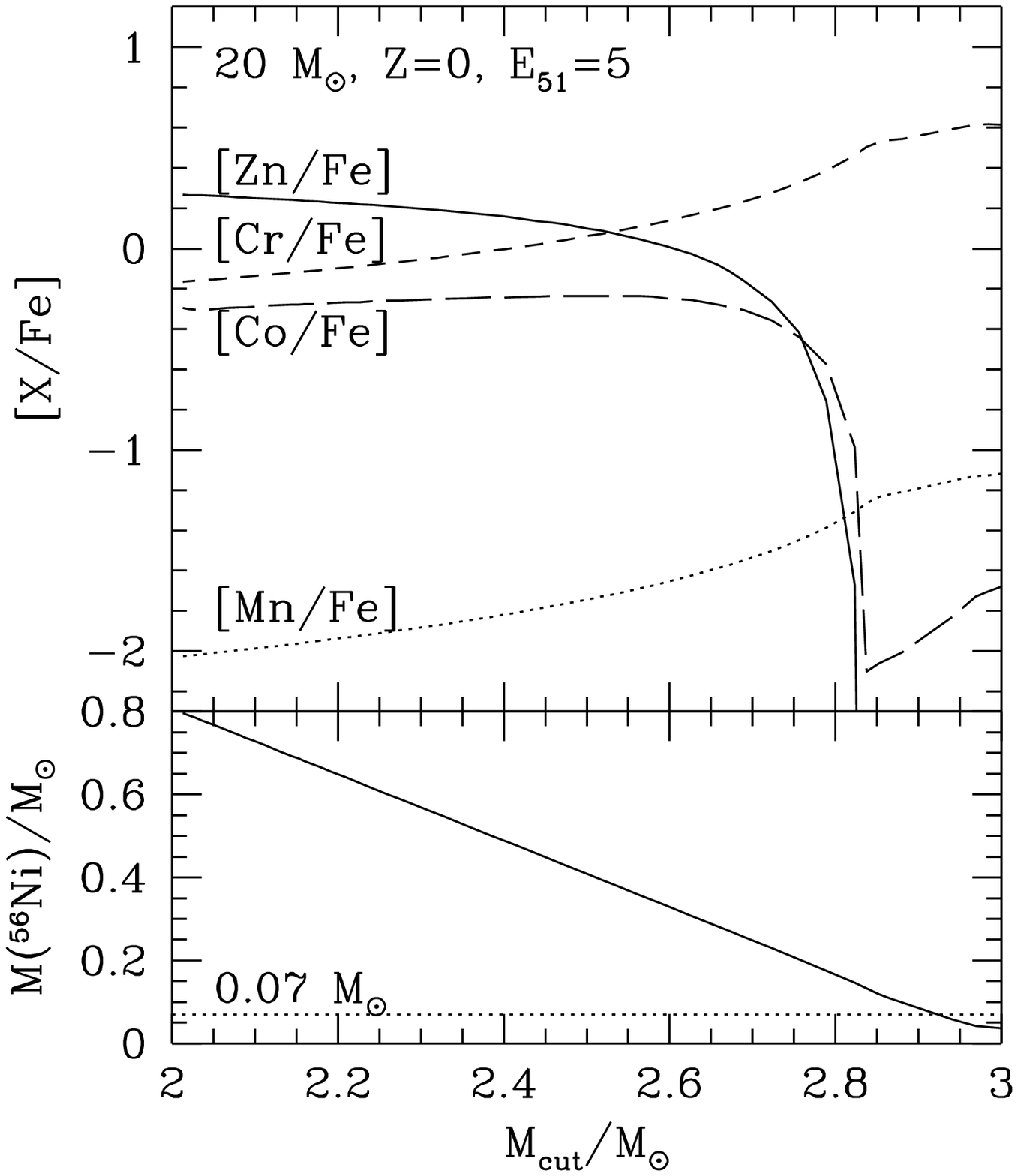}{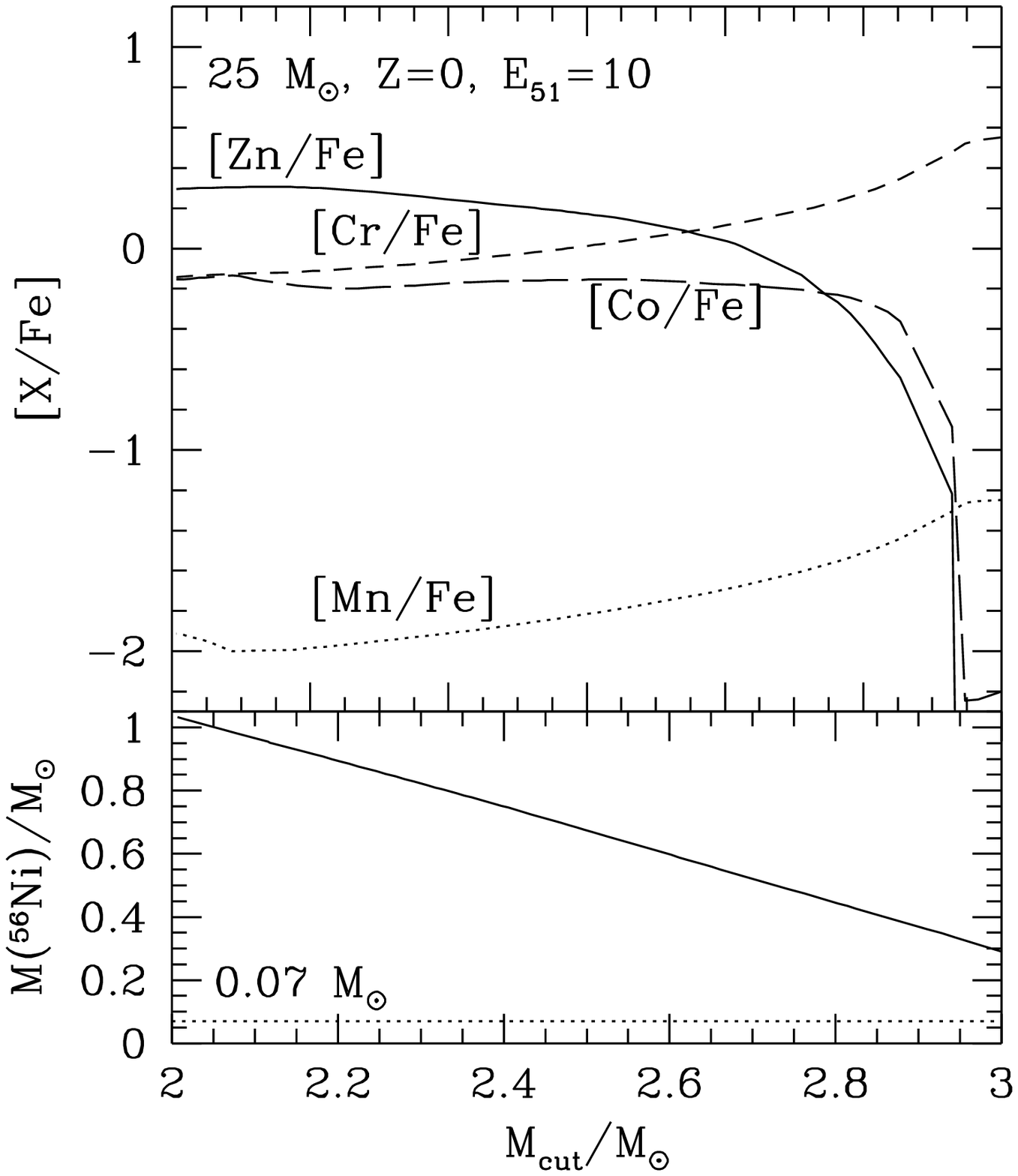}

\plottwo{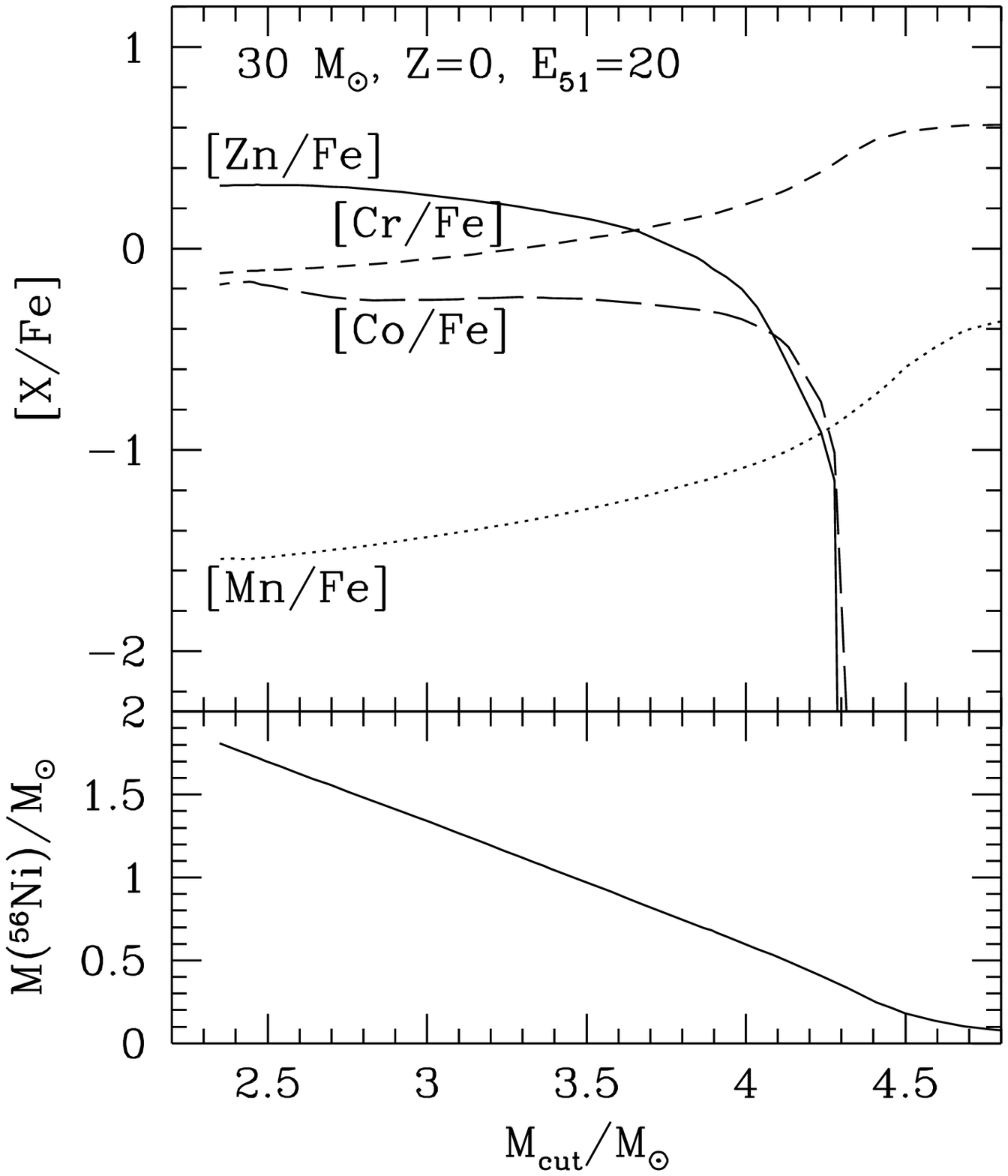}{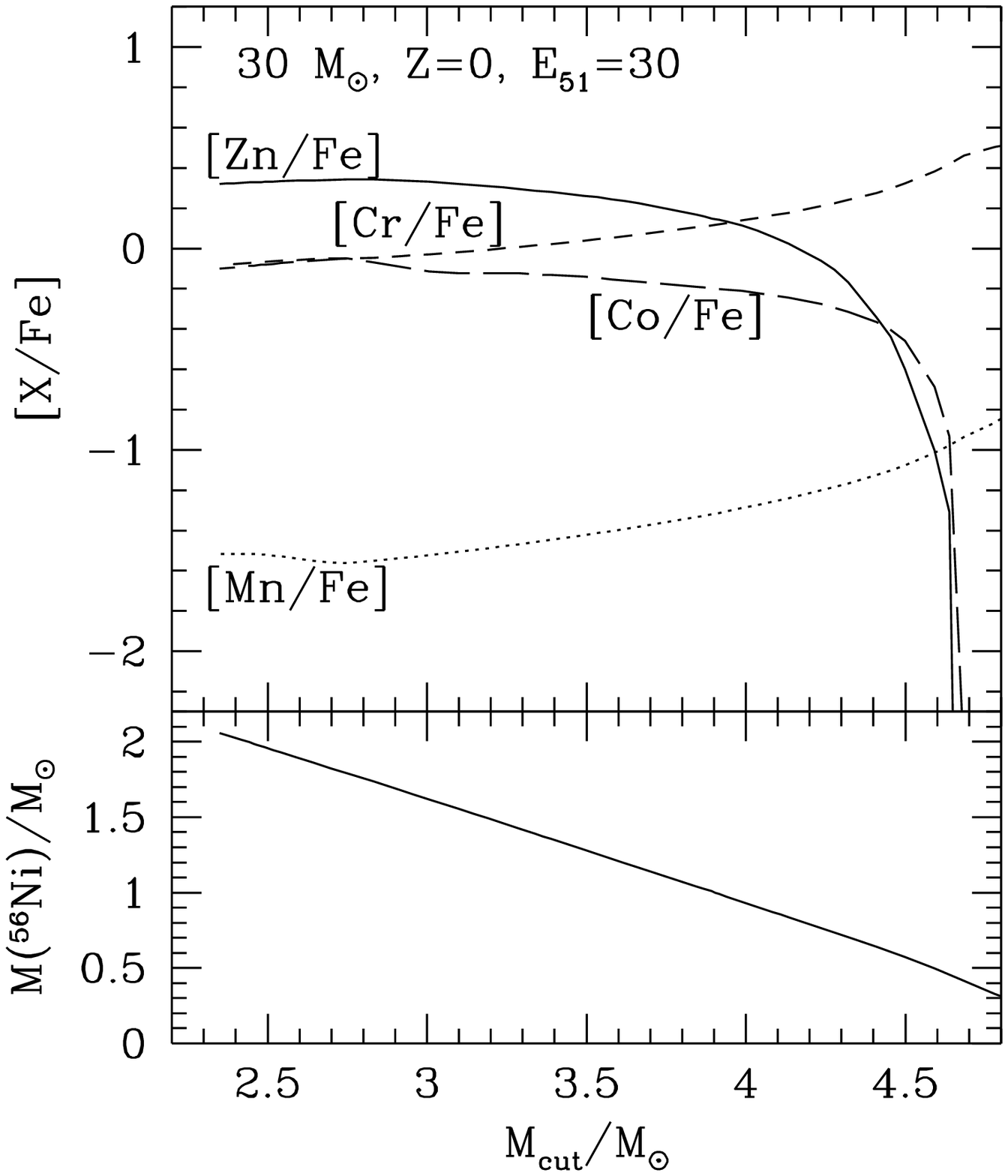}
\caption{Same as Figure 5, but for the energetic SNe II of $M = 20 -
30 M_\odot$.
\label{FIG9}}
\end{figure}

\placefigure{FIG9}

     The explosion energy also affects the local mass-fraction of
elements. For more energetic explosions, the temperature during the
explosion is higher for the same density (Nakamura et
al. 2001).  Figures 6 and 8 show that for a higher energy $X$(Co) and
$X$(Zn) are enhanced in complete-Si burning 
and $X$(Mn) is reduced in incomplete-Si burning. 
$X$(Cr) in incomplete-Si burning is
almost unchanged.

\begin{figure}
\epsfxsize=8cm
\hskip 2cm
\epsfbox{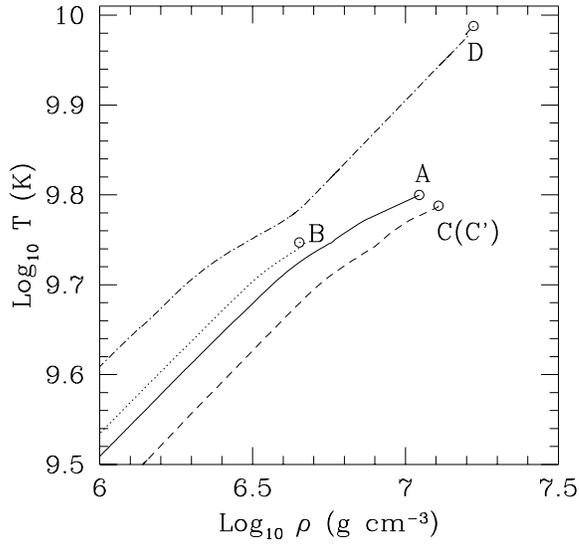}
\caption{ Density - temperature tracks during explosive Si-burning
for representative cases.
Here, the parameters of Cases A, B, C, C' and D are summarized in
Table 3.
\label{fig9}}
\end{figure}

     In order to show the parameter dependences of $X$(Zn), we plot in
Figure \ref{fig9} the density - temperature track during explosive
complete Si-burning for four representative cases following the maximum
temperature (open circles).  The parameters of the four models and
the mass fractions of $^{56}$Ni and Zn are summarized in Table
3. Cases B \& D produce large $X$(Zn), while A \& C are the
cases with smaller $X$(Zn).  Case C$'$ is the same as case C except
for the modifications of $Y_e$.  The general trend is that $X$(Zn)
is larger if $Y_e$ is closer to 0.5 and an explosion is more energetic
to produce larger specific radiation entropy 
$(4a/3) (T^3/\rho)$, where $a$ denotes the radiation constant.  
Cases A and C have low $X$(Zn) because
of relatively low $Y_e$.  Case C$'$ has the same density - temperature
history as case C, but it yields larger $X$(Zn) due to the larger
$Y_e$.  Case D has the same $Y_e$ as case C, but it yields larger
$X$(Zn) due to larger $T^3/\rho$. Case B yields large $X$(Zn) because
of the relatively large $Y_e$ and $T^3/\rho$.

\placefigure{fig9}

\begin{figure}
\epsfxsize=8cm
\hskip 2cm
\epsfbox{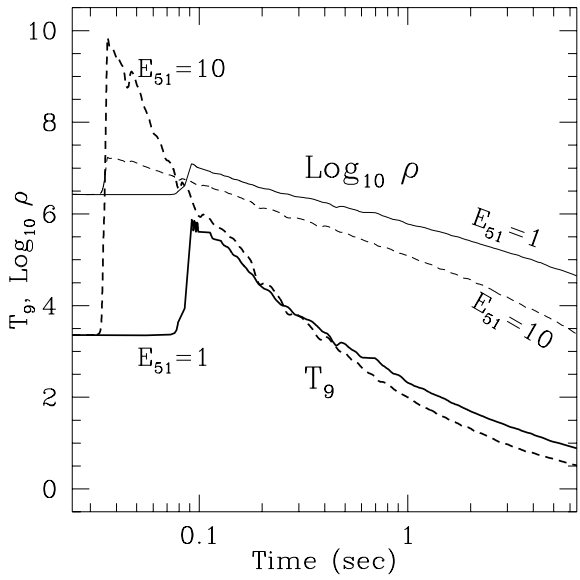}
\caption{Density and Temperature evolution for the 
cases C, C$'$ ($E_{51}$=1) and
D ($E_{51}$=10) in Figure \ref{fig9}. Here $T_9 = T/10^9$(K), and
$\rho$ in $g cm^{-3}$.
\label{fig10}}
\end{figure}

\begin{figure}
\plotone{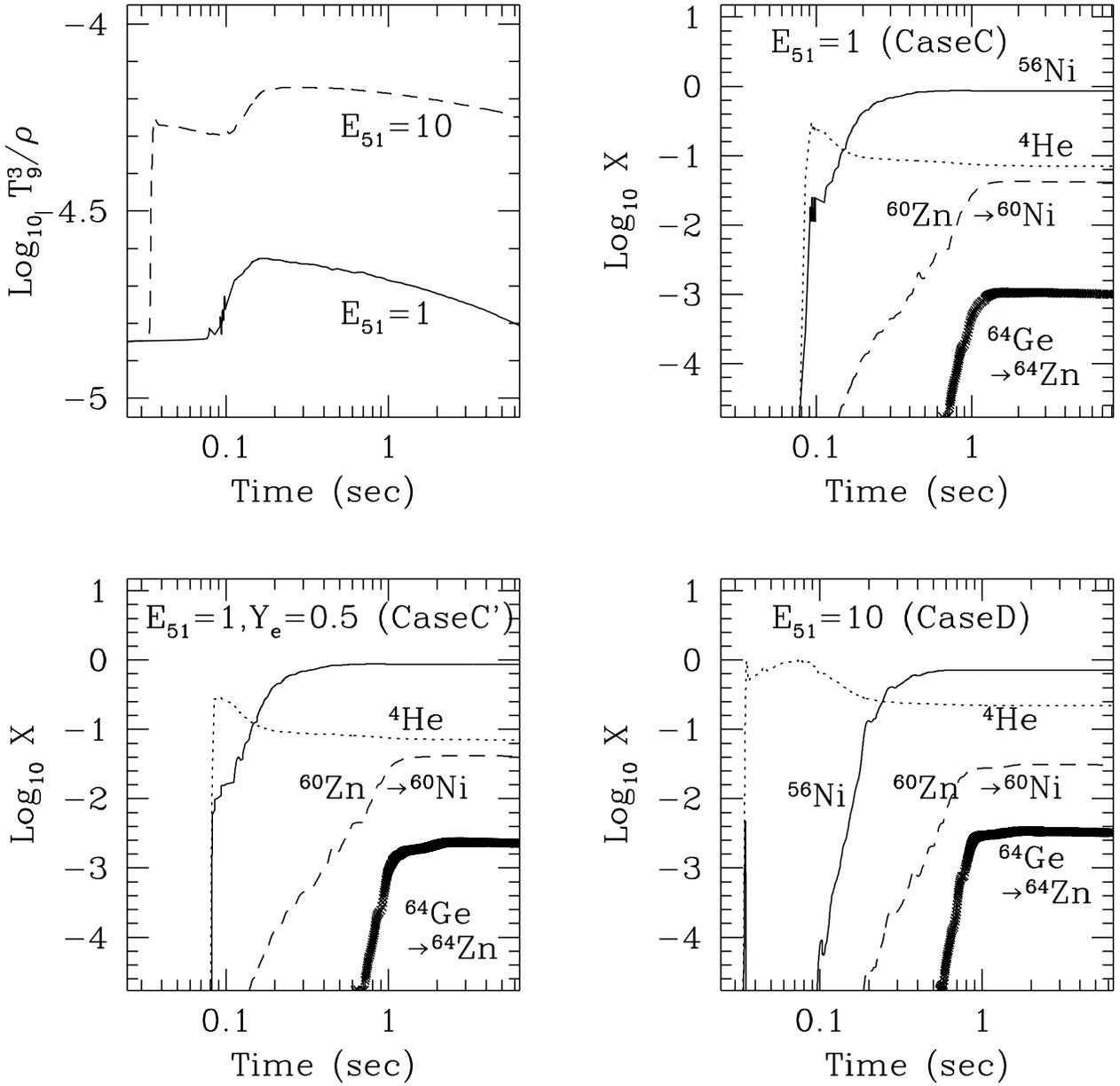}
\caption{Time evolution of $T^3/\rho$ for the cases 
C, C$'$ ($E_{51}$=1) and D ($E_{51}$=10) in Figure \ref{fig9} (left top panel),
and time evolution of mass fraction ratios of some elements most relevant
to Zn synthesis for these cases. \label{fig11}
}
\end{figure}

     In Figures \ref{fig10} and \ref{fig11}, we show the time
evolution of density, temperature, 
$T^3/\rho$, and mass fraction
ratios of some elements that are relevant to Zn synthesis for cases C
(C$'$) and D.  These figures show that for larger $T^3/\rho$ the mass
fraction of $^4$He is larger and thus the 
$\alpha$-rich freezout is
enhanced.  Then the larger fractions of $^{56}$Ni and $^{60}$Zn are
converted to $^{64}$Ge, which enhances the Zn mass fraction.

\placefigure{fig10}

\placefigure{fig11}

     We note that the trend that large $E$ gives high $T^3/\rho$
during an $\alpha$-rich freezeout can be further enhanced in
non-spherical explosions.  This is because the shock in the
jet-direction is stronger than the shock in the spherical model with
the same $E$ (e.g., Maeda \etal 2002; Nagataki et al. 1997).

\subsection{Mixing and Fall-back}

     We have shown that large [Zn, Co/Fe] and small [Mn, Cr/Fe] can be
obtained simultaneously if $M_{\rm cut}$ is sufficiently small.
However, the ejected $^{56}$Ni mass is larger for smaller $M_{\rm
cut}$, and $M(^{56}$Ni) required to get [Zn/Fe]$\sim 0.5$ appears to
be too large to be compatible with observations [O/Fe]$\sim 0 - 0.5$.

     Here we consider a possible process that realizes effectively
smaller mass-cuts without changing the $^{56}$Ni mass.  In SNe II,
when the rapidly expanding core hits the H and He envelopes, a reverse
shock forms and decelerates core expansion.  The deceleration
induces Rayleigh-Taylor instabilities at the composition interfaces of
H/He, He/C+O, and O/Si as has been found in SN 1987A (e.g., Ebisuzaki,
Shigeyama, \& Nomoto 1989; Arnett et al. 1989).  
Therefore, mixing can take place between
the complete and incomplete Si burning regions according to the recent
two dimensional numerical simulations (Kifonidis et
al. 2000; Kifonidis 2001).  
The reverse shock can further induce matter fall-back onto the compact
remnant (e.g., Chevalier 1989).

     Based on these earlier findings, we propose that the following
``mixing fall-back'' process takes place in most SNe II.  

(1) Burned material is uniformly mixed between the ``initial''
mass-cut ($M_{\rm cut}(i)$) and the top of the incomplete Si-burning
region at $M_r = M_{\rm Si}$.  Then [Zn/Fe] in the mixed region
becomes as large as $\sim$ 0.5.

(2) Afterwards the mixed materials below $M_{\rm cut}(f)$ ($> M_{\rm
cut}(i)$) fall-back onto the compact remnant, and  $M_{\rm cut}(f)$
becomes the final mass-cut. Then $M(^{56}$Ni)
becomes smaller while the mass ratios (Zn, Co, Mn)/Fe remain the same
compared with the values determined by $M_{\rm cut}(i)$ in 
Figures 5 and 9.

 We emphasize that the mixing has to take place across
the $M_{\rm cut}(f)$ in order to enhance the fractions of complete
Si-burning products. Otherwise, Zn and Co are underproduced, or
too much $^{56}$Ni is ejected as also seen in previous works such
as Nakamura et al. (1999) and WW95.

     The adopted model parameters of SNe II ($M, E, M_{\rm cut}(i),
M_{\rm cut}$(f)), $M_{\rm Si}$, and the resultant [O/Fe] and [Zn/Fe]
are summarized in Table 4.  Here the initial mass cuts $M_{\rm
cut}(i)$ are chosen to give maximum [Zn/Fe].  For $M_{\rm
cut}(f)$ we consider two cases that give [O/Fe] $\sim 0.3-0.5$ and
0.0, respectively (for $M \geq 20M_\odot$).  Here $M_{\rm cut}(f)$ is
chosen to eject no less than 0.07$M_\odot$ of $^{56}$Ni.
Note that the ratio
[Zn/Fe] is independent of $M_{\rm cut}(f)$.  Note also that larger $E$
leads to larger [Zn/Fe] as discussed in \S 3.4.
Metallicity dependence of  $M_{\rm cut}(f)$ will be discussed
elsewhere, but $M_{\rm cut}(f)$ tends to be smaller for Z=0.02,
being consistent with the observed neutron star masses.

 We note that the occurrence of the mixing has been demonstrated by 
the multi-D simulations of
SN1987A and SNe Ib (e.g., Arnett et al. 1989; Hachisu et al. 1990, 1991; 
Kifonidis et al. 2000), but the fall-back simulations has been done
only in 1D (WW95). Therefore, we need multi-D simulations of 
fall-back to confirm the occurrence of the ``mixing and fall-back''
process and the resulting modification of the ejecta composition,
which has not been done. Only when the mixing takes place across the 
``final mass-cut'', the SN yields are modified by the mixing, which
 has not been taken into account in previous SN yields.

     This ``mixing and fall-back'' effect may also be effectively
realized in non-spherical explosions accompanying energetic jets
(e.g., Maeda et al. 2002; Khokhlov et al. 1999;
Nagataki et al. 1997).  Compared with the
spherical model with the same $M_{\rm cut}(i)$ and $E$, the shock is
stronger (weaker) and thus temperatures are higher (lower) in the jet
(equatorial) direction.  As a result, a larger amount of complete
Si-burning products are ejected in the jet direction, while only
incomplete Si-burning products are ejected in the equatorial
direction.  In total, complete Si-burning elements can be 
enhanced (Maeda 2001; Nomoto et al. 2001).

\section{Conclusions}

     We have calculated nucleosynthesis in massive Pop III stars using
our recent metal-free progenitor models, and compared the results with
the abundances of metal-poor halo stars to constrain the  
explosion models of Pop III stars.  
In particular, we have found through the following parameter study
that the explosion has to be
energetic, and explosively synthesized matter needs to be mixed across the
final mass-cut unless the explosion is highly anisotropic. 

\subsection{Parameter Dependence}

In the present work, we have
focussed on iron-peak elements and, in particular, explored the
parameter ranges ($M_{\rm cut}(i)$, $Y_e$, $M$, and $E$) to reproduce
[Zn/F] $\sim$ 0.5 observed in extremely metal-poor stars.  Our main
results are summarized as follows.

1) The interesting trends of the observed ratios [(Zn, Co, Mn, Cr)/Fe]
can be understood in terms of the variation of the mass ratio between
the complete Si burning region and the incomplete Si burning region.
The large Zn and Co abundances observed in very metal-poor stars are
obtained in our models
if the mass-cut is deep enough (i.e., if $M_{\rm cut}(i)$ is
small enough in Figures 5 and 9), or equivalently if deep material from
the complete Si-burning region is ejected by mixing or aspherical effects
(\S 3.1).  Vanadium also appears to be abundant at low [Fe/H] (e.g.,
Goswami \& Prantzos 2000).  Since V is also produced mainly in the
complete Si-burning region (Fig. 6, 8), this trend can be explained in
the same way as those of Zn and Co.

2) The mass of the incomplete Si burning region is sensitive to the
progenitor mass $M$, being smaller for smaller $M$.  Thus [Zn/Fe]
tends to be larger for less massive stars for the same $E$ (\S 3.2).

3) The production of Zn and Co is sensitive to the value of $Y_e$,
being larger as $Y_e$ is closer to 0.5, especially for the case of
a normal explosion energy ($E_{51}\sim 1$) (\S 3.3).

4) A large explosion energy $E$ results in the enhancement of the
local mass fractions of Zn and Co, while Cr and Mn are not enhanced
(Fig. 9).  This is because larger $E$ produces larger entropy and thus
a stronger $\alpha$-rich freeze-out (\S 3.4).

5) To be consistent with the observed [O/Fe] $\sim$ 0 - 0.5 as
well as with [Zn/Fe] $\sim$ 0.5 in metal-poor stars, we propose that the
``mixing and fall-back" process or aspherical effects are significant
in the explosion of relatively massive stars (\S 3.5).

\subsection{Hypernova Scenario}

\begin{figure}
\hskip 2cm
\epsfxsize=12cm
\epsfbox{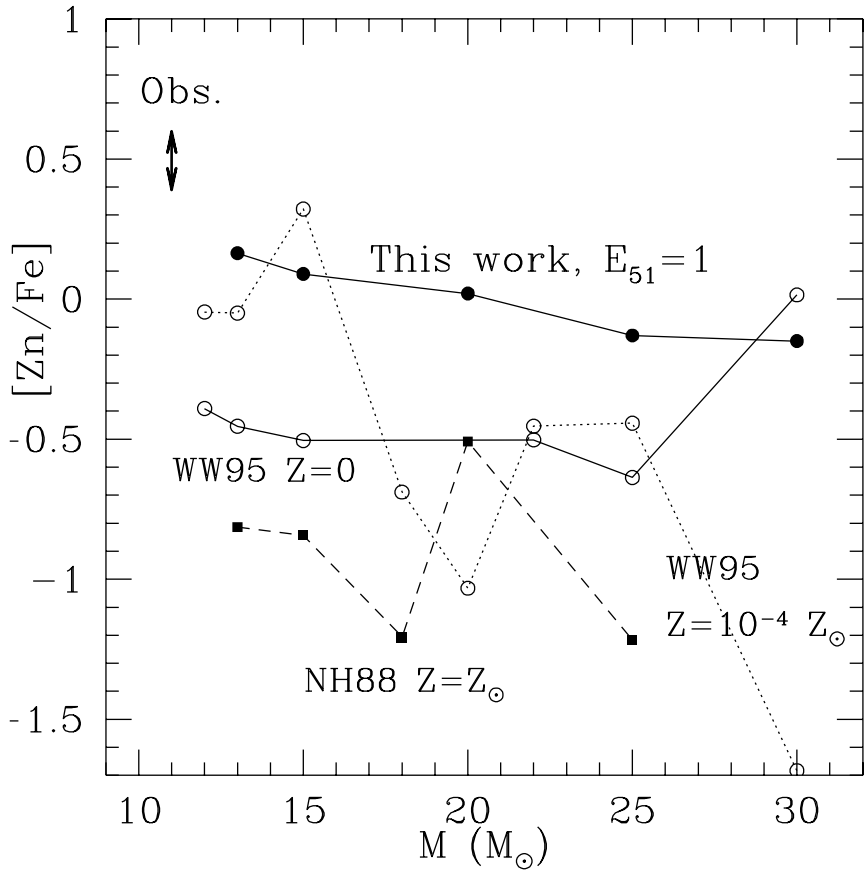}
\caption{[Zn/Fe] ratios of the present and previous works as a function of
stellar mass. Here WW95 denotes Woosley \& Weaver (1995) and NH88
denotes Nomoto \& Hashimoto (1998) models. The observed large [Zn/Fe] ratio
in very low-metal stars ([Fe/H] $<-2.6$)
found in Primas et al. (2000) and Blake et al.
(2001) are represented by a thick arrow.
The [Zn/Fe] ratios of the present work shown here
correspond to the maximum values. If the mass cut is larger 
the [Zn/Fe] ratios become smaller.
\label{fig12}}
\end{figure}

    The dependence of [Zn/Fe] on $M$ and $E$ is summarized in
Figures 13 and 14 as follows.

a) In Figure 13, we compare the [Zn/Fe] ratios in our $E_{51}
=1$ models with previous models for various progenitor masses $M$.
The [Zn/Fe] ratio depends also on $M_{\rm cut}(i)$, and our values in
Figure \ref{fig12} correspond to the maximum values for $E_{51} =1$.
The difference in $Y_e$ is the primary reason why Zn production is
much smaller when previous progenitor models by Nomoto \& Hashimoto
(1998) are used (Thielemann et al. 1996).  (Note that, for
hypernova-like explosion energies, Zn is abundantly produced even if
$Y_e$ is smaller while Zn production is suppressed otherwise.)
Differences from WW95 likely stem from the differences in $M_{\rm
cut}(i)$. Limongi, Straniero \& Chieffi (2000) has also shown 
their yields for some Z = 0 models. However, their nuclear reaction
network is not large enough to calculate $^{64}$Zn synthesis
(Limongi 2001, private communication).
 
b) In Figure 14, [Zn/Fe] is shown as a function of $M$ and
$E$, where the plotted ratios correspond to the maximum values for
given $E$.  We have found that models with $E_{51} =$ 1 do not produce
sufficiently large [Zn/Fe].  To be compatible with the observations of
[Zn/Fe] $\sim 0.5$, the explosion energy must be much larger, i.e.,
$E_{51} \gsim 2$ for $M \sim 13 M_\odot$ and $E_{51} \gsim 20$ for $M
\gsim 20 M_\odot$.

\placefigure{fig12}

\begin{figure}
\hskip 2cm
\epsfxsize=12cm
\epsfbox{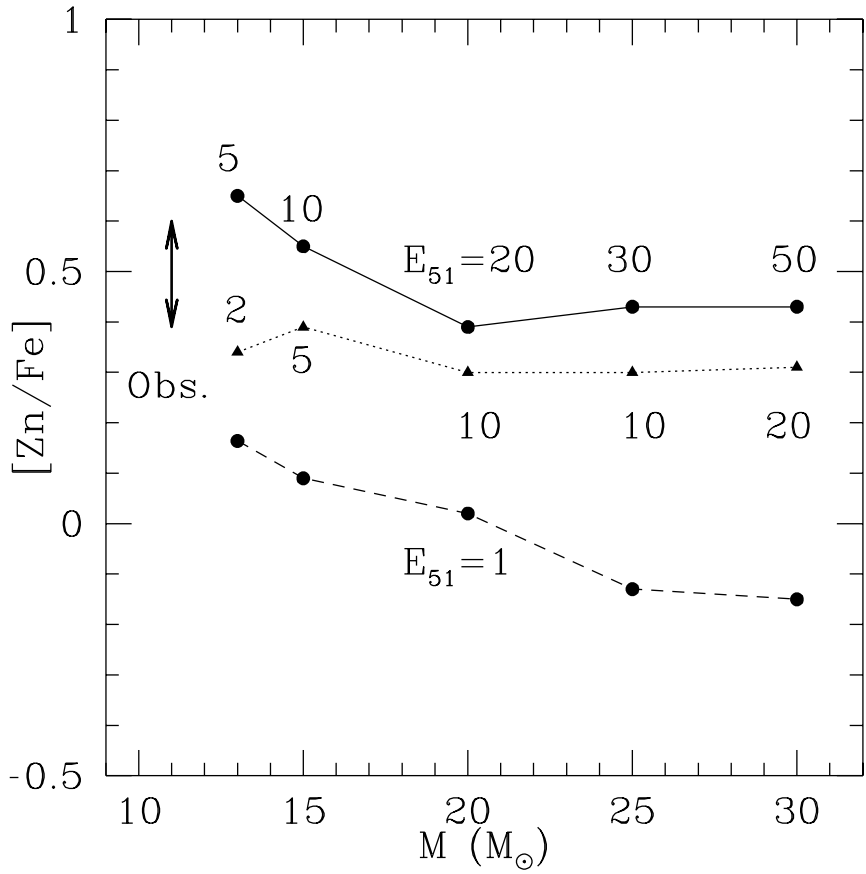}
\caption{The maximum [Zn/Fe] ratios as a function of $M$ and $E_{51}$.
The arrow (obs.) indicates the range of observed high [Zn/Fe] values
at [Fe/H]$<-2.6$ (Fig.1).
\label{fig13}}
\end{figure}

\placefigure{fig13}

     Observationally, the requirement of the large $E$ 
might suggest that large $M$ stars are responsible for large [Zn/Fe],
because $E$ and $M$ can be constrained from the observed brightness
and light curve shape of supernovae as follows.  
[The uncertainties in theoretical models
for gravitational collapse are still too large  to
determine $E$ (e.g., Mezzacappa \etal 2000; Rampp and Janka 2000).]
The recent supernovae 1987A, 1993J, and 1994I indicate that the
progenitors of these normal SNe are 13 - 20 $M_\odot$ stars and
$E_{51} \sim$ 1 - 1.5 (Nomoto et al. 1993, 1994; Shigeyama et
al. 1994; Blinnikov et al. 2000).  On the other hand, the masses of
the progenitors of hypernovae with $E_{51} >$ 10 (SNe 1998bw, 1997ef,
and 1997cy) are estimated to be $M \gsim 25 M_\odot$ (Nomoto et
al. 2000; Iwamoto et al. 1998, 2000; Woosley et al. 1999; Turatto et
al. 2000).  This could be related to the stellar mass dependence of
the explosion mechanisms and the formation of compact remnant, i.e.,
less massive stars form neutron stars, while more massive stars tend
to form black holes.

     To explain the observed relation between [Zn/Fe] and [Fe/H], we
further need to know how $M$ and $E$ of supernovae and [Fe/H] of
metal-poor halo stars are related.  In the early galactic epoch when
the galaxy is not yet chemically well-mixed, [Fe/H] may well be
determined by the first generation of SNe. The
formation of metal-poor stars has been suggested to be
driven by a supernova
shock, so that [Fe/H] is determined by the ejected Fe mass and the
amount of circumstellar hydrogen swept-up by the shock wave 
(Ryan et al. 1996).

     Explosions with the following two combinations of $M$ and $E$
may be responsible for the formation of stars with very small [Fe/H]: 

i) Energetic explosions of massive stars ($M \gsim 25 M_\odot$): For
these massive progenitors, the supernova shock wave tends to propagate
further out because of the large explosion energy and large
Str\"omgren sphere of the progenitors (Nakamura et al. 1999).  The
effect of $E$ may be important since the hydrogen mass swept up by the
supernova shock is roughly proportional to $E$ (e.g., Ryan et al 1996;
Shigeyama \& Tsujimoto 1998).

ii) Normal supernova explosions of less massive stars ($M \sim 13
M_\odot$): These supernovae are assumed to eject a rather small mass
of Fe (Shigeyama \& Tsujimoto 1998), and most SNe are assumed to explode
with normal $E$ irrespective of $M$.

     The above relations lead to the following two possible scenarios
to explain [Zn/Fe] $\sim 0.5$ observed in metal-poor stars.

i) Hypernova-like explosions of massive stars ($M \gsim 25
M_\odot$) with $E_{51} > 10$:  Contribution of highly asymmetric
explosions in these stars may also be important.  The question is 
what fraction of such massive stars explode as hypernovae; the
IMF-integrated yields must be consistent with [Zn/Fe] $\sim$ 0 at
[Fe/H] $\gsim -2.5$.

ii) Explosion of less massive stars ($M \lsim 13
M_\odot$) with $E_{51} \gsim 2$  or a large asymmetry:
This scenario, after integration over the IMF, might
reproduce the observed abundance pattern for [Fe/H]$\gsim -2$
(Tsujimoto \& Shigeyama 1998).  However, the Fe yield has to be very
small in order to satisfy the observed [O/Fe] value ($\gsim 0.5$) for
the metal-poor stars.  For example, the $^{56}$Ni mass yield of our
13$M_\odot$ model has to be less than 0.006$M_\odot$, which appears
to be inconsistent with the observed luminosities (and thus the $^{56}$Ni
mass) of core-collapse SNe of SNe 1993J and 1994I, whose progenitor
masses are estimated to be  13 - 15 $M_\odot$
(see, e.g., Figure 10 of Iwamoto et al. 2000).

     It seems that the [O/Fe] ratio of metal-poor stars and the
$E$-$M$ relations from supernova observations favor the massive
energetic explosion scenario for enhanced [Zn/Fe].  However, we need
to construct detailed galactic chemical evolution models to
distinguish between the two scenarios for [Zn/Fe].  For that purpose, in
Figures 15-17, we show the overall abundance pattern in the ejecta
(after radioactive decays) for the models which yield [Zn/Fe] = 0.3 -
0.6 as a result of the ``mixing and fall-back'' process (except for the 13
$M_\odot$ model).  The mass-cuts are chosen to eject 0.07$M_\odot$
$^{56}$Ni for $M=$ 13 and 15 $M_\odot$, and to realize [O/Fe] = 0 
for $M\gsim 25M_\odot$.  The yields are also listed in Tables 5-13.  Here
large [O/Fe] $(\sim 0.3-0.5)$ is assumed for $M\gsim 20M_\odot$ to
reproduce the majority of the observed abundance pattern.

\begin{figure}
\vskip -3cm
\hskip 2cm
\epsfxsize=12cm
\epsfbox{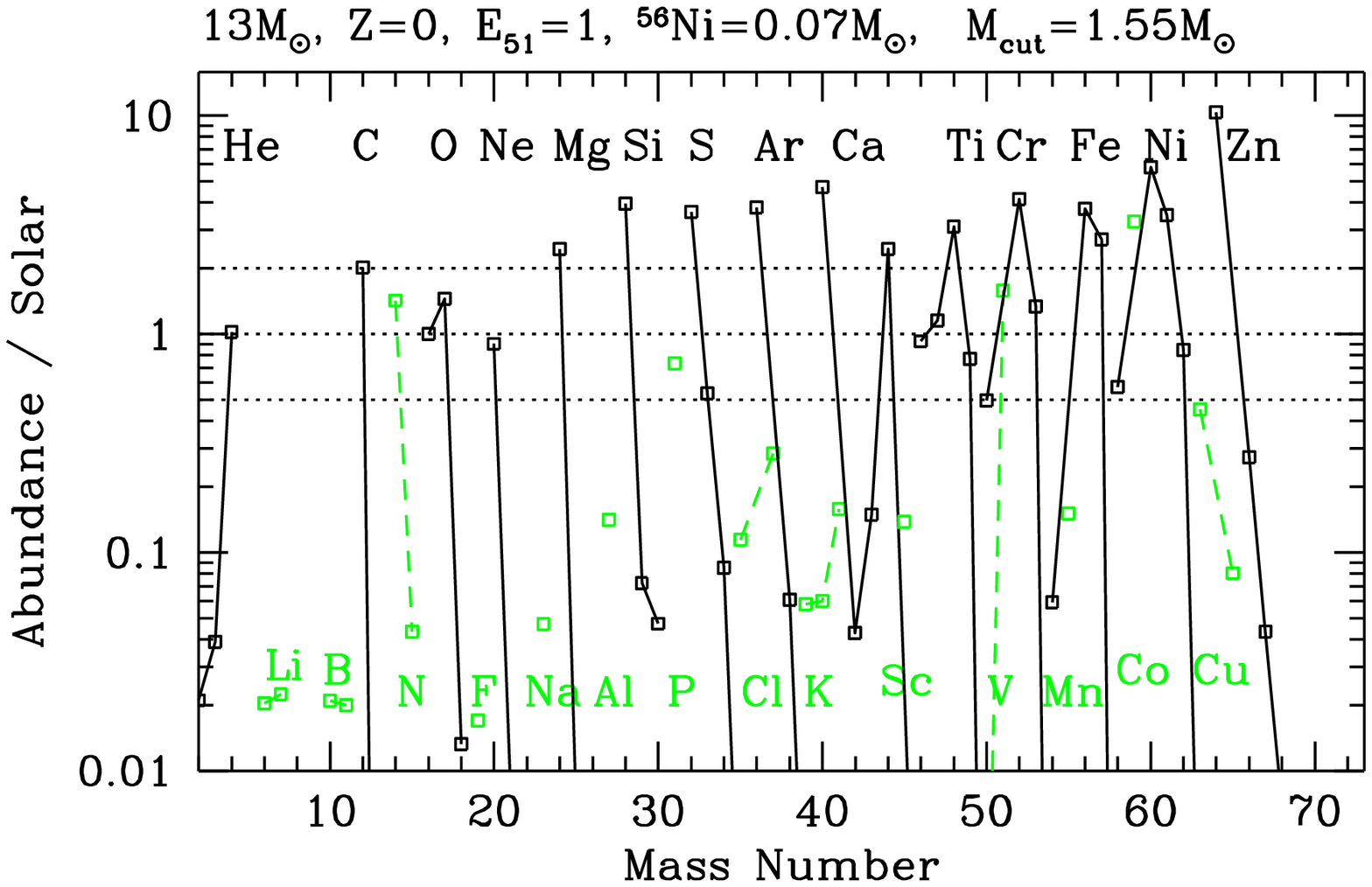}
\vskip -3cm
\hskip 2cm
\epsfxsize=12cm
\epsfbox{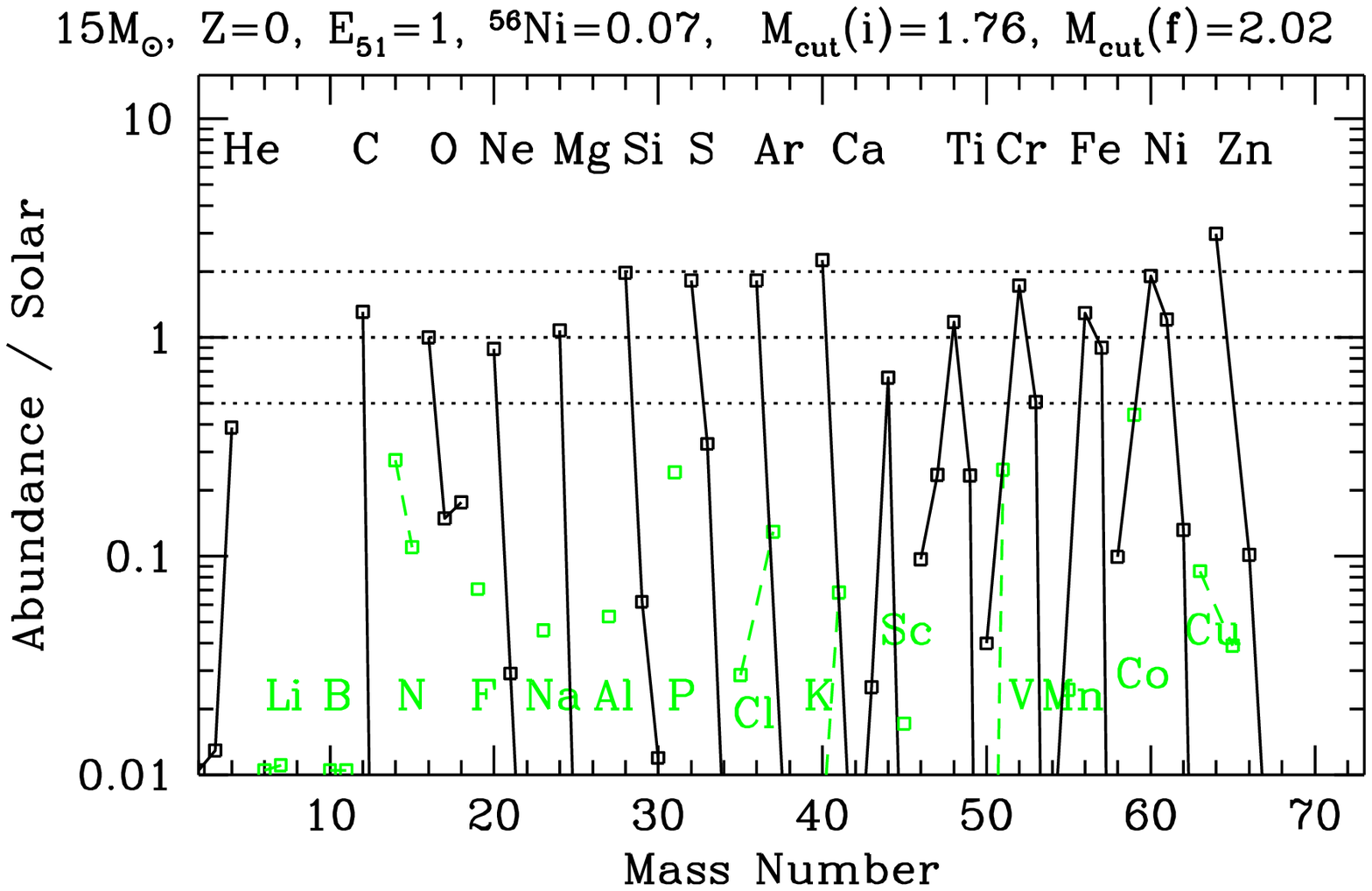}


\caption{Abundance pattern in the ejecta (after radio-active decay)
for the 13$M_\odot$ and 15$M_\odot$ models normalized by the solar
abundances of $^{16}$O. 
Mixing and fall-back is assumed for 15$M_\odot$ but not
for 13$M_\odot$.  The mass-cut is
chosen to eject 0.07 $M_\odot$ $^{56}$Ni.
\label{fig14}}
\end{figure}

\placefigure{fig14}

\begin{figure}
\vskip -3cm
\hskip 2cm
\epsfxsize=12cm
\epsfbox{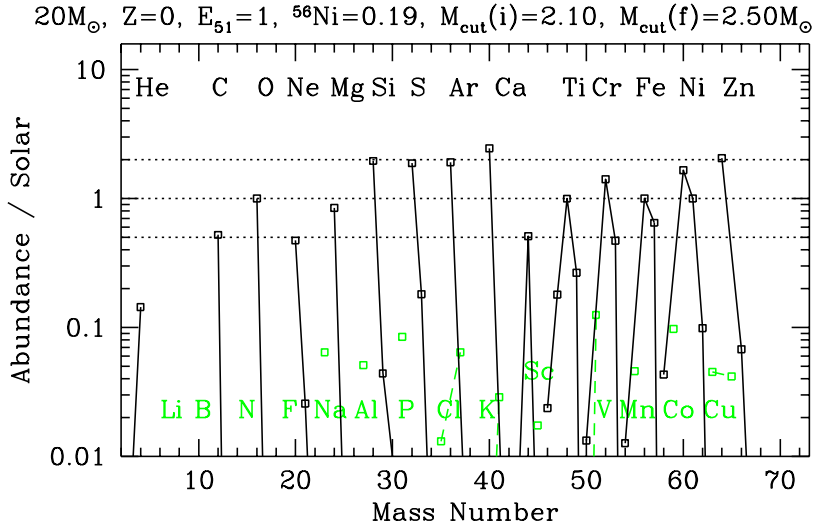}

\vskip -3cm
\hskip 2cm
\epsfxsize=12cm
\epsfbox{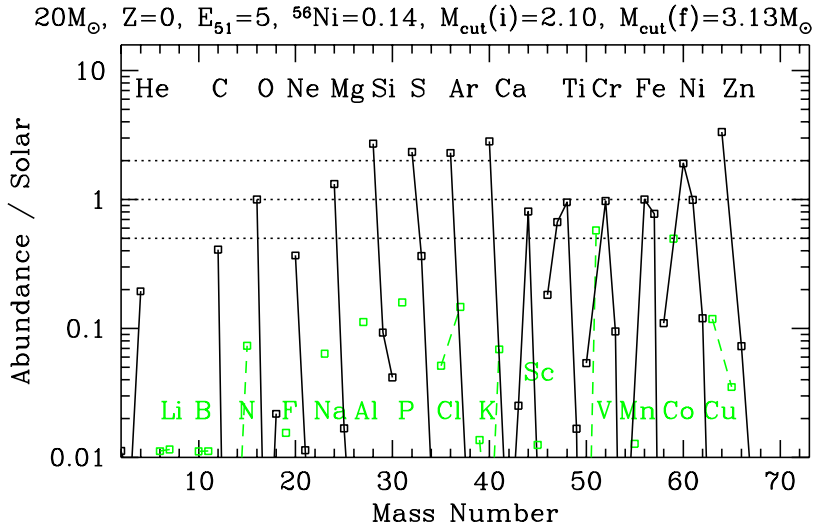}

\caption{Abundance pattern in the ejecta normalized by the solar $^{16}$O
abundances for the mixing fall-back 20$M_\odot$ model with $E_{51}=1$ and 5.
The mass-cuts are chosen to give large [Zn/Fe] and [O/Fe]=0.
\label{fig15}}
\end{figure}

\placefigure{fig15}

\begin{figure}
\vskip -3cm
\hskip 2cm
\epsfxsize=12cm
\epsfbox{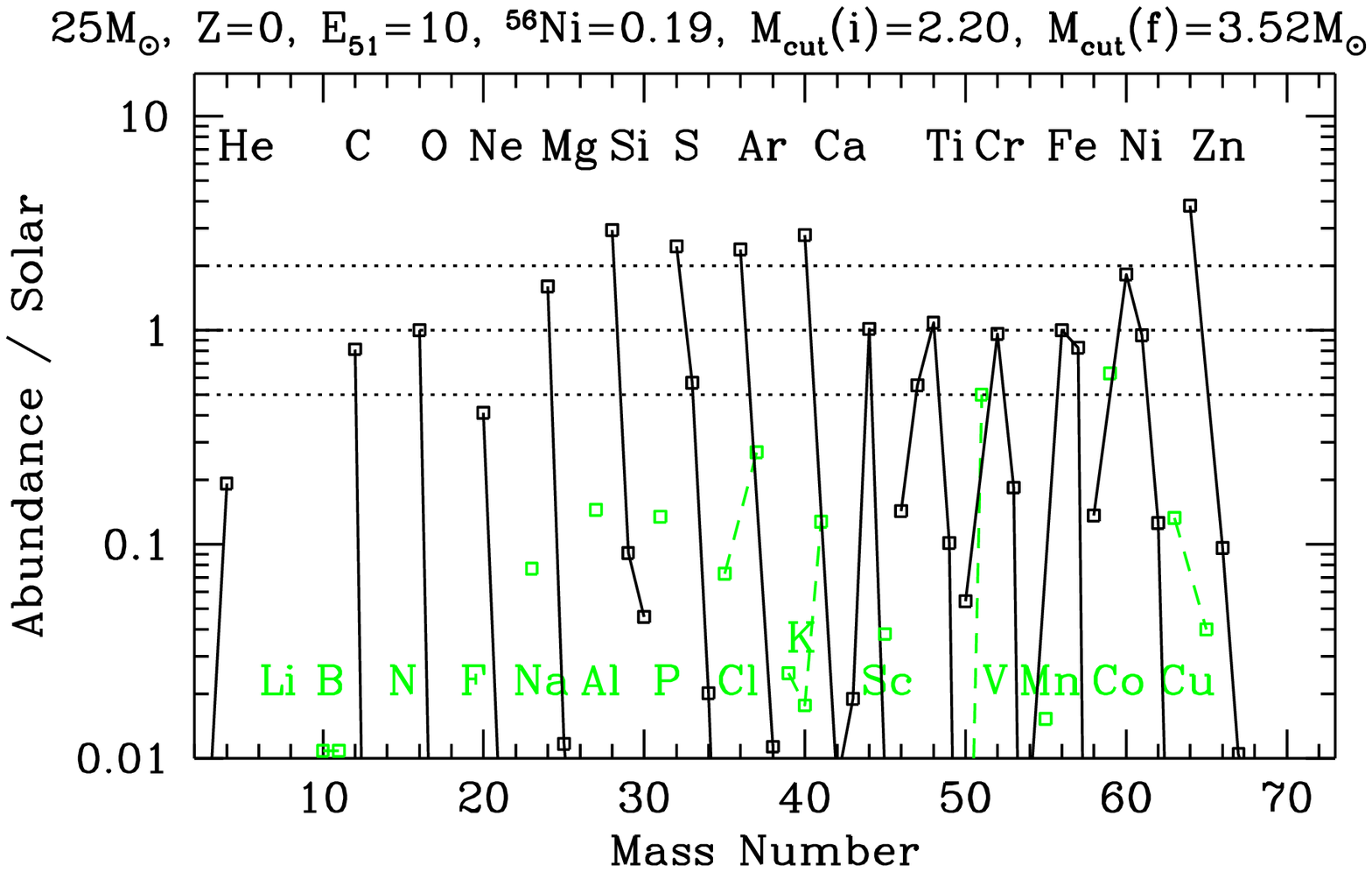}

\vskip -3cm
\hskip 2cm
\epsfxsize=12cm
\epsfbox{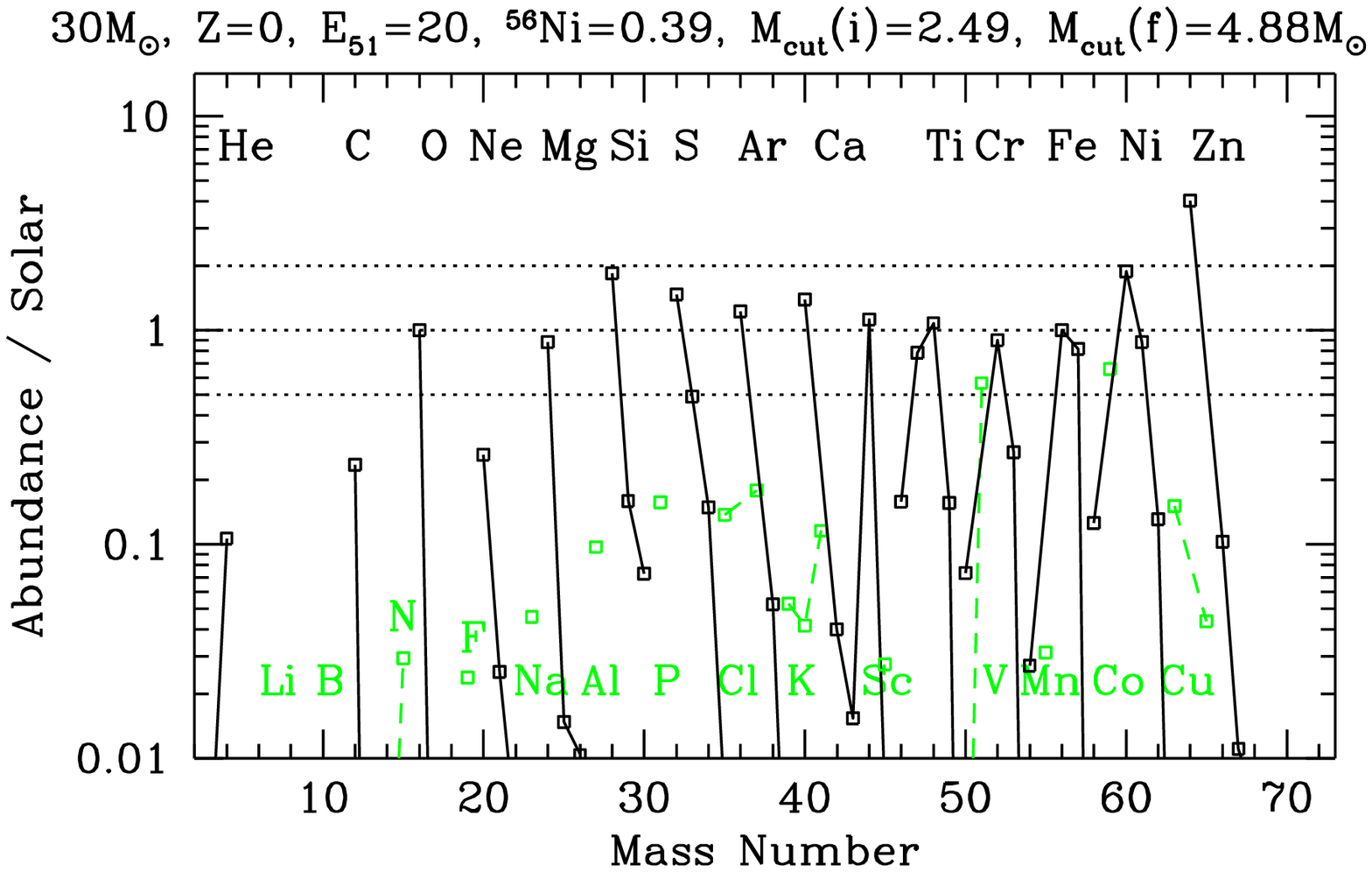}

\caption{Abundance pattern in the ejecta normalized by the solar $^{16}$O
abundances for the  mixing fall-back (25$M_\odot$, $E_{51}=10$)  and 
(30$M_\odot$, $E_{51}=20$) models.
The mass-cuts are chosen to give large [Zn/Fe] and [O/Fe]=0.
\label{fig16}}
\end{figure}

\placefigure{fig16}

\section{Discussion}

\subsection{Mn }

 We note that our Mn yields are roughly a factor of 10 smaller than in
Nomoto et al. (1997) and Nakamura et al. (1999), and Cr is slightly
overproduced. Main source of the differences is that $Y_e$ 
in the incomplete Si-burning region of our models are larger than that
of previous models.  If $Y_e$ in the incomplete Si-burning layers is 
slightly reduced from 0.49996 to 0.49977, for example, Mn
yield is enhanced by a factor of $\sim$ 10 and Cr yield is slightly reduced.
Smaller $Y_e$ leads to a smaller Zn mass fraction.  For 
$Y_e=0.49977$, however, the produced amount of Zn is almost the same
as in our current models as far as $E \gsim 10^{51}$ erg.
 Such $Y_e$  would be obtained for models with very low but non-zero
metallicity. In other words, the Mn abundance may be a good indicator
of real Pop III ejecta.

\subsection{ Co }

Our Co/Fe ratios as in all previous works
(Nomoto et al. 1997; WW95; Nakamura et al. 1999) are at least a factor of
3-5 smaller than the observed ones. However, we consider that the
deficiency of Co is not as serious as Zn by the following reasons.
First, we have not included neutrino processes yet,
but they might enhace the Co yield (WW95). Second, Co is the
decay product of odd-Z element $^{59}$Cu, and its yield
depends on uncertain reaction rates involving proton and neutrons.
On the other hand, Zn is mainly the decay product of even-Z element
$^{64}$Ge, and its abundance is mostly determined by the
less uncertain Q-values and partition functions of $\alpha$ - nuclei.

\subsection{Ni}

 The observed trend of another iron-peak element, Ni, is also
interesting. Unlike the elements we have focused on, [Ni/Fe] 
of metal-poor stars shows no clear trend (see e.g., Ryan et al. 1996;
Nakamura et al. 1999). Theoretically, this is understood as the
fact that Ni is produced abundantly by both complete and incomplete
Si-burning. Recently, Elliison, Ryan \& Prochaska (2001)
observed DLA abundance and found [Co/Fe] $>0$, 
which is similar to the metal-poor halo stars. 
They, on the other hand, did not find oversolar [Ni/Fe].
Similar results have also been found by Norris et al. (2001), who
observed abundances
of five halo stars with [Fe/H] $\lsim -3.5$.
They discussed that the results are inconsistent with the predictions
of Nakamura et al. (1999), where the enhancement of [Co/Fe] 
appears to be accompanied by the enhancement of [Ni/Fe].

 We note that the increase in [Ni/Fe]
along with the increase in [Co/Fe] is not significant in the
models shown in this paper. Let us compare the small [Co/Fe]
model in Table 6
(20$M_\odot$, $E_{51}=1$, $M(^{56}$Ni)=0.07, [Zn/Fe]=0.02) and
the high [Co/Fe] model
in Table 13 (30$M_\odot$, $E_{51}=50$, $M(^{56}$Ni)=0.087,
[Zn/Fe]=0.43). In the former model, the ejected masses of
Co and Ni (two most abundant isotopes are $^{60}$Ni and  $^{58}$Ni)
are ($^{59}$Co, $^{60}$Ni, $^{58}$Ni)=(2.0E-5, 2.0E-3,
1.2E-4)$M_\odot$. In the latter model,
($^{59}$Co, $^{60}$Ni, $^{58}$Ni)=(3.4E-4, 2.8E-3,
8.5E-4)$M_\odot$, so that Co is larger by a
factor of 17, while Ni is larger only by a factor of 1.8
than in the former model. 

 The apparent difference from the results in Nakamura et al. (1999)
can be understood as follows. In  Nakamura et al. (1999), the
explosion energy was fixed to be $E_{51}=1$. They obtained the larger
Co/Fe ratio for more massive SNe II
by assuming ``deeper'' mass-cuts so that 
 $Y_e$ in the explosive burning region is 
smaller $(Y_e \simeq 0.495$).
For smaller $Y_e$, the Co abundance is larger, but the
abundance of Ni (especially  $^{58}$Ni) is enhanced by a larger factor
than Co. Therefore, the increase in [Ni/Fe] with [Co/Fe] was
unavoidable, unless neutrinos substantially enhance $Y_e$ in the
deep complete Si burning region. 

In our present models, mass-cuts of the
larger Co (and Zn) models are not deeper in $M_r$
(see Table 4)
and $Y_e$ in the complete Si burning region  is not small.
This is because, we assume larger explosion energy for more
massive stars, which shifts the mass-cut outwards in $M_r$.
As a result, the dominant Ni isotope is $^{60}$Ni. 
In our model, with increasing $E$, 
the abundance of Co, Zn, and $^{58}$Ni increase
more than that of $^{60}$Ni. Therefore, Co and Zn abundances can be enhanced
without appreciable increase in the Ni abundance.
In this sense, 
the abundance trends of very metal-poor stars is better 
explained with hypernova models rather than the
simple ``deep'' mass-cut models (Nakamura et al. 1999).

\subsection{Pair Instability Supernovae ?}

     One may wonder whether the abundance anomaly of iron-peak elements
discussed in this paper may be related to the peculiar IMF of Pop III
stars.  It is quite likely that the IMF of Pop III stars is different
from that of Pop I and II stars, 
and that more massive stars are abundant for Pop III 
(e.g., Nakamura \& Umemura 1999; Omukai \& Nishi 1999; Bromm, Coppi
\& Larson 1999).
They have discussed that the IMF of Pop
III and very low metal stars may have a peak at even larger masses,
around approximately one hundred to a few hundred solar masses.
If $M\lsim 130M_\odot$, then these stars are likely to form
black holes either without explosion or with energetic explosions.
The nucleosynthesis of the latter case may not be so different from the
models considered here.  This might favor the scenario that invokes
the hypernova-like explosions for large [Zn/Fe].

     If stars are even more massive than $\sim 150 M_\odot$, these stars
become pair-instability SNe (PISNe)
and their nucleosynthesis is different
from core-collapse SNe as summarized in Appendix A
(Barkat, Rakavy \& Sack 1967; Ober, El Eid \& Fricke 1983; 
Woosley \& Weaver 1982). 
In paricular, PISNe produce [Zn/Fe] $< -1.5$, 
because in PISNe, iron peak elements are 
mostly produced by incomplete Si burning so that the mass fraction of
complete Si burning elements is much smaller than that of
SNe II (Fig. 18 \& 19). We thus conclude that PISNe are unlikely
to produce a large enough Zn/Fe ratio to explain the observations.

\begin{figure}
\vskip -3cm
\hskip 2cm
\epsfxsize=12cm
\epsfbox{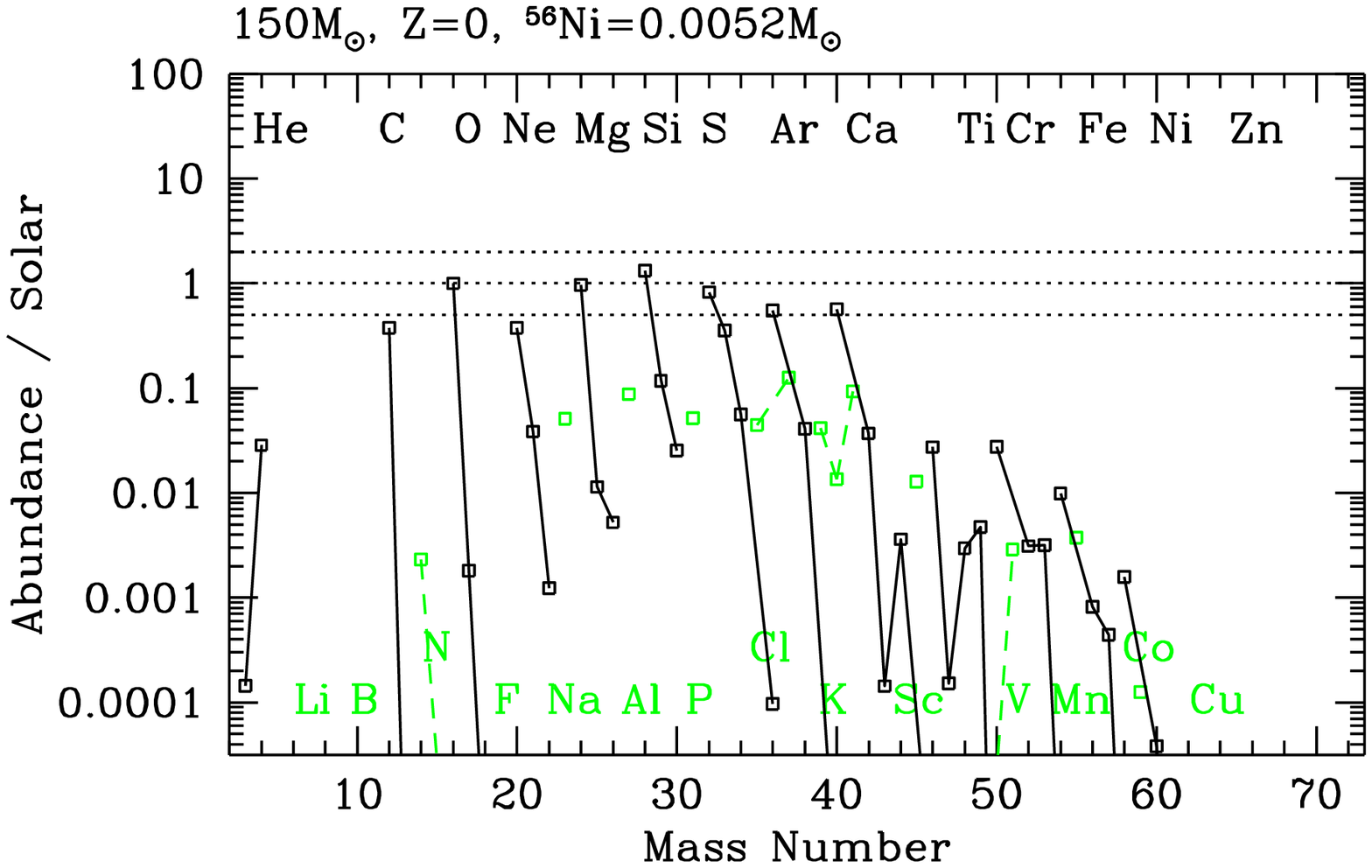}
\vskip -3cm
\hskip 2cm
\epsfxsize=12cm
\epsfbox{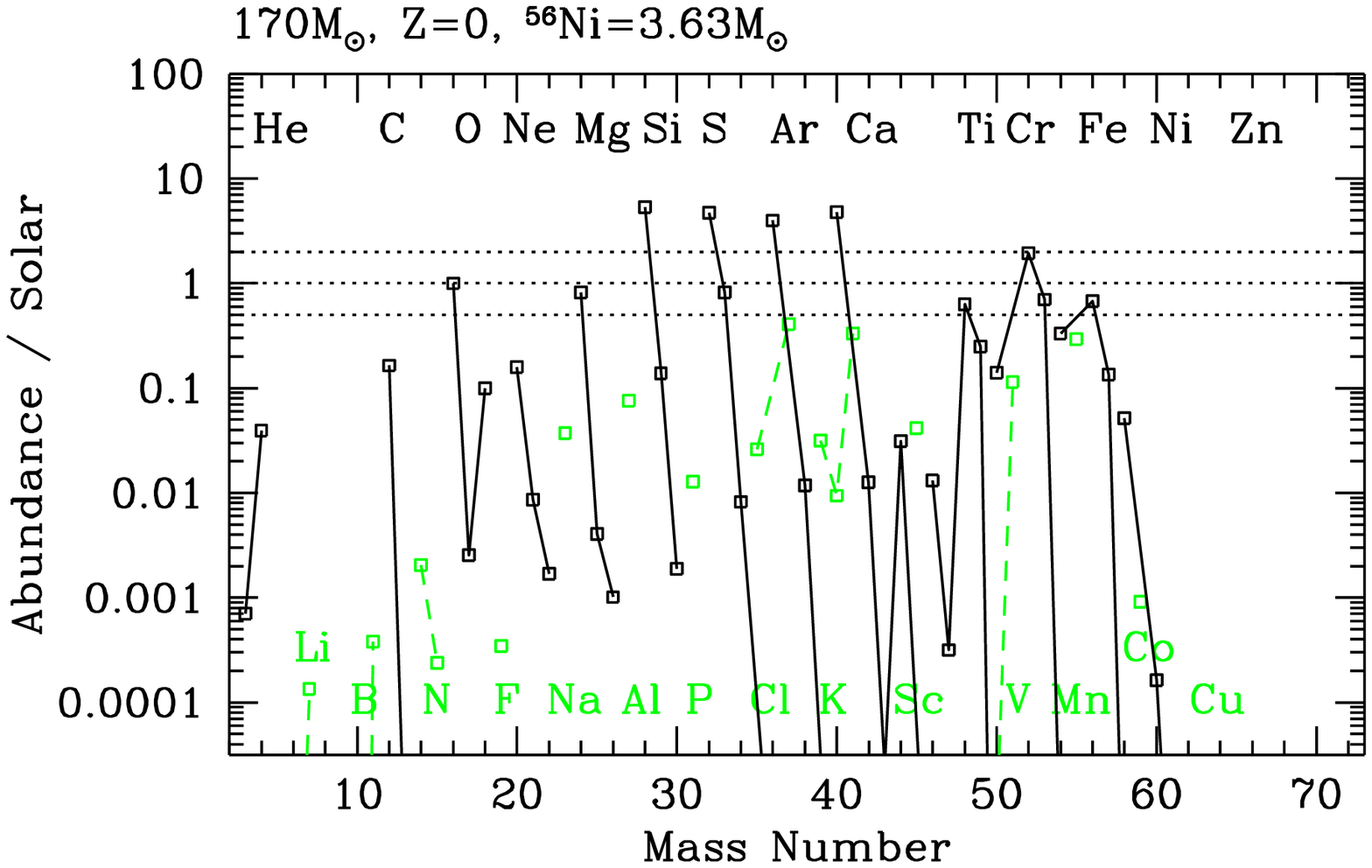}
\caption{Abundance pattern in the ejecta (after radio-active decay)
for the 150$M_\odot$ and 170$M_\odot$ PISN models normalized by the solar
abundances of $^{16}$O. 
\label{figA1}}
\end{figure}

\placefigure{figA1}

\begin{figure}
\vskip -3cm
\hskip 2cm
\epsfxsize=12cm
\epsfbox{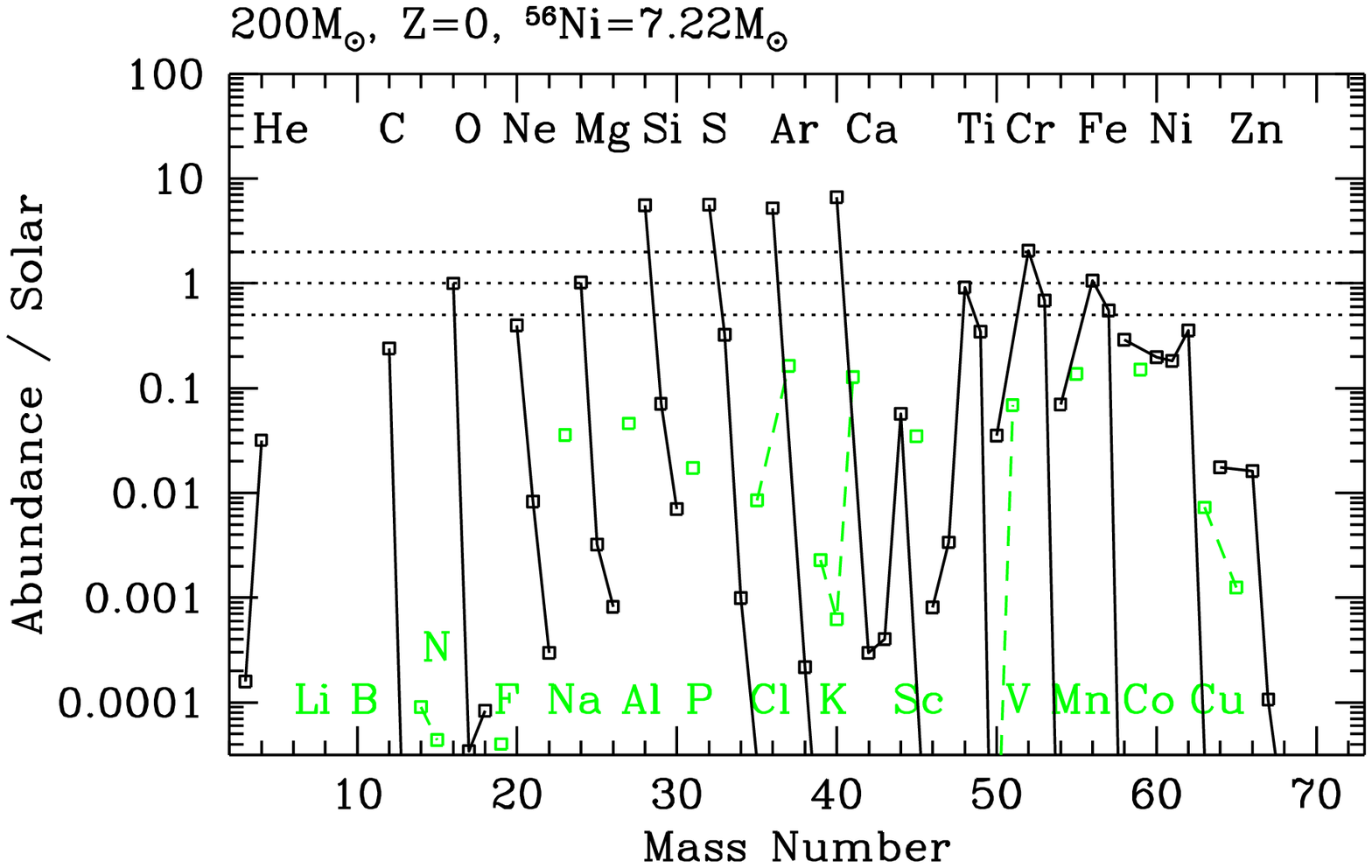}
\vskip -3cm
\hskip 2cm
\epsfxsize=12cm
\epsfbox{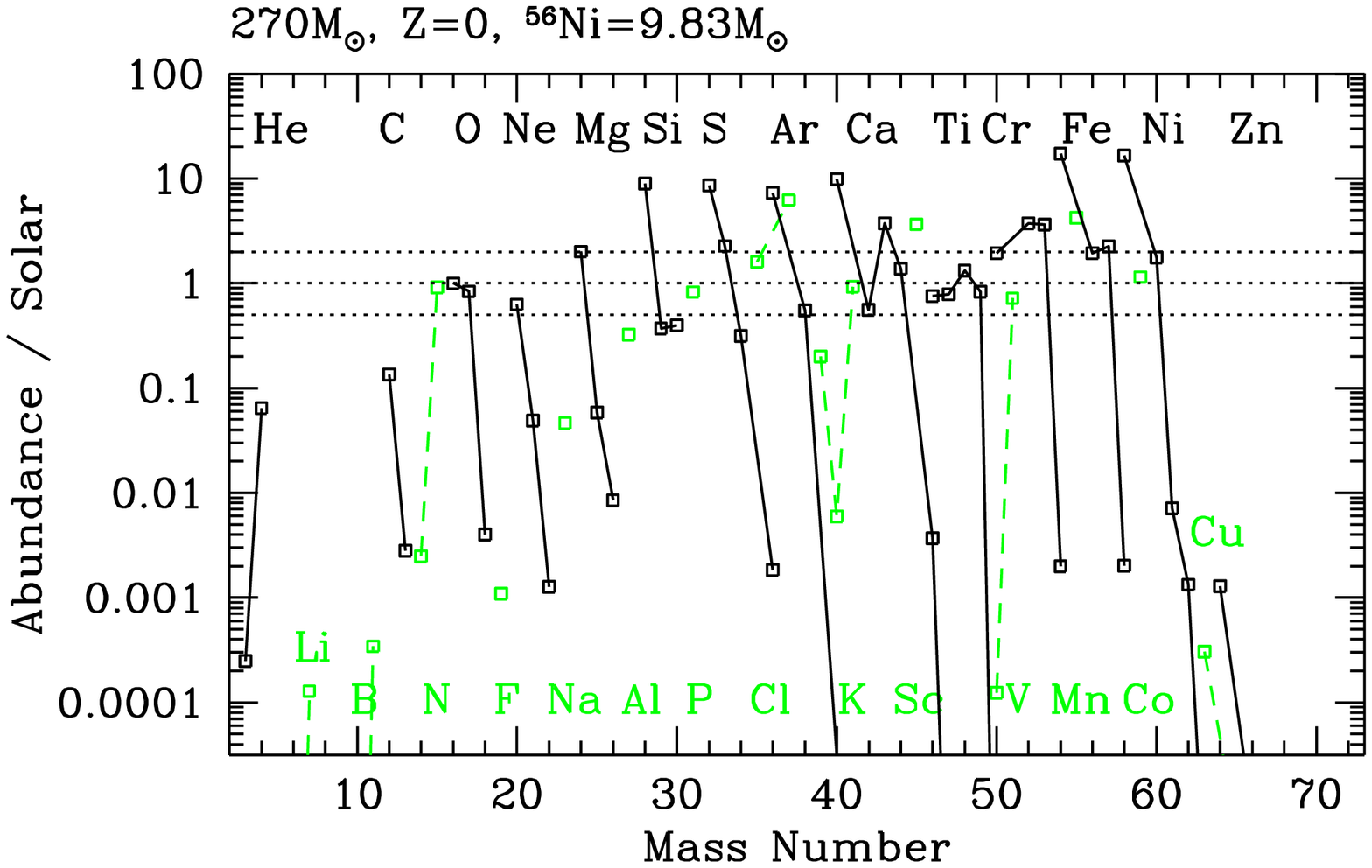}
\caption{Abundance pattern in the ejecta (after radio-active decay)
for the 200$M_\odot$ and 270$M_\odot$ PISN models normalized by the solar
abundances of $^{16}$O. 
\label{figA2}}
\end{figure}

\placefigure{figA2}

\subsection{Concluding Remarks}

      In this paper, we have shown that such a large Zn abundance as
[Zn/Fe] $\sim$ 0.5 observed in metal-poor stars can be realized in
certain supernova models.  This implies that the assumption of [Zn/Fe]
$\sim 0$ usually adopted in the DLA abundance analyses may not be well
justified.  Rather [Zn/Fe] may provide important information on the
IMF and/or the age of the DLA systems.

We have considered only a few elements to constrain the
nucleosynthesis of Pop III stars, because their trends are most clear. 
Data for other elements show less clear trends or currently have
relatively large error bars. However, additional information will be very
useful.  For example, [S/Fe] and [C/O] may be important to distinguish
the scenarios of $M \lsim 13 M_\odot$ and $M \gsim 20M_\odot$.  Also
mass-cut independent ratios [Ca, S, Si/Mg] will be important to
constrain the explosion energies of SNe.

\bigskip

We would like to thank S. Ryan, C. Kobayashi, T. Nakamura, K. Maeda,
M. Shirouzu, and K. Kifonidis   
for useful discussion.  We also thank the 
referee R.D. Hoffman for useful comments
to improve the paper. This work has been
supported in part by the grant-in-Aid for COE Scientific Research
(07CE2002, 12640233) of the Ministry of Education, Science, Culture,
and Sports in Japan.



\bigskip
\bigskip
\bigskip

\centerline{\it Appendix: Pair Instability Supernova yields}

\bigskip

 In this appendix we present the yields of our 
Pop III Pair Instability SN (PISN) models with the initial
masses $M =$ 150, 170, 200 and 270 $M_\odot$. These stars enter
into the electron-positron pair-instability region during the central
oxygen burning stages, and contract quasi-dynamically.
Then the central temperature increases to $3-6  \times 10^9 $K,
which is so high that central oxygen-burning
takes place explosively, being much
faster than neutrino energy losses. The generated 
nuclear energy is 
large enough for internal energy to exceed the gravitaional binding
energy. Then the stars disrupt completely without leaving 
compact remnants and become PISNe.
In our $M = 300M_\odot$ model, the total energy of the star does
not become positive after central oxygen and Si burnings and
hence the star collapse into a black hole. 

 The main purpose of
this appendix is to show that large [Zn/Fe] ratio is not realized
in these SNe. 
More detailed explanaton will be given elsewhere. 
In Table 14, we summarize the initial, He core, and C-O core masses
and the ejected $^{56}$Ni masses. 
We note that the amount of ejected $^{56}$Ni masses are
quite sensitive to the central temperature at bounce, which depends
on the ratio between the
kinetic energy and the internal energy generated by nuclear burning. 
Since the hydrodynamical behavior is rather sensitive to those factors, 
the results shown here are still preliminary; 
nevertheless this uncertainty does not affect the conclusion
on [Zn/Fe] described below.
In these models convective mixing parameter is
chosen to be $f_k=0.1$. The yields are shown in Table 15-18 and
the abundance patterns are shown in Figures 18 \& 19.
 
 Most striking feature is the small [Zn, Co/Fe]
ratios. This is because in PISNe the mass ratio between the complete
and incomplete Si burning regions are much smaller than
core-collapse SNe. A large amount of Zn and Co could be produced
if the central temperature at bounce is higher. However, in this
case incomplete Si-burning region is also extended. Therefore,
the small [Zn, Co/Fe] ratios are inevitable for PISN
models and this conclusion is independent of any possible 
uncertainties. Therefore, we can conclude that the
abundance pattern seen in the very metal-poor halo stars were
not resulted from the pattern of PISNe.

\clearpage


\clearpage


\clearpage

\clearpage

\begin{table}
\begin{center}
\renewcommand{\arraystretch}{1.2}
\begin{tabular}{cccccc} \hline\hline
Z & \multicolumn{5}{c}{Initial Mass ($M_\odot$)}  \cr 
 & 13 & 15 & 20& 25& 30  \cr 
\hline
0 & 1.29& 1.38& 1.52& 1.70 &1.77 \cr 
$10^{-4}$ & 1.32& ---& 1.51& --- & --- \cr 
0.02 & 1.27& 1.34& 1.52& 1.67 & ---\cr 
 \hline 
\end{tabular}
\end{center}
\caption{``Fe''-core masses in $M_\odot$ defined as a region with 
$Y_e \leq 0.49$ for
the progenitor models with $Z=0, 10^{-4}$ and 0.02.\label{fecore}}
\end{table}

\begin{table}
\begin{center}
\renewcommand{\arraystretch}{1.2}
\begin{tabular}{cccccc} \hline\hline
($M$, $E_{51}$) & $M_r$ (incomplete Si-b.) & $\Delta M(^{56}$Ni)  \cr 
\hline
(13, 1) & 1.60 -- 1.67 & 0.022  \cr 
(15, 1) & 1.87 -- 2.06 & 0.052  \cr 
(20, 1) & 2.38 -- 2.88 & 0.14  \cr 
(20, 5) & 2.82 -- 3.38 & 0.15  \cr 
(25, 1) & 2.47 -- 3.00 & 0.12  \cr 
(25, 10) & 3.13 -- 3.87 & 0.18  \cr 
(30, 1) & 2.70 -- 3.67 & 0.35  \cr 
(30, 20) & 4.28 -- 5.58 & 0.36  \cr 
(30, 30) & 4.64 -- 6.32 & 0.45  \cr 
 \hline 
\end{tabular}
\end{center}
\caption{The mass coordinates in $M_\odot$ of the incomplete
Si-burning regions for models with several initial masses and
explosion energies. The upper and lower bounds of the regions are
defined by $X(^{56}$Ni) = $10^{-3}$ and $X(^{28}$Si) = $10^{-4}$,
respectively. The $^{56}$Ni mass in $M_\odot$ in these regions,
$\Delta M(^{56}$Ni), are also shown.}
\end{table}

\begin{table}
\begin{center}
\renewcommand{\arraystretch}{1.2}
\begin{tabular}{ccccc} \hline\hline
Case& ($M /M_\odot$, $E_{51}$, $M_r /M_\odot$)& 
 $Y_e$ & $X$($^{56}$Ni)& $X$(Zn) \cr 
A& (13, 1, 1.52)&   0.4996 & 7.65E-01 & 6.04E-04 \cr 
B& (13, 1, 1.57)&   0.4999 & 7.61E-01 & 3.46E-03 \cr 
C& (25, 1, 2.2)&  0.4998 & 8.48E-01 & 8.89E-04 \cr 
C$'$& (25, 1, 2.2)&  0.5000 & 8.55E-01 & 2.01E-03 \cr 
D& (25, 10, 2.2)&  0.4998 & 7.12E-01 & 3.36E-03 \cr 
 \hline 
\end{tabular}
\end{center}
\caption{Mass fractions of $^{56}$Ni and 
Zn (decay products of $^{64}$Ge) after explosive nucleosynthesis
for representative cases shown in Figure 10.
\label{tab4}}
\end{table}

\begin{table}
\begin{center}
\renewcommand{\arraystretch}{1.2}
\begin{tabular}{cccccccc} \hline\hline
($M$, $E_{51}$) &  $M_{\rm cut}(i)$ & $M_{\rm Si}$ &
$M_{\rm cut}(f)$ & $M(^{56}$Ni) & $M(^{44}$Ti) & [O/Fe] & [Zn/Fe]  \cr 
\hline
(13, 1) & 1.54& 1.67 & 1.54  & 0.070 & 5.6E-5 &-0.54 & 0.16\cr 
(15, 1) & 1.76& 2.06 & 2.02  & 0.070 & 5.6E-5&-0.08 & 0.09\cr 
(20, 1) & 2.10& 2.88 & 2.73 (2.50) & 0.070 (0.19)&4.4E-5 (1.2E-4)& 0.43 (0.0)& 0.02\cr 
(20, 5) & 2.10& 3.38 & 3.27 (3.13) & 0.070 (0.14)&6.9E-5 (1.4E-4)& 0.31 (0.0)& 0.24\cr 
(25, 1) & 2.20& 3.00 & 2.80 (2.40) & 0.089 (0.27)&7.1E-5 (1.7E-4)& 0.50 (0.0)& -0.13\cr 
(25, 10) & 2.20& 3.86 & 3.74 (3.52)& 0.070 (0.19)&8.7E-5 (2.3E-4)& 0.45 (0.0)& 0.30\cr 
(25, 30) & 2.50& 4.52 & 4.39 (4.27)& 0.070 (0.14)&1.3E-4 (2.5E-4)& 0.28 (0.0)& 0.43\cr 
(30, 1) & 2.26& 3.66 & 3.29 (2.57) & 0.20 (0.58)&1.5E-4 (4.4E-4)& 0.50 (0.0)& -0.18\cr 
(30, 20) & 2.49& 5.58 & 5.36 (4.87)& 0.12 (0.39)&1.6E-7 (5.3E-4)& 0.50 (0.0)& 0.32\cr 
(30, 30) & 2.83& 6.01 & 5.81 (5.36)& 0.11 (0.36)&1.7E-4 (5.4E-4)& 0.50 (0.0)& 0.34\cr 
(30, 50) & 3.15& 6.92 & 6.74 (6.35)& 0.087 (0.28)&1.7E-4 (5.1E-4)& 0.50 (0.0)& 0.43\cr 
 \hline 
\end{tabular}
\end{center}
\caption{Ejected $^{56}$Ni mass, $^{44}$Ti mass in $M_\odot$, 
[Zn/Fe] and [O/Fe] ratios in the
ejecta of mixing fall-back models. In these models, 
the matter is first uniformly mixed between $M_r=M_{\rm cut}(i)$ and
top of the incomplete Si-burning region  $M_r= M_{\rm Si}$, then
the matter below $M_r= M_{\rm cut}(f)$ is fallen back. For the
models with $M \geq 20M_\odot$ two choices
of $M_{\rm cut}(f)$ are shown, that gives relatively large [O/Fe]
$(\sim 0.3-0.5)$ and small [O/Fe] ($\sim 0)$ ratios.
Here $M_{\rm cut}(f)$ is
chosen to eject no less than 0.07$M_\odot$ of $^{56}$Ni.}
\end{table}

\begin{table}
\begin{center}
\renewcommand{\arraystretch}{1.2}
\begin{tabular}{rcrcrcrcrc} \hline\hline
     p   &    6.666E+00&  d       &    7.754E-06& $^3$He   &    2.139E-05& $^4$He   &    4.317E+00& $^6$Li   &    2.092E-10\\ 
$^7$Li   &    3.338E-09& $^9$Be   &    5.416E-20& $^{10}$B &    3.504E-10& $^{11}$B &    1.501E-09& $^{12}$C &    9.763E-02\\ 
$^{13}$C &    1.625E-09& $^{14}$N &    2.519E-02& $^{15}$N &    3.034E-06& $^{16}$O &    1.535E-01& $^{17}$O &    9.014E-05\\ 
$^{18}$O &    4.601E-06& $^{19}$F &    1.103E-07& $^{20}$Ne&    2.336E-02& $^{21}$Ne&    5.655E-07& $^{22}$Ne&    7.747E-07\\ 
$^{23}$Na&    2.516E-05& $^{24}$Mg&    2.020E-02& $^{25}$Mg&    7.049E-06& $^{26}$Mg&    2.237E-06& $^{26}$Al&    5.961E-07\\ 
$^{27}$Al&    1.311E-04& $^{28}$Si&    4.122E-02& $^{29}$Si&    3.958E-05& $^{30}$Si&    1.777E-05& $^{31}$P &    1.575E-05\\ 
$^{32}$S &    2.296E-02& $^{33}$S &    2.761E-05& $^{34}$S &    2.536E-05& $^{36}$S &    2.814E-11& $^{35}$Cl&    4.627E-06\\ 
$^{37}$Cl&    3.885E-06& $^{36}$Ar&    4.715E-03& $^{38}$Ar&    1.492E-05& $^{40}$Ar&    6.183E-12& $^{39}$K &    3.212E-06\\ 
$^{40}$K &    4.259E-10& $^{41}$K &    6.640E-07& $^{40}$Ca&    4.517E-03& $^{42}$Ca&    2.878E-07& $^{43}$Ca&    2.140E-07\\ 
$^{44}$Ca&    5.556E-05& $^{46}$Ca&    9.666E-12& $^{48}$Ca&    2.223E-12& $^{45}$Sc&    8.608E-08& $^{46}$Ti&    3.314E-06\\ 
$^{47}$Ti&    3.842E-06& $^{48}$Ti&    1.070E-04& $^{49}$Ti&    2.014E-06& $^{50}$Ti&    1.842E-11& $^{50}$V &    1.065E-11\\ 
$^{51}$V &    9.538E-06& $^{50}$Cr&    5.917E-06& $^{52}$Cr&    9.887E-04& $^{53}$Cr&    3.686E-05& $^{54}$Cr&    6.145E-11\\ 
$^{55}$Mn&    3.211E-05& $^{54}$Fe&    6.735E-05& $^{56}$Fe&    7.000E-02& $^{57}$Fe&    1.242E-03& $^{58}$Fe&    6.629E-11\\ 
$^{59}$Co&    1.759E-04& $^{58}$Ni&    4.522E-04& $^{60}$Ni&    1.815E-03& $^{61}$Ni&    4.806E-05& $^{62}$Ni&    3.758E-05\\ 
$^{64}$Ni&    1.499E-11& $^{63}$Cu&    4.167E-06& $^{65}$Cu&    3.403E-07& $^{64}$Zn&    1.643E-04& $^{66}$Zn&    2.560E-06\\ 
$^{67}$Zn&    6.092E-08& $^{68}$Zn&    4.653E-08& $^{70}$Zn&    1.434E-11& $^{69}$Ga&    5.899E-09& $^{71}$Ga&    4.674E-11\\
$^{70}$Ge&    1.281E-08& $^{72}$Ge&    3.285E-11& $^{73}$Ge&    4.584E-11& $^{74}$Ge&    1.863E-11& &\\
 \hline 
\end{tabular}
\end{center}
\caption{Yields in the ejecta in $M_\odot$ after radio active 
decay (except $^{26}$Al) for the 13$M_\odot$ $E_{51}=1$ 
model shown in Table 4. ($M_{\rm cut}(i)$,  $M_{\rm Si}$, 
$M_{\rm cut}(f)$)=(1.54, 1.67, 1.54), [O/Fe]=$-0.54$ and [Zn/Fe]=0.16. 
}
\end{table}

\begin{table}
\begin{center}
\renewcommand{\arraystretch}{1.2}
\begin{tabular}{rcrcrcrcrc} \hline\hline
     p   &    7.581E+00&  d       &    1.117E-05& $^3$He   &    2.052E-05& $^4$He   &    4.727E+00& $^6$Li   &    3.100E-10\\ 
$^7$Li   &    4.748E-09& $^9$Be   &    2.242E-21& $^{10}$B &    5.057E-10& $^{11}$B &    2.277E-09& $^{12}$C &    1.839E-01\\ 
$^{13}$C &    8.899E-09& $^{14}$N &    1.404E-02& $^{15}$N &    2.219E-05& $^{16}$O &    4.434E-01& $^{17}$O &    2.676E-05\\ 
$^{18}$O &    1.767E-04& $^{19}$F &    1.324E-06& $^{20}$Ne&    6.639E-02& $^{21}$Ne&    5.543E-06& $^{22}$Ne&    4.458E-06\\ 
$^{23}$Na&    7.051E-05& $^{24}$Mg&    2.570E-02& $^{25}$Mg&    7.822E-06& $^{26}$Mg&    4.186E-06& $^{26}$Al&    2.424E-07\\ 
$^{27}$Al&    1.419E-04& $^{28}$Si&    5.974E-02& $^{29}$Si&    9.778E-05& $^{30}$Si&    1.299E-05& $^{31}$P &    1.102E-05\\ 
$^{32}$S &    3.330E-02& $^{33}$S &    4.855E-05& $^{34}$S &    4.504E-06& $^{36}$S &    7.750E-12& $^{35}$Cl&    3.337E-06\\ 
$^{37}$Cl&    5.086E-06& $^{36}$Ar&    6.509E-03& $^{38}$Ar&    1.460E-06& $^{40}$Ar&    1.588E-13& $^{39}$K &    1.316E-06\\ 
$^{40}$K &    1.093E-10& $^{41}$K &    8.288E-07& $^{40}$Ca&    6.247E-03& $^{42}$Ca&    3.242E-08& $^{43}$Ca&    1.040E-07\\ 
$^{44}$Ca&    4.292E-05& $^{46}$Ca&    1.043E-13& $^{48}$Ca&    9.289E-14& $^{45}$Sc&    3.088E-08& $^{46}$Ti&    9.959E-07\\ 
$^{47}$Ti&    2.265E-06& $^{48}$Ti&    1.166E-04& $^{49}$Ti&    1.762E-06& $^{50}$Ti&    2.016E-13& $^{50}$V &    2.425E-13\\ 
$^{51}$V &    4.311E-06& $^{50}$Cr&    1.368E-06& $^{52}$Cr&    1.184E-03& $^{53}$Cr&    4.033E-05& $^{54}$Cr&    2.895E-12\\ 
$^{55}$Mn&    1.505E-05& $^{54}$Fe&    1.292E-05& $^{56}$Fe&    7.000E-02& $^{57}$Fe&    1.188E-03& $^{58}$Fe&    1.390E-11\\ 
$^{59}$Co&    6.883E-05& $^{58}$Ni&    2.275E-04& $^{60}$Ni&    1.729E-03& $^{61}$Ni&    4.813E-05& $^{62}$Ni&    1.689E-05\\ 
$^{64}$Ni&    3.083E-13& $^{63}$Cu&    2.269E-06& $^{65}$Cu&    4.774E-07& $^{64}$Zn&    1.368E-04& $^{66}$Zn&    2.761E-06\\ 
$^{67}$Zn&    2.314E-08& $^{68}$Zn&    3.672E-08& $^{70}$Zn&    2.685E-13& $^{69}$Ga&    4.957E-09& $^{71}$Ga&    8.216E-13\\ 
$^{70}$Ge&    1.152E-08& $^{72}$Ge&    1.664E-12& $^{73}$Ge&    9.740E-13& $^{74}$Ge&    1.081E-12& &\\
 \hline 
\end{tabular}
\end{center}
\caption{Yields in the ejecta in $M_\odot$ after radio active 
decay (except $^{26}$Al) for the 15$M_\odot$ $E_{51}=1$ 
model shown in Table 4. ($M_{\rm cut}(i)$,  $M_{\rm Si}$, 
$M_{\rm cut}(f)$)=(1.76, 2.06, 2.02), [O/Fe]=$-0.08$ and [Zn/Fe]=0.09. 
}
\end{table}

\begin{table}
\begin{center}
\renewcommand{\arraystretch}{1.2}
\begin{tabular}{rcrcrcrcrc} \hline\hline
      p   &    9.396E+00&  d       &    2.768E-05& $^3$He   &    2.556E-05& $^4$He   &    6.258E+00& $^6$Li   &    7.648E-10\\ 
$^7$Li   &    1.154E-08& $^9$Be   &    4.661E-19& $^{10}$B &    1.252E-09& $^{11}$B &    5.613E-09& $^{12}$C &    2.569E-01\\ 
$^{13}$C &    4.559E-08& $^{14}$N &    2.688E-04& $^{15}$N &    1.769E-06& $^{16}$O &    1.550E+00& $^{17}$O &    2.385E-07\\ 
$^{18}$O &    4.878E-06& $^{19}$F &    3.298E-07& $^{20}$Ne&    1.240E-01& $^{21}$Ne&    1.721E-05& $^{22}$Ne&    1.177E-05\\ 
$^{23}$Na&    3.468E-04& $^{24}$Mg&    7.069E-02& $^{25}$Mg&    3.046E-05& $^{26}$Mg&    1.878E-05& $^{26}$Al&    9.127E-07\\ 
$^{27}$Al&    4.785E-04& $^{28}$Si&    9.815E-02& $^{29}$Si&    2.423E-04& $^{30}$Si&    3.682E-05& $^{31}$P &    2.750E-05\\ 
$^{32}$S &    4.118E-02& $^{33}$S &    8.979E-05& $^{34}$S &    2.003E-06& $^{36}$S &    3.230E-12& $^{35}$Cl&    3.677E-06\\ 
$^{37}$Cl&    7.208E-06& $^{36}$Ar&    6.904E-03& $^{38}$Ar&    4.062E-07& $^{40}$Ar&    7.280E-13& $^{39}$K &    9.696E-07\\ 
$^{40}$K &    3.596E-11& $^{41}$K &    7.139E-07& $^{40}$Ca&    6.124E-03& $^{42}$Ca&    9.075E-09& $^{43}$Ca&    2.692E-08\\ 
$^{44}$Ca&    4.355E-05& $^{46}$Ca&    1.022E-12& $^{48}$Ca&    2.924E-13& $^{45}$Sc&    3.251E-08& $^{46}$Ti&    3.250E-07\\ 
$^{47}$Ti&    2.290E-06& $^{48}$Ti&    1.205E-04& $^{49}$Ti&    2.106E-06& $^{50}$Ti&    3.776E-13& $^{50}$V &    3.424E-13\\ 
$^{51}$V &    2.800E-06& $^{50}$Cr&    5.588E-07& $^{52}$Cr&    1.190E-03& $^{53}$Cr&    4.325E-05& $^{54}$Cr&    2.661E-12\\ 
$^{55}$Mn&    3.223E-05& $^{54}$Fe&    4.456E-05& $^{56}$Fe&    7.000E-02& $^{57}$Fe&    1.120E-03& $^{58}$Fe&    5.539E-12\\ 
$^{59}$Co&    2.009E-05& $^{58}$Ni&    1.154E-04& $^{60}$Ni&    1.995E-03& $^{61}$Ni&    5.286E-05& $^{62}$Ni&    1.684E-05\\ 
$^{64}$Ni&    1.030E-12& $^{63}$Cu&    1.591E-06& $^{65}$Cu&    6.779E-07& $^{64}$Zn&    1.249E-04& $^{66}$Zn&    2.438E-06\\ 
$^{67}$Zn&    9.175E-09& $^{68}$Zn&    9.031E-08& $^{70}$Zn&    7.743E-13& $^{69}$Ga&    9.441E-09& $^{71}$Ga&    3.297E-12\\ 
$^{70}$Ge&    1.064E-08& $^{72}$Ge&    6.328E-12& $^{73}$Ge&    3.331E-12& $^{74}$Ge&    2.699E-12& &\\
\hline 
\end{tabular}
\end{center}
\caption{Yields in the ejecta in $M_\odot$ after radio active 
decay (except $^{26}$Al) for the 20$M_\odot$ $E_{51}=1$ 
model shown in Table 4. ($M_{\rm cut}(i)$,  $M_{\rm Si}$, 
$M_{\rm cut}(f)$)=(2.10, 2.88, 2.73), [O/Fe]=0.43 and [Zn/Fe]=0.02.  
}
\end{table}

\begin{table}
\begin{center}
\renewcommand{\arraystretch}{1.2}
\begin{tabular}{rcrcrcrcrc} \hline\hline
     p   &    1.081E+01&  d       &    6.294E-17& $^3$He   &    3.929E-05& $^4$He   &    7.929E+00& $^6$Li   &    7.899E-22\\ 
$^7$Li   &    3.344E-10& $^9$Be   &    1.223E-19& $^{10}$B &    6.319E-10& $^{11}$B &    2.534E-09& $^{12}$C &    6.138E-01\\ 
$^{13}$C &    4.218E-08& $^{14}$N &    3.583E-04& $^{15}$N &    8.050E-08& $^{16}$O &    2.177E+00& $^{17}$O &    1.707E-07\\ 
$^{18}$O &    1.964E-06& $^{19}$F &    4.055E-10& $^{20}$Ne&    2.034E-01& $^{21}$Ne&    2.129E-05& $^{22}$Ne&    1.302E-05\\ 
$^{23}$Na&    9.229E-04& $^{24}$Mg&    1.172E-01& $^{25}$Mg&    8.523E-05& $^{26}$Mg&    6.545E-05& $^{26}$Al&    2.251E-06\\ 
$^{27}$Al&    1.115E-03& $^{28}$Si&    1.631E-01& $^{29}$Si&    4.945E-04& $^{30}$Si&    9.131E-05& $^{31}$P &    6.875E-05\\ 
$^{32}$S &    6.761E-02& $^{33}$S &    1.790E-04& $^{34}$S &    4.339E-05& $^{36}$S &    8.673E-11& $^{35}$Cl&    1.369E-05\\ 
$^{37}$Cl&    1.429E-05& $^{36}$Ar&    1.093E-02& $^{38}$Ar&    1.826E-05& $^{40}$Ar&    1.641E-12& $^{39}$K &    5.902E-06\\ 
$^{40}$K &    6.253E-10& $^{41}$K &    1.605E-06& $^{40}$Ca&    9.370E-03& $^{42}$Ca&    3.438E-07& $^{43}$Ca&    8.692E-08\\ 
$^{44}$Ca&    7.142E-05& $^{46}$Ca&    4.504E-13& $^{48}$Ca&    4.037E-13& $^{45}$Sc&    5.783E-08& $^{46}$Ti&    1.360E-07\\ 
$^{47}$Ti&    3.508E-07& $^{48}$Ti&    1.714E-04& $^{49}$Ti&    3.020E-06& $^{50}$Ti&    3.396E-13& $^{50}$V &    1.741E-12\\ 
$^{51}$V &    1.412E-06& $^{50}$Cr&    1.614E-06& $^{52}$Cr&    1.558E-03& $^{53}$Cr&    6.067E-05& $^{54}$Cr&    4.511E-11\\ 
$^{55}$Mn&    7.805E-05& $^{54}$Fe&    2.121E-04& $^{56}$Fe&    8.941E-02& $^{57}$Fe&    1.640E-03& $^{58}$Fe&    7.253E-11\\ 
$^{59}$Co&    6.672E-06& $^{58}$Ni&    1.092E-04& $^{60}$Ni&    2.385E-03& $^{61}$Ni&    8.696E-05& $^{62}$Ni&    7.567E-05\\ 
$^{64}$Ni&    4.090E-12& $^{63}$Cu&    9.255E-07& $^{65}$Cu&    1.211E-06& $^{64}$Zn&    1.094E-04& $^{66}$Zn&    4.751E-06\\ 
$^{67}$Zn&    3.077E-09& $^{68}$Zn&    1.742E-08& $^{70}$Zn&    1.304E-12& $^{69}$Ga&    3.980E-09& $^{71}$Ga&    4.263E-12\\ 
$^{70}$Ge&    1.597E-08& $^{72}$Ge&    1.003E-11& $^{73}$Ge&    4.500E-12& $^{74}$Ge&    1.840E-12& &\\
\hline 
\end{tabular}
\end{center}
\caption{Yields in the ejecta in $M_\odot$ after radio active 
decay (except $^{26}$Al) for the 25$M_\odot$ $E_{51}=1$ 
model shown in Table 4. ($M_{\rm cut}(i)$,  $M_{\rm Si}$, 
$M_{\rm cut}(f)$)=(2.20, 3.00, 2.80), [O/Fe]=0.50 and [Zn/Fe]=$-0.13$.  
}
\end{table}

\begin{table}
\begin{center}
\renewcommand{\arraystretch}{1.2}
\begin{tabular}{rcrcrcrcrc} \hline\hline
     p   &    1.083E+01&  d       &    4.890E-16& $^3$He   &    3.931E-05& $^4$He   &    7.899E+00& $^6$Li   &    2.211E-18\\ 
$^7$Li   &    3.320E-10& $^9$Be   &    1.189E-21& $^{10}$B &    6.579E-10& $^{11}$B &    2.955E-09& $^{12}$C &    3.987E-01\\ 
$^{13}$C &    4.442E-08& $^{14}$N &    7.193E-04& $^{15}$N &    3.131E-06& $^{16}$O &    1.534E+00& $^{17}$O &    1.598E-07\\ 
$^{18}$O &    2.115E-05& $^{19}$F &    5.244E-07& $^{20}$Ne&    1.075E-01& $^{21}$Ne&    4.190E-06& $^{22}$Ne&    6.660E-06\\ 
$^{23}$Na&    4.155E-04& $^{24}$Mg&    1.329E-01& $^{25}$Mg&    1.276E-04& $^{26}$Mg&    2.536E-05& $^{26}$Al&    1.960E-06\\ 
$^{27}$Al&    1.354E-03& $^{28}$Si&    1.981E-01& $^{29}$Si&    5.011E-04& $^{30}$Si&    1.711E-04& $^{31}$P &    9.133E-05\\ 
$^{32}$S &    7.179E-02& $^{33}$S &    2.893E-04& $^{34}$S &    5.153E-05& $^{36}$S &    6.776E-11& $^{35}$Cl&    2.316E-05\\ 
$^{37}$Cl&    3.447E-05& $^{36}$Ar&    1.028E-02& $^{38}$Ar&    2.580E-05& $^{40}$Ar&    1.052E-11& $^{39}$K &    1.057E-05\\ 
$^{40}$K &    1.268E-09& $^{41}$K &    4.468E-06& $^{40}$Ca&    6.155E-03& $^{42}$Ca&    5.070E-07& $^{43}$Ca&    1.042E-07\\ 
$^{44}$Ca&    8.673E-05& $^{46}$Ca&    2.959E-11& $^{48}$Ca&    1.654E-11& $^{45}$Sc&    9.992E-08& $^{46}$Ti&    2.076E-06\\ 
$^{47}$Ti&    7.047E-06& $^{48}$Ti&    1.334E-04& $^{49}$Ti&    4.576E-07& $^{50}$Ti&    1.791E-11& $^{50}$V &    2.659E-11\\ 
$^{51}$V &    1.148E-05& $^{50}$Cr&    2.797E-06& $^{52}$Cr&    8.058E-04& $^{53}$Cr&    1.356E-05& $^{54}$Cr&    1.085E-10\\ 
$^{55}$Mn&    8.665E-06& $^{54}$Fe&    2.672E-05& $^{56}$Fe&    7.000E-02& $^{57}$Fe&    1.422E-03& $^{58}$Fe&    1.808E-10\\ 
$^{59}$Co&    1.297E-04& $^{58}$Ni&    3.767E-04& $^{60}$Ni&    2.190E-03& $^{61}$Ni&    5.001E-05& $^{62}$Ni&    2.136E-05\\ 
$^{64}$Ni&    4.488E-11& $^{63}$Cu&    4.671E-06& $^{65}$Cu&    6.493E-07& $^{64}$Zn&    2.319E-04& $^{66}$Zn&    3.467E-06\\ 
$^{67}$Zn&    5.623E-08& $^{68}$Zn&    1.101E-07& $^{70}$Zn&    2.816E-11& $^{69}$Ga&    1.237E-08& $^{71}$Ga&    9.127E-11\\ 
$^{70}$Ge&    1.412E-08& $^{72}$Ge&    9.611E-11& $^{73}$Ge&    1.201E-10& $^{74}$Ge&    5.454E-11& &\\
\hline 
\end{tabular}
\end{center}
\caption{Yields in the ejecta in $M_\odot$ after radio active 
decay (except $^{26}$Al) for the 25$M_\odot$ $E_{51}=10$ 
model shown in Table 4. ($M_{\rm cut}(i)$,  $M_{\rm Si}$, 
$M_{\rm cut}(f)$)=(2.20, 3.86, 3.74), [O/Fe]=0.45 and [Zn/Fe]=0.30. 
}
\end{table}

\begin{table}
\begin{center}
\renewcommand{\arraystretch}{1.2}
\begin{tabular}{rcrcrcrcrc} \hline\hline
     p   &    1.083E+01&  d       &    1.268E-16& $^3$He   &    3.931E-05& $^4$He   &    7.783E+00& $^6$Li   &    2.978E-19\\ 
$^7$Li   &    3.319E-10& $^9$Be   &    3.498E-22& $^{10}$B &    6.588E-10& $^{11}$B &    2.963E-09& $^{12}$C &    1.659E-01\\ 
$^{13}$C &    7.908E-09& $^{14}$N &    2.099E-03& $^{15}$N &    1.527E-04& $^{16}$O &    1.030E+00& $^{17}$O &    5.606E-08\\ 
$^{18}$O &    8.921E-05& $^{19}$F &    2.381E-06& $^{20}$Ne&    5.912E-02& $^{21}$Ne&    1.319E-06& $^{22}$Ne&    2.799E-06\\ 
$^{23}$Na&    2.001E-04& $^{24}$Mg&    1.226E-01& $^{25}$Mg&    3.482E-05& $^{26}$Mg&    4.364E-06& $^{26}$Al&    5.487E-05\\ 
$^{27}$Al&    1.284E-03& $^{28}$Si&    3.412E-01& $^{29}$Si&    1.363E-03& $^{30}$Si&    3.504E-04& $^{31}$P &    3.646E-04\\ 
$^{32}$S &    1.571E-01& $^{33}$S &    3.029E-04& $^{34}$S &    5.753E-05& $^{36}$S &    1.978E-11& $^{35}$Cl&    5.579E-05\\ 
$^{37}$Cl&    6.283E-05& $^{36}$Ar&    2.352E-02& $^{38}$Ar&    2.906E-05& $^{40}$Ar&    8.835E-12& $^{39}$K &    1.701E-05\\ 
$^{40}$K &    1.299E-09& $^{41}$K &    1.001E-05& $^{40}$Ca&    1.500E-02& $^{42}$Ca&    8.096E-07& $^{43}$Ca&    2.856E-07\\ 
$^{44}$Ca&    1.286E-04& $^{46}$Ca&    9.054E-11& $^{48}$Ca&    1.528E-11& $^{45}$Sc&    5.722E-07& $^{46}$Ti&    3.519E-06\\ 
$^{47}$Ti&    1.033E-05& $^{48}$Ti&    1.758E-04& $^{49}$Ti&    2.926E-06& $^{50}$Ti&    1.574E-10& $^{50}$V &    9.977E-11\\ 
$^{51}$V &    2.236E-05& $^{50}$Cr&    7.384E-06& $^{52}$Cr&    8.749E-04& $^{53}$Cr&    4.865E-06& $^{54}$Cr&    3.043E-10\\ 
$^{55}$Mn&    1.670E-05& $^{54}$Fe&    2.406E-05& $^{56}$Fe&    7.000E-02& $^{57}$Fe&    1.241E-03& $^{58}$Fe&    5.608E-10\\ 
$^{59}$Co&    3.471E-04& $^{58}$Ni&    8.285E-04& $^{60}$Ni&    2.258E-03& $^{61}$Ni&    3.926E-05& $^{62}$Ni&    2.805E-05\\ 
$^{64}$Ni&    1.465E-10& $^{63}$Cu&    9.108E-06& $^{65}$Cu&    6.024E-07& $^{64}$Zn&    3.128E-04& $^{66}$Zn&    4.406E-06\\ 
$^{67}$Zn&    2.284E-07& $^{68}$Zn&    1.078E-07& $^{70}$Zn&    8.656E-11& $^{69}$Ga&    1.440E-08& $^{71}$Ga&    8.630E-10\\ 
$^{70}$Ge&    1.696E-08& $^{72}$Ge&    8.892E-10& $^{73}$Ge&    6.904E-10& $^{74}$Ge&    3.390E-10& &\\
\hline 
\end{tabular}
\end{center}
\caption{Yields in the ejecta in $M_\odot$ after radio active 
decay (except $^{26}$Al) for the 25$M_\odot$ $E_{51}=30$ 
model shown in Table 4.  ($M_{\rm cut}(i)$,  $M_{\rm Si}$, 
$M_{\rm cut}(f)$)=(2.50, 4.52, 4.39), [O/Fe]=0.28 and [Zn/Fe]=0.43.  
}
\end{table}

\begin{table}
\begin{center}
\renewcommand{\arraystretch}{1.2}
\begin{tabular}{rcrcrcrcrc} \hline\hline
     p   &    1.167E+01&  d       &    2.042E-16& $^3$He   &    2.106E-05& $^4$He   &    8.776E+00& $^6$Li   &    6.170E-19\\ 
$^7$Li   &    2.938E-10& $^9$Be   &    6.545E-18& $^{10}$B &    5.577E-15& $^{11}$B &    3.274E-15& $^{12}$C &    3.561E-01\\ 
$^{13}$C &    1.733E-08& $^{14}$N &    1.944E-04& $^{15}$N &    1.412E-05& $^{16}$O &    4.792E+00& $^{17}$O &    7.218E-07\\ 
$^{18}$O &    4.632E-05& $^{19}$F &    1.416E-05& $^{20}$Ne&    2.720E-01& $^{21}$Ne&    4.170E-05& $^{22}$Ne&    5.711E-04\\ 
$^{23}$Na&    3.664E-04& $^{24}$Mg&    2.379E-01& $^{25}$Mg&    3.205E-04& $^{26}$Mg&    1.495E-04& $^{26}$Al&    2.903E-06\\ 
$^{27}$Al&    5.031E-03& $^{28}$Si&    2.450E-01& $^{29}$Si&    1.304E-03& $^{30}$Si&    9.735E-04& $^{31}$P &    3.842E-04\\ 
$^{32}$S &    1.011E-01& $^{33}$S &    2.925E-04& $^{34}$S &    5.197E-04& $^{36}$S &    1.018E-08& $^{35}$Cl&    5.177E-05\\ 
$^{37}$Cl&    1.380E-05& $^{36}$Ar&    1.711E-02& $^{38}$Ar&    7.218E-05& $^{40}$Ar&    7.808E-11& $^{39}$K &    9.494E-06\\ 
$^{40}$K &    2.788E-09& $^{41}$K &    1.520E-06& $^{40}$Ca&    1.634E-02& $^{42}$Ca&    1.014E-06& $^{43}$Ca&    6.962E-08\\ 
$^{44}$Ca&    1.488E-04& $^{46}$Ca&    7.124E-12& $^{48}$Ca&    1.401E-14& $^{45}$Sc&    8.049E-08& $^{46}$Ti&    2.160E-07\\ 
$^{47}$Ti&    4.430E-07& $^{48}$Ti&    3.153E-04& $^{49}$Ti&    5.405E-06& $^{50}$Ti&    9.707E-13& $^{50}$V &    1.146E-11\\ 
$^{51}$V &    3.209E-06& $^{50}$Cr&    4.571E-06& $^{52}$Cr&    2.804E-03& $^{53}$Cr&    1.173E-04& $^{54}$Cr&    1.326E-10\\ 
$^{55}$Mn&    2.122E-04& $^{54}$Fe&    7.244E-04& $^{56}$Fe&    1.968E-01& $^{57}$Fe&    3.241E-03& $^{58}$Fe&    1.246E-10\\ 
$^{59}$Co&    9.233E-06& $^{58}$Ni&    2.733E-04& $^{60}$Ni&    5.399E-03& $^{61}$Ni&    2.044E-04& $^{62}$Ni&    1.344E-04\\ 
$^{64}$Ni&    3.041E-13& $^{63}$Cu&    8.295E-07& $^{65}$Cu&    3.492E-06& $^{64}$Zn&    2.159E-04& $^{66}$Zn&    9.159E-06\\ 
$^{67}$Zn&    3.475E-09& $^{68}$Zn&    6.713E-08& $^{70}$Zn&    2.637E-13& $^{69}$Ga&    1.018E-08& $^{71}$Ga&    7.251E-13\\ 
$^{70}$Ge&    2.994E-08& $^{72}$Ge&    1.351E-12& $^{73}$Ge&    1.038E-12& $^{74}$Ge&    5.457E-13& &\\
\hline 
\end{tabular}
\end{center}
\caption{Yields in the ejecta in $M_\odot$ after radio active 
decay (except $^{26}$Al) for the 30$M_\odot$ $E_{51}=1$ 
model shown in Table 4.  ($M_{\rm cut}(i)$,  $M_{\rm Si}$, 
$M_{\rm cut}(f)$)=(2.26, 3.66, 3.29), [O/Fe]=0.50 and [Zn/Fe]=$-0.18$.  
}
\end{table}

\begin{table}
\begin{center}
\renewcommand{\arraystretch}{1.2}
\begin{tabular}{rcrcrcrcrc} \hline\hline
     p   &    2.148E+02&  d       &    1.664E-15& $^3$He   &    1.821E-03& $^4$He   &    7.543E+01& $^6$Li   &    1.752E-17\\ 
$^7$Li   &    2.966E-10& $^9$Be   &    2.780E-21& $^{10}$B &    9.314E-15& $^{11}$B &    1.612E-12& $^{12}$C &    1.704E-01\\ 
$^{13}$C &    3.273E-08& $^{14}$N &    9.203E-05& $^{15}$N &    4.360E-05& $^{16}$O &    2.761E+00& $^{17}$O &    1.554E-07\\ 
$^{18}$O &    9.132E-05& $^{19}$F &    1.098E-06& $^{20}$Ne&    1.011E-01& $^{21}$Ne&    2.124E-05& $^{22}$Ne&    1.144E-04\\ 
$^{23}$Na&    4.131E-04& $^{24}$Mg&    1.303E-01& $^{25}$Mg&    2.990E-04& $^{26}$Mg&    2.649E-04& $^{26}$Al&    1.703E-05\\ 
$^{27}$Al&    1.134E-03& $^{28}$Si&    3.708E-01& $^{29}$Si&    1.170E-03& $^{30}$Si&    3.732E-04& $^{31}$P &    3.798E-04\\ 
$^{32}$S &    2.081E-01& $^{33}$S &    4.947E-04& $^{34}$S &    7.353E-04& $^{36}$S &    3.626E-09& $^{35}$Cl&    2.263E-04\\ 
$^{37}$Cl&    5.457E-05& $^{36}$Ar&    3.456E-02& $^{38}$Ar&    2.861E-04& $^{40}$Ar&    9.876E-11& $^{39}$K &    4.137E-05\\ 
$^{40}$K &    6.983E-09& $^{41}$K &    8.493E-06& $^{40}$Ca&    2.449E-02& $^{42}$Ca&    5.354E-06& $^{43}$Ca&    1.699E-07\\ 
$^{44}$Ca&    1.745E-04& $^{46}$Ca&    3.572E-11& $^{48}$Ca&    7.367E-12& $^{45}$Sc&    4.161E-07& $^{46}$Ti&    4.628E-06\\ 
$^{47}$Ti&    1.373E-05& $^{48}$Ti&    2.510E-04& $^{49}$Ti&    3.173E-06& $^{50}$Ti&    7.185E-11& $^{50}$V &    1.023E-10\\ 
$^{51}$V &    2.285E-05& $^{50}$Cr&    7.364E-06& $^{52}$Cr&    1.489E-03& $^{53}$Cr&    5.253E-05& $^{54}$Cr&    4.736E-10\\ 
$^{55}$Mn&    6.651E-05& $^{54}$Fe&    6.336E-04& $^{56}$Fe&    1.127E-01& $^{57}$Fe&    2.227E-03& $^{58}$Fe&    3.411E-10\\ 
$^{59}$Co&    2.753E-04& $^{58}$Ni&    1.038E-03& $^{60}$Ni&    3.353E-03& $^{61}$Ni&    6.804E-05& $^{62}$Ni&    3.645E-05\\ 
$^{64}$Ni&    7.030E-11& $^{63}$Cu&    9.326E-06& $^{65}$Cu&    1.050E-06& $^{64}$Zn&    4.082E-04& $^{66}$Zn&    6.260E-06\\ 
$^{67}$Zn&    1.546E-07& $^{68}$Zn&    1.949E-07& $^{70}$Zn&    7.280E-11& $^{69}$Ga&    2.222E-08& $^{71}$Ga&    2.880E-10\\ 
$^{70}$Ge&    2.167E-08& $^{72}$Ge&    1.853E-10& $^{73}$Ge&    3.036E-10& $^{74}$Ge&    1.292E-10& &\\
\hline 
\end{tabular}
\end{center}
\caption{Yields in the ejecta in $M_\odot$ after radio active 
decay (except $^{26}$Al) for the 30$M_\odot$ $E_{51}=30$ 
model shown in Table 4.  ($M_{\rm cut}(i)$,  $M_{\rm Si}$, 
$M_{\rm cut}(f)$)=(2.83, 6.01, 5.81), [O/Fe]=0.50 and [Zn/Fe]=0.34.  
}
\end{table}

\begin{table}
\begin{center}
\renewcommand{\arraystretch}{1.2}
\begin{tabular}{rcrcrcrcrc} \hline\hline
     p   &    1.157E+01&  d       &    9.904E-16& $^3$He   &    2.000E-05& $^4$He   &    8.755E+00& $^6$Li   &    8.284E-18\\ 
$^7$Li   &    2.985E-10& $^9$Be   &    0.000E+00& $^{10}$B &    4.046E-16& $^{11}$B &    2.323E-12& $^{12}$C &    1.136E-01\\ 
$^{13}$C &    1.958E-08& $^{14}$N &    4.340E-05& $^{15}$N &    4.765E-05& $^{16}$O &    2.119E+00& $^{17}$O &    1.400E-07\\ 
$^{18}$O &    1.152E-04& $^{19}$F &    1.921E-06& $^{20}$Ne&    5.777E-02& $^{21}$Ne&    6.114E-06& $^{22}$Ne&    6.427E-05\\ 
$^{23}$Na&    2.285E-04& $^{24}$Mg&    8.563E-02& $^{25}$Mg&    2.349E-04& $^{26}$Mg&    1.961E-04& $^{26}$Al&    2.037E-05\\ 
$^{27}$Al&    6.201E-04& $^{28}$Si&    2.406E-01& $^{29}$Si&    5.870E-04& $^{30}$Si&    2.531E-04& $^{31}$P &    3.291E-04\\ 
$^{32}$S &    1.575E-01& $^{33}$S &    4.397E-04& $^{34}$S &    2.807E-04& $^{36}$S &    3.732E-09& $^{35}$Cl&    4.499E-04\\ 
$^{37}$Cl&    8.097E-05& $^{36}$Ar&    4.998E-02& $^{38}$Ar&    1.899E-04& $^{40}$Ar&    6.076E-11& $^{39}$K &    2.726E-04\\ 
$^{40}$K &    6.351E-09& $^{41}$K &    1.433E-05& $^{40}$Ca&    1.292E-02& $^{42}$Ca&    5.425E-06& $^{43}$Ca&    7.554E-07\\ 
$^{44}$Ca&    1.674E-04& $^{46}$Ca&    1.666E-11& $^{48}$Ca&    4.226E-12& $^{45}$Sc&    5.962E-07& $^{46}$Ti&    4.129E-06\\ 
$^{47}$Ti&    1.076E-05& $^{48}$Ti&    2.311E-04& $^{49}$Ti&    4.454E-06& $^{50}$Ti&    5.385E-11& $^{50}$V &    6.394E-11\\ 
$^{51}$V &    2.065E-05& $^{50}$Cr&    9.525E-06& $^{52}$Cr&    1.083E-03& $^{53}$Cr&    3.418E-05& $^{54}$Cr&    6.767E-10\\ 
$^{55}$Mn&    3.088E-05& $^{54}$Fe&    2.110E-04& $^{56}$Fe&    8.679E-02& $^{57}$Fe&    1.705E-03& $^{58}$Fe&    1.006E-09\\ 
$^{59}$Co&    3.387E-04& $^{58}$Ni&    8.460E-04& $^{60}$Ni&    2.854E-03& $^{61}$Ni&    5.241E-05& $^{62}$Ni&    3.304E-05\\ 
$^{64}$Ni&    8.406E-11& $^{63}$Cu&    1.003E-05& $^{65}$Cu&    9.195E-07& $^{64}$Zn&    3.909E-04& $^{66}$Zn&    6.189E-06\\ 
$^{67}$Zn&    2.463E-07& $^{68}$Zn&    1.167E-07& $^{70}$Zn&    3.716E-11& $^{69}$Ga&    1.707E-08& $^{71}$Ga&    4.113E-10\\ 
$^{70}$Ge&    1.906E-08& $^{72}$Ge&    3.417E-10& $^{73}$Ge&    2.320E-10& $^{74}$Ge&    2.510E-10& &\\
\hline 
\end{tabular}
\end{center}
\caption{Yields in the ejecta in $M_\odot$ after radio active 
decay (except $^{26}$Al) for the 30$M_\odot$ $E_{51}=50$ 
model shown in Table 4. ($M_{\rm cut}(i)$,  $M_{\rm Si}$, 
$M_{\rm cut}(f)$)=(2.83, 6.01, 5.81), [O/Fe]=0.50 and [Zn/Fe]=0.43.  
}
\end{table}

\clearpage

\begin{table}   
\begin{center}
\renewcommand{\arraystretch}{1.2}
\begin{tabular}{ccccc} \hline\hline
$M$ & 150 & 170 & 200& 270  \cr 
$M$(He) & 70.0 & 82.3 & 117& 129  \cr 
$M$(CO) & 62.2 & 72.7 & 109& 121  \cr 
$M(^{56}$Ni) & 0.0 & 3.6 & 7.2  & 9.8  \cr 
 \hline 
\end{tabular}
\end{center}
\caption{The initial, He core and C-O core masses of the progenitors
of PISN models. The ejected $^{56}$Ni masses are also shown.
Units are all in $M_\odot$.}
\end{table}

\begin{table}
\begin{center}
\renewcommand{\arraystretch}{1.2}
\begin{tabular}{rcrcrcrcrc} \hline\hline
     p   &    4.041E+01&  d       &    3.232E-16& $^3$He   &    3.089E-05& $^4$He   &    4.727E+01& $^6$Li   &    2.159E-19\\ 
$^7$Li   &    3.363E-10& $^9$Be   &    9.930E-22& $^{10}$B &    1.428E-14& $^{11}$B &    7.713E-14& $^{12}$C &    7.168E+00\\ 
$^{13}$C &    3.693E-07& $^{14}$N &    1.614E-02& $^{15}$N &    7.005E-07& $^{16}$O &    6.013E+01& $^{17}$O &    4.440E-05\\ 
$^{18}$O &    2.863E-07& $^{19}$F &    1.184E-08& $^{20}$Ne&    3.835E+00& $^{21}$Ne&    9.979E-04& $^{22}$Ne&    1.008E-03\\ 
$^{23}$Na&    1.062E-02& $^{24}$Mg&    3.127E+00& $^{25}$Mg&    4.840E-03& $^{26}$Mg&    2.544E-03& $^{26}$Al&    8.115E-05\\ 
$^{27}$Al&    3.181E-02& $^{28}$Si&    5.401E+00& $^{29}$Si&    2.520E-02& $^{30}$Si&    3.765E-03& $^{31}$P &    2.568E-03\\ 
$^{32}$S &    2.046E+00& $^{33}$S &    7.220E-03& $^{34}$S &    6.565E-03& $^{36}$S &    5.677E-08& $^{35}$Cl&    7.060E-04\\ 
$^{37}$Cl&    6.713E-04& $^{36}$Ar&    2.664E-01& $^{38}$Ar&    3.950E-03& $^{40}$Ar&    2.340E-10& $^{39}$K &    9.059E-04\\ 
$^{40}$K &    3.787E-08& $^{41}$K &    1.532E-04& $^{40}$Ca&    2.134E-01& $^{42}$Ca&    9.721E-05& $^{43}$Ca&    8.050E-08\\ 
$^{44}$Ca&    3.212E-05& $^{46}$Ca&    2.451E-11& $^{48}$Ca&    4.988E-12& $^{45}$Sc&    3.113E-06& $^{46}$Ti&    3.836E-05\\ 
$^{47}$Ti&    1.984E-07& $^{48}$Ti&    4.018E-05& $^{49}$Ti&    4.823E-06& $^{50}$Ti&    5.084E-12& $^{50}$V &    1.164E-10\\ 
$^{51}$V &    6.846E-06& $^{50}$Cr&    1.284E-04& $^{52}$Cr&    2.899E-04& $^{53}$Cr&    3.416E-05& $^{54}$Cr&    4.551E-09\\ 
$^{55}$Mn&    3.104E-04& $^{54}$Fe&    4.432E-03& $^{56}$Fe&    5.988E-03& $^{57}$Fe&    7.971E-05& $^{58}$Fe&    6.077E-09\\ 
$^{59}$Co&    2.640E-06& $^{58}$Ni&    4.926E-04& $^{60}$Ni&    4.680E-06& $^{61}$Ni&    1.076E-09& $^{62}$Ni&    1.772E-10\\ 
$^{64}$Ni&    2.065E-11& $^{63}$Cu&    8.241E-11& $^{65}$Cu&    9.884E-11& $^{64}$Zn&    1.496E-10& $^{66}$Zn&    2.046E-10\\ 
$^{67}$Zn&    9.381E-11& $^{68}$Zn&    1.758E-10& $^{70}$Zn&    2.721E-11& $^{69}$Ga&    9.451E-11& $^{71}$Ga&    3.892E-11\\
$^{70}$Ge&    1.726E-10& $^{72}$Ge&    6.959E-11& $^{73}$Ge&    3.882E-11& $^{74}$Ge&    2.440E-11\\ 
\hline 
\end{tabular}
\end{center}
\caption{Yields in the ejecta in $M_\odot$ after radio active 
decay (except $^{26}$Al) for the 150$M_\odot$ PISN model.}
\end{table}

\begin{table}
\begin{center}
\renewcommand{\arraystretch}{1.2}
\begin{tabular}{rcrcrcrcrc} \hline\hline
     p   &    4.081E+01&  d       &    3.667E-05& $^3$He   &    1.122E-04& $^4$He   &    4.795E+01& $^6$Li   &    3.928E-15\\ 
$^7$Li   &    5.766E-09& $^9$Be   &    1.291E-14& $^{10}$B &    6.640E-22& $^{11}$B &    8.212E-09& $^{12}$C &    2.297E+00\\ 
$^{13}$C &    3.849E-07& $^{14}$N &    1.039E-02& $^{15}$N &    4.813E-06& $^{16}$O &    4.423E+01& $^{17}$O &    4.546E-05\\ 
$^{18}$O &    9.985E-03& $^{19}$F &    6.489E-07& $^{20}$Ne&    1.190E+00& $^{21}$Ne&    1.632E-04& $^{22}$Ne&    1.021E-03\\ 
$^{23}$Na&    5.762E-03& $^{24}$Mg&    1.943E+00& $^{25}$Mg&    1.264E-03& $^{26}$Mg&    3.624E-04& $^{26}$Al&    1.248E-05\\ 
$^{27}$Al&    2.038E-02& $^{28}$Si&    1.616E+01& $^{29}$Si&    2.192E-02& $^{30}$Si&    2.043E-04& $^{31}$P &    4.027E-04\\ 
$^{32}$S &    8.663E+00& $^{33}$S &    1.221E-02& $^{34}$S &    7.065E-04& $^{36}$S &    6.776E-10& $^{35}$Cl&    3.045E-04\\ 
$^{37}$Cl&    1.608E-03& $^{36}$Ar&    1.419E+00& $^{38}$Ar&    8.338E-04& $^{40}$Ar&    1.036E-11& $^{39}$K &    5.029E-04\\ 
$^{40}$K &    1.932E-08& $^{41}$K &    4.032E-04& $^{40}$Ca&    1.319E+00& $^{42}$Ca&    2.452E-05& $^{43}$Ca&    1.136E-08\\ 
$^{44}$Ca&    2.037E-04& $^{46}$Ca&    1.413E-12& $^{48}$Ca&    1.539E-16& $^{45}$Sc&    7.492E-06& $^{46}$Ti&    1.350E-05\\ 
$^{47}$Ti&    3.035E-07& $^{48}$Ti&    6.251E-03& $^{49}$Ti&    1.890E-04& $^{50}$Ti&    4.479E-14& $^{50}$V &    3.071E-11\\ 
$^{51}$V &    1.992E-04& $^{50}$Cr&    4.818E-04& $^{52}$Cr&    1.335E-01& $^{53}$Cr&    5.560E-03& $^{54}$Cr&    6.950E-09\\ 
$^{55}$Mn&    1.795E-02& $^{54}$Fe&    1.089E-01& $^{56}$Fe&    3.630E+00& $^{57}$Fe&    1.768E-02& $^{58}$Fe&    1.137E-08\\ 
$^{59}$Co&    1.422E-05& $^{58}$Ni&    1.180E-02& $^{60}$Ni&    1.475E-05& $^{61}$Ni&    2.915E-09& $^{62}$Ni&    9.618E-10\\ 
$^{64}$Ni&    5.968E-15& $^{63}$Cu&    2.862E-11& $^{65}$Cu&    3.317E-13& $^{64}$Zn&    1.117E-10& $^{66}$Zn&    1.885E-11\\ 
$^{67}$Zn&    1.742E-13& $^{68}$Zn&    3.459E-12& $^{70}$Zn&    2.282E-15& $^{69}$Ga&    9.034E-14& $^{71}$Ga&    7.908E-15\\ 
$^{70}$Ge&    3.578E-11& $^{72}$Ge&    1.514E-12& $^{73}$Ge&    9.534E-14& $^{74}$Ge&    1.042E-14\\ 
\hline 
\end{tabular}
\end{center}
\caption{Same as Table 15, but for the 170$M_\odot$ PISN.
}
\end{table}

\begin{table}
\begin{center}
\renewcommand{\arraystretch}{1.2}
\begin{tabular}{rcrcrcrcrc} \hline\hline

\hline      p   &    3.725E+01&  d       &    2.038E-16& $^3$He   &    3.183E-05& $^4$He   &    4.903E+01& $^6$Li   &    1.583E-20\\ 
$^7$Li   &    5.789E-10& $^9$Be   &    2.838E-18& $^{10}$B &    2.027E-14& $^{11}$B &    1.441E-12& $^{12}$C &    4.241E+00\\ 
$^{13}$C &    3.491E-07& $^{14}$N &    5.839E-04& $^{15}$N &    1.114E-06& $^{16}$O &    5.595E+01& $^{17}$O &    7.757E-07\\ 
$^{18}$O &    1.054E-05& $^{19}$F &    9.477E-08& $^{20}$Ne&    3.748E+00& $^{21}$Ne&    2.001E-04& $^{22}$Ne&    2.251E-04\\ 
$^{23}$Na&    6.999E-03& $^{24}$Mg&    3.075E+00& $^{25}$Mg&    1.268E-03& $^{26}$Mg&    3.690E-04& $^{26}$Al&    3.775E-05\\ 
$^{27}$Al&    1.557E-02& $^{28}$Si&    2.122E+01& $^{29}$Si&    1.412E-02& $^{30}$Si&    9.665E-04& $^{31}$P &    7.493E-04\\ 
$^{32}$S &    1.310E+01& $^{33}$S &    6.068E-03& $^{34}$S &    1.085E-04& $^{36}$S &    4.821E-10& $^{35}$Cl&    1.266E-04\\ 
$^{37}$Cl&    8.106E-04& $^{36}$Ar&    2.361E+00& $^{38}$Ar&    1.949E-05& $^{40}$Ar&    4.287E-12& $^{39}$K &    4.590E-05\\ 
$^{40}$K &    1.618E-09& $^{41}$K &    1.947E-04& $^{40}$Ca&    2.316E+00& $^{42}$Ca&    7.254E-07& $^{43}$Ca&    2.025E-07\\ 
$^{44}$Ca&    4.717E-04& $^{46}$Ca&    4.228E-12& $^{48}$Ca&    2.751E-12& $^{45}$Sc&    8.101E-06& $^{46}$Ti&    9.282E-07\\ 
$^{47}$Ti&    3.490E-06& $^{48}$Ti&    1.148E-02& $^{49}$Ti&    3.354E-04& $^{50}$Ti&    2.769E-12& $^{50}$V &    8.221E-12\\ 
$^{51}$V &    1.308E-04& $^{50}$Cr&    1.232E-04& $^{52}$Cr&    1.795E-01& $^{53}$Cr&    6.882E-03& $^{54}$Cr&    5.898E-11\\ 
$^{55}$Mn&    8.977E-03& $^{54}$Fe&    2.332E-02& $^{56}$Fe&    7.249E+00& $^{57}$Fe&    8.856E-02& $^{58}$Fe&    8.554E-10\\ 
$^{59}$Co&    2.465E-03& $^{58}$Ni&    6.394E-02& $^{60}$Ni&    2.270E-02& $^{61}$Ni&    8.263E-04& $^{62}$Ni&    4.245E-03\\ 
$^{64}$Ni&    1.358E-11& $^{63}$Cu&    2.161E-05& $^{65}$Cu&    1.872E-06& $^{64}$Zn&    1.419E-04& $^{66}$Zn&    4.107E-05\\ 
$^{67}$Zn&    4.517E-08& $^{68}$Zn&    9.072E-09& $^{70}$Zn&    1.398E-11& $^{69}$Ga&    3.048E-09& $^{71}$Ga&    2.864E-11\\ 
$^{70}$Ge&    5.377E-08& $^{72}$Ge&    3.197E-11& $^{73}$Ge&    2.767E-11& $^{74}$Ge&    1.363E-11& &\\
\hline 
\end{tabular}
\end{center}
\caption{Same as Table 15, but for the 200$M_\odot$ PISN.
}
\end{table}

\begin{table}
\begin{center}
\renewcommand{\arraystretch}{1.2}
\begin{tabular}{rcrcrcrcrc} \hline\hline

\hline

     p   &    6.582E+01&  d       &    1.838E-13& $^3$He   &    3.955E-05&
$^4$He   &    7.901E+01& $^6$Li   &    9.328E-18\\
$^7$Li   &    5.522E-09& $^9$Be   &    4.396E-21& $^{10}$B &    3.622E-16&
$^{11}$B &    7.461E-09& $^{12}$C &    1.886E+00\\
$^{13}$C &    4.724E-04& $^{14}$N &    1.263E-02& $^{15}$N &    1.844E-02&
$^{16}$O &    4.431E+01& $^{17}$O &    1.503E-02\\
$^{18}$O &    4.010E-04& $^{19}$F &    2.038E-06& $^{20}$Ne&    4.697E+00&
$^{21}$Ne&    9.380E-04& $^{22}$Ne&    7.623E-04\\
$^{23}$Na&    7.159E-03& $^{24}$Mg&    4.778E+00& $^{25}$Mg&    1.821E-02&
$^{26}$Mg&    3.029E-03& $^{26}$Al&    2.504E-03\\
$^{27}$Al&    8.621E-02& $^{28}$Si&    2.695E+01& $^{29}$Si&    5.876E-02&
$^{30}$Si&    4.331E-02& $^{31}$P &    3.101E-02\\
$^{32}$S &    1.578E+01& $^{33}$S &    3.382E-02& $^{34}$S &    2.719E-02&
$^{36}$S &    7.981E-07& $^{35}$Cl&    1.874E-02\\
$^{37}$Cl&    2.464E-02& $^{36}$Ar&    2.602E+00& $^{38}$Ar&    3.902E-02&
$^{40}$Ar&    3.527E-09& $^{39}$K &    3.196E-03\\
$^{40}$K &    1.229E-08& $^{41}$K &    1.126E-03& $^{40}$Ca&    2.752E+00&
$^{42}$Ca&    1.085E-03& $^{43}$Ca&    1.544E-03\\
$^{44}$Ca&    9.011E-03& $^{46}$Ca&    4.766E-08& $^{48}$Ca&    1.307E-14&
$^{45}$Sc&    6.597E-04& $^{46}$Ti&    7.756E-04\\
$^{47}$Ti&    7.576E-04& $^{48}$Ti&    1.316E-02& $^{49}$Ti&    6.297E-04&
$^{50}$Ti&    1.593E-11& $^{50}$V &    5.282E-10\\
$^{51}$V &    1.253E-03& $^{50}$Cr&    6.641E-03& $^{52}$Cr&    2.561E-01&
$^{53}$Cr&    2.902E-02& $^{54}$Cr&    4.016E-06\\
$^{55}$Mn&    2.613E-01& $^{54}$Fe&    5.676E+00& $^{56}$Fe&    1.046E+01&
$^{57}$Fe&    2.978E-01& $^{58}$Fe&    3.447E-05\\
$^{59}$Co&    1.778E-02& $^{58}$Ni&    3.785E+00& $^{60}$Ni&    1.593E-01&
$^{61}$Ni&    2.826E-05& $^{62}$Ni&    1.708E-05\\
$^{64}$Ni&    1.203E-11& $^{63}$Cu&    8.094E-07& $^{65}$Cu&    8.436E-09&
$^{64}$Zn&    5.897E-06& $^{66}$Zn&    2.257E-08\\
$^{67}$Zn&    6.145E-09& $^{68}$Zn&    6.217E-09& $^{70}$Zn&    1.692E-12&
$^{69}$Ga&    3.784E-10& $^{71}$Ga&    2.183E-11\\
$^{70}$Ge&    1.174E-08& $^{72}$Ge&    1.222E-11& $^{73}$Ge&    4.086E-11&
$^{74}$Ge&    2.641E-12& &\\

\hline 
\end{tabular}
\end{center}
\caption{Same as Table 15, but for the 270$M_\odot$ PISN.
}
\end{table}


\begin{thebibliography}{}

\bibitem{}
Arnett, W.D. 1996, Supernovae and Nucleosynthesis 
(New Jersey: Princeton University Press)

\bibitem{}
Arnett, W.D., Bahcall, J.N., Kirshner, R.P., \& Woosley, S.E. 1989,
 ARA\&A, 27, 629 

\bibitem{}
Audouze, J., \& Silk, J. 1995, ApJ, 451, L49

\bibitem{}
Barkat, Z., Rakavy, G., \& Sack, N., 1967, Phys. Rev. Letters, 18, 379

\bibitem{}
Blake, L. A. J., Ryan S. G., Norris, J. E., \& Beers, T. C. 2001,
Nucl.Phys.A., 688, 502 

\bibitem{}
Blinnikov, S., Lundqvist, P., Bartunov, O., Nomoto, K.,
\& Iwamoto, K., 2000, ApJ, 532

\bibitem{}
B\"ohringer, H., \& Hensler, G. 1989, A\&A, 215, 147

\bibitem{}
Bromm, V., Coppi, P. S., \& Larson, R. B. 1999, ApJ, 527, L5


\bibitem{}
Castellani, V., Chieffi A., \& Tornamb\'e A. 1983, ApJ, 272, 249

\bibitem{}
Chevalier, R.A. 1989, ApJ, 346, 847

\bibitem{}
Ebisuzaki, T, Shigeyama, T., \& Nomoto, K. 1989, ApJ, 344, 65

\bibitem{}
Ellison, S. L., Ryan, S. G., \& Prochaska, J. X. 2001, MNRAS,
326, 628

\bibitem{}
Ezer, D., \& Cameron, A. G. W. 1971, Astrophys. Space Sci., 14, 399

\bibitem{}
Fuller, G. M., Fowler, W. A., \& Newman, M. J., 1980, ApJS, 42 447

\bibitem{}
----- 1982, ApJS, 48, 279

\bibitem{}
Galama, T. et al. 1998,  Nature, 395, 

\bibitem{}
Goswami, A. \& Prantzos, N. 2000, A\&A, 359, 191

\bibitem{}
Hachisu, I., Matsuda, T., Nomoto, K. 
\& Shigeyama, T. 1990, \apj, 358, L57

\bibitem{}
----- 1991, ApJ, 368, L27

\bibitem{}
Heger, A., Woosley, S. E., Mart\'inez-Pinedo, G., \&
Langanke, K. 2000, ApJ, submitted (astro-ph/0011507)

\bibitem{}
Hashimoto, M., Nomoto, K., Shigeyama, T. 1989, A\&A, 210, L5

\bibitem{}
Hix, W. R. \& Thielemann, F.-K. 1996, \apj, 460, 869

\bibitem{}
Hoffman, R. D., Woosley, S. E., Fuller, G. M., Meyer, B. S.,
\& Meyer, B. S. 1996, \apj, 460, 478

\bibitem{}
Hou, J.L., Boissier, S., \& Prantzos, N. 2001, 
A\&A, 370, 23


\bibitem{}
Iwamoto, K., Mazzali, P.A., Nomoto, K., Umeda, H., et al. 1998,
 Nature, 395, 672

\bibitem{}
Iwamoto, K., Nakamura, T., Nomoto, K. et al. 2000, ApJ, 534, 660

\bibitem{}
Khokhlov, A. M., H\"oflich, P. A., Oran, E. S., Wheeler, J. C., Wang, L.,
\& Chtchelkanova, A. Y. 1999, ApJ, 524, 107

\bibitem{}
Kifonidis, K. 2001, Ph.D. thesis, Max-Planck-Institut f\"ur Astrophysik  

\bibitem{}
Kifonidis, K., Plewa, T., Janka, H-Th, \& M\"uller, E. 2000,
 \apj, 531, L123

\bibitem{}
Langanke, K., \& Mart\'inez-Pinedo, G. 2000, Nucl. Phys. A, 673, 481

\bibitem{}
Limongi, M., Chieffi, A., \& Straniero, O. 1998, in Nuclei in the
Cosmos V, ed. N. Prantzos, \& S. Harissopulos 
(Paris: Editions Frontieres), 144

\bibitem{}
Limongi, M., Straniero, O., \& Chieffi, A. 2000, ApJS, 129, 625

\bibitem{}
Lu, L., Sargent, W. L. W., Barlow, T. A., Churchill,
C. W., \& Vogt, Steven S. 1996, ApJS, 107, 405

\bibitem{}
Maeda, K., Nakamura, T., Nomoto, K., Mazzali, P.A., Patat, F.,
\& Hachisu, I. 2002, \apj, 565, in press (astro-ph/0011003)

\bibitem{}
Maeda, K. 2001, Master's thesis, University of Tokyo

\bibitem{}
Matteucci, F. 2001, The Chemical Evolution of the Galaxy
(Dordrecht: Kluwer) 

\bibitem{}
Mezzacappa, A., Liebendoerfer, M., Messer, O. E. B., Hix, W. R.,
Thielemann, F.-K., \& Bruenn, S. W. 2000, 
Phys. Rev. Lett., 86, 1935

\bibitem{}
McWilliam, A., Preston, G. W., Sneden, C., \& Searle, L.
1995, \aj, 109, 2757

\bibitem{}
Molaro, P., Bonifacio, P., Centurion, M., D'Odorico, S., Vladilo, G., 
Santin, P., \& Di Marcantonio, P. 2000, ApJ, 541, 54

\bibitem{}
Nagataki, S., Hashimoto, M., Sato, K., \& Yamada, S. 1997, \apj, 
486, 1026

\bibitem{}
Nakamura, F., \& Umemura, M. 1999, \apj, 515, 239

\bibitem{}
Nakamura, T., Umeda, H., Nomoto, K., Thielemann, F.-K.,
\& Burrows, A. 1999, \apj, 
517, 193

\bibitem{}
Nakamura, T., Umeda, H., Iwamoto, K., Nomoto, K., Hashimoto, M.,
Hix, W.R., \& Thielemann, F.-K. 2001, \apj, 550, 880 

\bibitem{}
Nomoto, K. \& Hashimoto, M. 1988, Phys. Rep., 256, 173

\bibitem{}
Nomoto, K., Suzuki, T., Shigeyama, T., Kumagai, S., Yamaoka, H.,
 \& Saio, H. 1993, Nature, 364, 507

\bibitem{}
Nomoto, K., Yamaoka, H., Pols, O.R., van den Heuvel, E.P.J.,
 Iwamoto, K., Kumagai, S., \& Shigeyama, T. 1994, Nature, 371, 227

\bibitem{}
Nomoto, K., Hashimoto, M., Tsujimoto, T., Thielemann, F.-K.,
Kishimoto, N., Kubo, Y., \& Nakasato, N.
1997, Nucl. Phys. A616, 79

\bibitem{}
Nomoto, K., et al. 2000, in Supernovae and Gamma Ray Bursts,
ed. M. Livio 
et al. (Cambridge: Cambridge University Press), 144
(astro-ph/0003077)

\bibitem{}
Nomoto, K., Maeda, K., Umeda, H., \& Nakamura, T. 2001, 
in The Influence of Binaries on Stellar Populations Studies,
ed. D. Vanbeveren (Kluwer), 507 (astro-ph/0105127)

\bibitem{}
Norris, J.E., Ryan S.G., \& Beers, T.C. 2001,
ApJ, 561, 1034

\bibitem{}
Ober, W. W., El Eid, M. F., \& Fricke, K. J. 1983, A\&A, 119, 61

\bibitem{}
Omukai, K., \& Nishi, R. 1999, ApJ, 518, 64

\bibitem{}
Pettini, M., Ellison, S. L., Steidel, C. C., Bowen, D. V. 1999, ApJ,
510, 576

\bibitem{}
Primas, F., Reimers, D., Wisotzki, L., Reetz, J.,
Gehren, T., \& Beers, T. C.  2000, in 
The First Stars, ed. A. Weiss, T. Abel,
\& V. Hill (Berlin: Springer), 51

\bibitem{}
Prochaska, J. X., \& Wolfe, A. M. 1999, ApJS, 211, 369

\bibitem{}
Rampp, M., \& Janka, H.-T. 2000, \apj, 539, L33

\bibitem{}
Ryan, S. G., Norris, J. E., \& Beers, T. C. 1996, \apj, 471, 254

\bibitem{}
Ryan, S. G. 2001, in The Influence of Binaries on Stellar Populations Studies,
ed. D. Vanbeveren (Dordrecht: Kluwer), 491

\bibitem{}
Shigeyama, T., Suzuki, T., Kumagai, S., Nomoto, K., Saio, H., \&
Yamaoka, H. 1994, ApJ, 420, 341

\bibitem{}
Shigeyama, T., \& Tsujimoto, T. 1998, \apj, 507, L135

\bibitem{}
Sneden, C., Gratton, R. G., \& Crocker, D. A. 1991, A\&A, 246, 354

\bibitem{}
Thielemann, F.-K., Nomoto, K., \& Hashimoto, M. 1996, ApJ, 460, 408

\bibitem{}
Timmes, F. X., Woosley, S. E., \& Weaver, T. A. 1995, \apj, 98, 617

\bibitem{}
Tsujimoto, T., \& Shigeyama, T. 1998, \apj, 508, L151

\bibitem{}
Turatto, M., Suzuki, T., Mazzali, P.A., Benetti, S., Cappellaro, E.,
Nomoto, K., Nakamura, T., Young, T., \& Patat, F. 2000, ApJ, 534, L57

\bibitem{}
Umeda, H., Nomoto, K., \& Nakamura, T. 2000, in 
The First Stars, ed. A. Weiss, 
T. Abel, \& V. Hill (Berlin: Springer), 150 (astro-ph/9912248)

\bibitem{}
Umeda, H., \& Nomoto, K. 2002, in preparation 

\bibitem{}
Wheeler, J.C., Sneden, C., \& Truran, J.W. 1989, ARA\&A, 27, 279 

\bibitem{}
Weiss, A., Abel, T., \& Hill, V. 2000, ed. The First Stars  (Berlin: Springer)

\bibitem{}
Woosley, S. E., Eastman, R. G., \& Schmidt, B. P. 1999, ApJ, 516, 788

\bibitem{}
Woosley, S. E., \& Weaver, T. A. 1982, in Supernovae: A survey of
Current Research, eds. M. J. Rees \& R. J. Stoneham (Dordrecht: Reidel), 79

\bibitem{}
Woosley, S. E., \& Weaver, T. A. 1995, ApJS, 101, 181 (WW95)

\bibitem{}
Woosley, S. E., Wilson, J. R., Mathews, G. J., Hoffman, R. D., 
\& Meyer, B. S. 1994, ApJ, 433, 229

\end{thebibliography}
\end{document}